\documentclass[12pt]{article}

\usepackage[margin=0.7in]{geometry}
 \usepackage[comma]{natbib}
\usepackage[usenames, dvipsnames]{color}
\usepackage{xcolor,color}
\usepackage{graphicx}
\usepackage{amsmath,amssymb,amsthm,amsopn,mathrsfs}
\usepackage[pdftex,bookmarks,colorlinks]{hyperref}
\usepackage{caption}
\usepackage{subcaption}
\usepackage[applemac]{inputenc}
\usepackage{booktabs}
\usepackage[makeroom]{cancel}
\usepackage{dsfont}
\usepackage{xcolor,color}
\usepackage{epstopdf}
\usepackage{arydshln} % For dashed lines in table/array

\definecolor{navy}{rgb}{0,0,0.502}

\hypersetup{colorlinks,%
citecolor=navy,%
filecolor=navy,%
linkcolor=navy,%
urlcolor=navy,%
}

\theoremstyle{definition}
\theoremstyle{plain}\newtheorem{prop}{Proposition}\newtheorem{lem}{Lemma}
\newtheorem{cor}{Corollary}
\theoremstyle{definition}

\def\prob{{\mathbb P}}
\def\cB{\mathcal{B}}
\def\cC{\mathcal{C}}
\def\cL{\mathcal{L}}
\def\cX{\mathcal{X}}
\def\cY{\mathcal{Y}}
\def\cS{\mathcal{S}}
\def\cT{\mathcal{T}}
\newcommand{\bs}{{\boldsymbol{s}}}
\newcommand{\bb}{{\boldsymbol{b}}}
\newcommand{\bx}{{\boldsymbol{x}}}
\newcommand{\bz}{{\boldsymbol{z}}}
\newcommand{\bB}{{\boldsymbol{B}}}
\newcommand{\bX}{{\boldsymbol{X}}}
\def\der{{\mathrm{d}}}
\def\real{{\mathbb R}}
\def\indic{\mathds{1}}
\def\diag{\mbox{diag}}

\begin{document}

\title{New models for symbolic data analysis}

\author{B. Beranger\footnote{School of Mathematics and Statistics, University of New South Wales, Sydney.},\, H. Lin$^*$ and S. A. Sisson\footnote{Communicating author: Scott.Sisson@unsw.edu.au}\:\,$^*$}

\date{}

\maketitle

\begin{abstract}
Symbolic data analysis (SDA) is an emerging area of statistics concerned with understanding and modelling data that takes distributional form (i.e.~{\em symbols}), such as random lists, intervals and histograms. It was developed under the premise that the statistical unit of interest is the symbol, and that inference is required at this level. Here we consider a different perspective, which opens a new research direction in the field of SDA. We assume that, as with a standard statistical analysis, inference is required at the level of individual-level data. However, the individual-level data are aggregated into symbols -- group-based distributional-valued summaries -- prior to the analysis. In this way, large and complex datasets can be reduced to a smaller number of distributional summaries, that may be analysed more efficiently than the original dataset. As such, we develop SDA techniques as a new approach for the analysis of big data. In particular we introduce a new general method for constructing likelihood functions for symbolic data based on a desired probability model for the underlying measurement-level data, while only observing the distributional summaries. This approach opens the door for new classes of symbol design and construction, in addition to developing SDA as a viable tool to enable and improve upon classical data analyses, particularly for very large and complex datasets. We illustrate this new direction for SDA research through several real and simulated data analyses.\\

\end{abstract}

\section{Introduction}

Symbolic data analysis (SDA) is an emerging area of statistics that has immense potential to become a standard inferential technique in the near future \citep{billard2003statistics}.
At its core, it  builds on the notion that exploratory analyses and statistical inferences are commonly required at a group level rather than at an individual level \citep{billard2011brief,billard2006}. This is the familiar notion behind hierarchical modelling \citep[e.g.][Chapter 5]{bda}. For example, the performance of school and higher level units in standardised testing exams is usually of interest rather than the performance of the individual students \citep{rodrigues+ns16,rubin81}. 

SDA explicitly embraces this idea by 
considering group level distributional summaries (i.e.~{\em symbols}) as the statistical unit of interest, and then analysing the data at this summary level \citep{billard2011brief,billard2006}.
The most common choice of these summaries  is the random interval (or the $d$-dimensional equivalent, the random rectangle), whereby for individual-level observations $X_1,\ldots,X_n\in\mathbb{R}$, the random interval is typically constructed as $S=(\min_i X_i, \max_i X_i)\subseteq\mathbb{R}$. Quantile-based intervals have only received little attention \citep[e.g.][]{hron+bf17}. Other common symbol types include random histograms \citep{dias2015,le2013principal} and categorical multi-valued variables \citep{billard2006}.
Under the SDA framework, the collection of group-level data summaries $S_1,\ldots,S_m\in{\mathcal S}$ are considered the new data ``points'', whereby each datum is a distribution of some kind with an internal distributional structure. Statistical inference is then performed at the level of the symbols directly, with reference to their distributional forms, and without any further reference to the underlying measurement-level data.
See e.g.~\cite{noirhomme2011far,billard2011brief} and \cite{billard2003statistics} for a comprehensive overview of symbolic data types and their analysis.

This approach is potentially extremely attractive given present technological trends requiring the analysis of increasingly large and complex datasets. 
SDA effectively states that for many analyses, the high level of computation required for 
e.g.~divide-and-recombine techniques \citep[e.g.][]{guha12,jordan+ly18,vono2019,rendell+jlw18} or subsampling-based techniques \citep{quiroz+tvk18,bardenet+dh14,quiroz+kvt19}, 
 is not necessary to make inference at the group level. By aggregating the individual-level data to a much smaller number of group level symbols $m$ (where $m\ll n$), `big data' analyses can be performed cheaply and effectively on low-end computing devices. 
Beyond data aggregation, distributional-valued observations can arise naturally through the data recording process. This can include observational rounding or truncation, which results in imprecise data known to lie within some interval \citep{heitjan1991,vardeman+l05}, the elicitation of distributions from experts thought to contain quantities of interest \citep{fisher+olmkc15,lin+cs17}, and the construction of particle size distributions of particulate matter, typically in histogram form \citep{wraith+marh14}. 
In this sense, \cite{schweizer84}'s often-quoted statement that ``distributions are the numbers of the future'' seems remarkably prescient.

Many SDA techniques for analysing distributional-valued random variables have been developed, including
regression models \citep{irpino2015,dias2015,Giordani2015},
principal component analysis \citep{choukria2000,lauro2005, chouakria2011, ichino2011quantile, makosso2012, le2013principal},
time series \citep{lin+g16,wang+zl16,teles2015,arroyo2011},
clustering and classification \citep{brito2015probabilistic, brito2015,whitaker2019b},
discriminant analysis \citep{duartesilva+b15} and
Bayesian hierarchical modelling \citep{lin+cs17}. 
Likelihood-based inference was introduced by \cite{lerademacher2011} and \cite{brito2012} with further development and application by
\cite{zhang2020,lin+cs17}.

While there have been many successes in the analysis of symbolic data, from a
statistical perspective there are many opportunities for methodological improvement.
Some of these opportunities relate to existing SDA approaches, under which the statistical unit of interest is the symbol, and where inference is required at this level.
For example, the large majority of SDA techniques are descriptive and do not permit statistical inference on model parameters. E.g.,~regression models tend to be fitted by symbolic variants of least squares. 
Other opportunities arise, as with the present work, by re-imagining how the ideas behind SDA can be used to solve modern statistical challenges.
Here we assume that, as with a standard statistical analysis, inference is required at the level of the individual-level data, but where we deliberately aggregate the individual-level data into symbols prior to the analysis. Hence, if we can develop a way to perform statistical inference on the individual-level data when only given the group-level summaries,  then we can potentially perform standard statistical inference for large and complex datasets more efficiently via these distributional summaries than when directly using the original data. This alternative perspective on the ideas underlying SDA methodology opens up a new research direction in the field of SDA. Here, we focus on likelihood-based inference.

The likelihood approach of \cite{lerademacher2011,brito2012} maps each symbol to a random vector that uniquely defines the symbol, and then models this via a
standard likelihood model.
E.g.,~suppose that $X_{ij}\in\mathbb{R}$ is the value of some process recorded on the $i$-th second, $i=1,\ldots,n=86400$, of the $j$-th day, $j=1,\ldots,m$. If interest is in modelling these data as, say, i.i.d~draws from a skew-normal distribution $X_{ij}\sim SN(\mu_0,\sigma_0,\alpha_0)$, the likelihood function $L(x|\theta)$, $\theta\in\Theta$, may then be easily constructed.
However, suppose that interval symbols are now constructed so that $S_j=(\min_i X_{ij}, \max_i X_{ij})\subseteq\mathbb{R}$ is the random interval describing the observed process range on day $j$.
Due to the equivalence of representing continuous subsets of $\mathbb{R}$ by the associated bivariate vector in this setting \citep{zhang2020}, the approach of \cite{lerademacher2011,brito2012} constructs a model for the vectorised symbols $S_1,\ldots,S_m$, perhaps after a reparameterisation. For example,
\[
S_j\sim SN_2(\mu,\Sigma,\alpha)\qquad\mbox{or}\qquad\tilde{S}_j\sim SN_2(\mu,\Sigma,\alpha),
\]
where $\tilde{S}_j=((a+b)/2,\log(b-a))$ is a typical reparameterisation of $S_j=(a,b)$ into a function of interval mid-point and log range \citep{brito2012}. 
While there is inferential value in models of these kind \citep[e.g.][]{brito2012,lin+cs17}, it is clear that if  there is interest in modelling the underlying $X_{ij}$ as skew-normal, it is difficult to construct even a loosely equivalent model at the level of the symbol $S_j$ (or $\tilde{S}_j)$. That is, while the analyst may intuitively construct complex statistical models at the level of the individual-level data, it is less obvious how to construct models at the symbolic level and for different symbolic forms.

By design, modelling symbols directly, without specifying a probabilistic model for the underlying micro-data, only permits inference and predictions at the symbol level. This is unsatisfactory
because predictive inference for the underlying micro-data is often of interest, even if primary focus is on group-level analysis, and
as we demonstrate in Section \ref{section:CreditAnalysis}, ignoring the structure of the micro-data can result in symbolic-level analyses producing  poorer inferential outcomes. 
Another clear and acknowledged problem \citep{kosmelj2014,cariou+b15} is that 
even though existing SDA techniques do not focus on the individual-level data, the distribution of this data within random intervals/rectangles and within histogram bins is typically assumed to be uniform. Alternatives include the triangular distribution \citep{lerademacher2011,dias2017}. When considering that random intervals are typically constructed 
by specifying $S_j=(\min_i X_{ij}, \max_i X_{ij})$, it is almost certain that the distribution of the underlying data within $S_j$ is non-uniform. This implies that any inferential procedure built on the uniformity assumption (i.e.~almost all current SDA methods) is likely to produce questionable results.

One principled difference between SDA and regular statistical analyses is that the analysed symbolic data can be constructed by the analyst. This raises the question of how this should be undertaken.
Intuitively, if looking to design, say,  a random interval $S_j$ to maximise information about a location parameter, using the sample maximum and minimum is likely a poor choice as these statistics are highly variable. A more useful alternative could use e.g.~sample quantiles to define the interval. While sample quantiles have been considered in SDA methods, they have only been used as a robust method to avoid outliers that would otherwise dominate the size of a random interval \citep{hron+bf17}. 
In general, little consideration has been given to the design of informative symbols.

In this paper we introduce a new general method for constructing likelihood functions for symbolic data based on specifying a standard statistical model $L(X|\theta)$ for the underlying measurement-level data and then deriving the implied model $L(S|\theta)$ at the symbolic level by considering how $S$ is constructed from $x$. 
This provides a way to fit the measurement-level data model $L(X|\theta)$ while only observing the symbol level data, $S$. It provides both a natural way of specifying models for symbolic data, while also  opening up SDA methods as a mainstream technique for the fast analysis of large and complex datasets.
This approach naturally avoids making the likely invalid assumption of within-symbol uniformity,
allows inference and predictions at both the measurement data and symbolic data levels, 
permits symbolic inference using multivariate symbols (a majority of symbolic analyses are based on vectors of univariate symbols), and can provide a higher quality of inference than standard SDA techniques.
The method recovers some known models in the statistical literature, as well as introducing several new ones, and reduces to standard likelihood-based inference for the measurement-level data (so that $L(S|\theta)\rightarrow L(X|\theta)$) when  $S\rightarrow X$.

As a result we demonstrate some weaknesses of current symbol construction techniques.
In particular we establish informational limits on random rectangles constructed from marginal minima/maxima or quantiles, and introduce a new class of quantile-based random rectangles. 
These new symbol variations produce more efficient analyses than existing symbol constructions, and permit the estimation of within-symbol multivariate dependencies that were not previously estimable.

The new symbolic likelihood function is presented in Section \ref{sec:gen_sym_lik} with specific results for random rectangles and histograms. 
All derivations are relegated to the Appendix. The performance of these models is demonstrated in Section \ref{sec:examples} through a meta-analysis of univariate histograms, a simulation study of the inferential performance of a new class of multivariate random rectangle constructions, and an analysis of a large loan dataset. In all cases, the existing state-of-the-art models and symbolic constructions are outperformed by the new symbolic model. 
Section \ref{sec:discussion} concludes with a discussion.

%%%%%%%%%%%%%%%%%%%%%%%%%%%%%%%%%%%%%%%%
%
% SECTION 2: GENERAL CONSTRUCTION
%
%%%%%%%%%%%%%%%%%%%%%%%%%%%%%%%%%%%%%%%%
%
\section{A general construction tool for symbolic likelihoods}
\label{sec:gen_sym_lik}

\subsection{Symbolic likelihood functions}
\label{ssec:lik}

Suppose  that $\Omega$ is a population of interest 
defined on a probability space $( \Omega, \mathcal{F}, \prob )$, and that
each individual in $\Omega$ is described by a measurable random variable $X$, 
defined by
$X: \Omega \rightarrow \cX,$ and 
$\prob (X \in \cY) = \prob ( \omega \in \Omega | X(\omega) \in \cY ), 
\textrm{ for } \cY \subset \cX.$
We follow the standard SDA construction of a {\em class} \citep{billard2003statistics}
and let the random variable $C: \Omega \rightarrow \cC$ denote the class to which an
individual belongs. For simplicity we assume that  
$\cC = \{ 1, \ldots, m\}$ is finite.
Consequently let $\Omega_c = \{ \omega \in \Omega \textrm{ s.t.}~C(\omega) =c \}$ be the 
set of individuals in the population that belong to class $c \in \cC$ such that 
$\cup_{c \in \cC} \Omega_c = \Omega$, and define $X_c: \Omega_c \rightarrow \cX_c \subseteq \cX$ 
as the random variable that describes them. 
We denote $\textrm{Card}(\Omega) = N$ and $\textrm{Card}(\Omega_c) = N_c$  
where $\sum_{c \in \cC} N_c = N$.

We now consider an interpretation  of SDA, where the symbolic random variable $S_c$ for class $c \in \cC$  is assumed to be the result of the aggregation of  the random variables $\bX_c = (X_{1,c}, \ldots, X_{N_c,c} )$ via some aggregation function $\pi_c$, so that 
$S_c = \pi_c (\bX_c):  [\cX_c]^{N_c}  \rightarrow \cS_c$ and
$\bx_c  \mapsto \pi (\bx_c).$
That is, a symbolic random variable represents a summary of the information brought
by measurement over individuals. 
The choice of this summary (and thus of the aggregation function) is critical and we explore this in later Sections.
In the following we refer to random variables of the measurement-level data $X$  as {\em classical} random 
variables.
By construction symbolic random variables require knowledge of the underlying classical random 
variables. Accordingly, this should also be true when dealing with likelihood functions, particularly if inference is required at both classical and symbolic levels, but when only information at the symbolic level is observed. 

To construct a symbolic likelihood function,
suppose that the classical random variable $X$ has probability density 
and distribution functions
$g_X (\,\cdot\,; \theta)$ and $G_X (\,\cdot\,; \theta)$ respectively, where $\theta \in \Theta$.
Consider a random classical data sample $\bx = (x_1, \ldots, x_n)$ of size $n<N$ from the population, 
and denote by $\bx_c = (x_{1,c}, \ldots, x_{n_c,c})$,  the 
collection of those in class $c$, where $\sum_{c \in \cC} n_c = n$.
Similarly let $s_c = \pi_c (\bx_c)$ be the resulting observed symbol obtained 
through the aggregate function $\pi_c$ and define the symbolic dataset to be the collection of symbols 
$ \bs = (s_c; c \in \cC)$.

%
% Begin Proposition
%

\begin{prop}
	\label{prop:symb_lik_fun}
	For the subset $\bx_c$ of $\bx$ associated with class $c \in \cC$,
	the likelihood function of the corresponding symbolic observation $s_c=\pi_c(\bx_c)$ is given by
	\begin{equation}
	\label{eq:symb_lik}
	L(s_c; \vartheta, \theta) \propto 
	\int_{\cX^n} f_{S_c|\bX_c = \bz_c}(s_c; \vartheta) 
	g_{\bX}( \bz; \theta) \der \bz, 
	\quad \forall c \in \cC, 
	\end{equation}
	where $\bz_c \in \cX_c^{n_c}$ is a subset of $\bz \in \cX^n$,  $f_{S_c | \bX_c}(\cdot; \vartheta)$ is the conditional density of $S_c$ given 
	$\bX_c$
	and $ g_{\bX}( \,\cdot\,; \theta)$ is the joint density of $\bX$.
\end{prop} 
%
% End Proposition
%
We refer to $L(s_c; \vartheta, \theta)$ given in \eqref{eq:symb_lik} as the symbolic likelihood function. 
A discrete version of \eqref{eq:symb_lik} is easily constructed.
Note that by writing the joint density $g_{\bX}(\,\cdot\,; \theta)=g_{\bX_c}(\,\cdot\,; \theta) g_{\bX_{-c}| \bX_c}(\,\cdot\,; \theta)$, where 
$\bX_{-c} = \bX \backslash \bX_c$, then after integration with respect to $\bx_{-c} = \bx 
\backslash \bx_c$, equation \eqref{eq:symb_lik} becomes
$$
L(s_c; \vartheta, \theta) \propto 
\int_{\cX_c^{n_c}} f_{S_c|\bX_c = \bz_c}(s_c; \vartheta) 
g_{\bX_c}( \bz_c; \theta) \der \bz_c.
$$
This construction method can easily be interpreted: the probability of observing a symbol $s_c$ is equal to the probability of generating a classical dataset under the classical data model that produces the observed symbol under the aggregation function $\pi_c$. That is, we have established a direct link between the user-specified classical likelihood function $L(\bx|\theta)\propto g_\bX(\bx;\theta)$ and the resulting probabilistic model on the derived symbolic data.
As a result we may directly estimate the parameters $\theta$ of the underlying classical data model, based only on observing the symbols $\bs$.

In the case where there is no aggregation of $\bx_c$ into a symbol, so that $\pi(\bx_c)=\bx_c$ and $\cS_c = [\cX_c]^{N_c}$, then
$ f_{S_c|\bX_c = \bz_c} (s_c) \equiv f_{\pi(\bX_c) | \bX_c = \bz_c} (\pi(\bx_c)) 
= f_{\bX_c | \bX_c = \bz_c} (\bx_c) = \delta_{\bz_c} (\bx_c)$,
where $\delta_{\bz_c} (\bx_c)$ is the Dirac delta function, taking the value 1 if $\bz_c = \bx_c$ and $0$ otherwise. 
As a result the symbolic likelihood function reduces to $g_{\bX_c}( \bx_c; \theta)$, the classical
likelihood contribution of class $c$.
Under the assumption that the classical data are independently distributed between classes, so that $g_\bX(\,\cdot\,; \theta) = \prod_{c \in \cC} g_{\bX_c}(\,\cdot\,; \theta)$, the associated symbols are
also independent and the likelihood of the symbolic dataset $\bs$ is given by
$$
L(\bs;\vartheta, \theta) 
= \prod_{c \in \cC} L(s_c; \vartheta, \theta) 
\propto \prod_{c \in \cC} \int_{\cX_c^{n_c}} f_{S_c|\bX_c = \bz_c}(s_c; \vartheta) 
g_{\bX_c}( \bz_c; \theta) \der \bz_c.
$$
If, further, the observations within a class $c \in \cC$ are independent and 
identically distributed, then in the scenario where $\pi(\bx_c) = \bx_c$ we have
$
L(\theta) = \prod_{i=1}^n g_X (x_i; \theta).
$
Because
$\text{Card}(\cC) = m$ and typically $m\ll n$, this implies that large computational savings can be made through the analysis of symbolic rather than classical data, depending on the complexity of the classical data likelihood function.
The method established in Proposition~\ref{prop:symb_lik_fun} specifies a probability model for the micro-data which, combined with knowledge of the aggregation process $\pi$, induces a likelihood function at the aggregate level. In contrast, the likelihood function defined by \citet{lerademacher2011,brito2012} specifies a probability model directly on the symbols.

In the following Subsections, 
we establish analytical expressions of the symbolic likelihood function based on various choices of the aggregation function $\pi$, which leads to different symbol types. The performance of each of these models will be examined in Section \ref{sec:examples}.
For clarity of presentation the class index $c$ is omitted in the remainder of this Section as the results presented are class specific.

%
% GENERATIVE MODEL INTERVAL VALUED SYMBOLS
%

\subsection{Modelling random intervals}
\label{ssec:interv}

The univariate random interval is the most common symbolic form, and is typically constructed as the range of the underlying classical data e.g.~$S=(\min_i X_i, \max_i X_i)$. Here we generalise this to order statistics $S=(X_{(l)},X_{(u)})$ for indices $l\leq u$ given their higher information content.
We define an interval-valued symbolic random variable to be constructed by the aggregation function $\pi$ where
\begin{equation}
\label{eq:pi_univ_int}
S = \pi (\bX): \real^{N} \rightarrow \cS = \{ (a_1, a_2) \in \real^2: a_1 \leq a_2 \} \times \natural 
\end{equation}
so that $\bx \mapsto (x_{(l)}, x_{(u)}, N)$,
where $x_{(k)}$ is the $k$-th order statistic of $\bx$ and $l,u \in \{ 1, \ldots, N \}, l \leq u$ are fixed. Taking $l=1, u=N$ corresponds to determining the range of the data. 
Note that this construction explicitly includes the number of underlying datapoints $N$ in the interval as part of the symbol, in direct contrast to almost all existing SDA techniques. This allows random intervals constructed using different numbers of underlying classical datapoints to contribute to the likelihood function in relation to the size of the data that they represent. This is not available in the construction of \cite{lerademacher2011,brito2012}.
\begin{lem}
	\label{lem:univ_int}
	Consider a univariate interval-valued random variable $S = (s_l, s_u, n) \in \cS$, obtained through \eqref{eq:pi_univ_int} and assume that $g_\bX(\bx; \theta) = \prod_{i=1}^n g_X(x_i; \theta), \bx \in \real^n$. The corresponding symbolic likelihood function is then given by
	\begin{align*}
	L(s_l,s_u, n; \theta)=
	\frac{n!}{ (l-1)! (u-l-1)! (n-u)! } 
	\left[ G_X(s_l; \theta) \right]&^{l-1} 
	\left[ G_X(s_u; \theta) - G_X(s_l; \theta) \right]^{u-l-1}\\
	& \times \left[ 1- G_X(s_u; \theta) \right]^{n-u} g_X(s_l; \theta) g_X(s_u; \theta).
	\end{align*}
\end{lem}
\noindent It is worth noting that this expression can also be obtained by evaluating 
$\prob (S_l \leq s_l, S_u \leq s_l) = \prob( X_{(l)} \leq s_l, X_{(u)} \leq s_u )$ and then taking derivatives with respect to $s_l$ and $s_u$, and corresponds to the joint distribution of order two statistics.
This model was previously established by \citet{zhang2020} as a generative model for random intervals built from i.i.d.~random variables.

%%%%%%%%%%%%%%%%%%%%%%%%%%%%%%%%%%%%%%%%%
\subsection{Modelling random rectangles}
%%%%%%%%%%%%%%%%%%%%%%%%%%%%%%%%%%%%%%%%%
\label{ssect:rectangles}

The typical method of constructing multivariate random rectangles from underlying $d$-dimensional data $X \in \real^d, d \in \mathbb{N}$ is by taking the cross product of each $d$ univariate random interval described by their marginal minima and maxima \citep[e.g.][]{neto+cc11, ichino2011quantile}. 
The number of datapoints underlying  this rectangle is often not used.
We improve on this scheme by making use of additional information available at the time of rectangle construction (Section \ref{sssect:minmax}), and then develop several alternative constructions for random rectangles based on marginal order statistics (Section \ref{ssect:order}).

%%%%%%%%%%%%%%%%%%%%%%%%%%%%%%%%%%%%%%%%%
\subsubsection{Using marginal maxima and minima}
%%%%%%%%%%%%%%%%%%%%%%%%%%%%%%%%%%%%%%%%%
\label{sssect:minmax}

While it is in principle possible to identify a small amount of information about the dependence between two variables summarised by a marginally constructed bounding box, this information content is very weak, and the direction of dependence is not identifiable \citep{zhang2020}. E.g.~if $n$ datapoints are generated from a multivariate distribution and the marginal minimum and maximum values recorded, what can be said about the correlation strength and direction? We propose that dependence information can be obtained if the locations of those datapoints involved in construction of the bounding rectangle, and the total number of points are known. For the examples in Figure \ref{NeedNewFig} (top), if the rectangle is generated from only two points (left panel) one can surmise stronger dependence than if three points are used (centre), with rectangle construction using four points (right) producing the weakest dependence. The exact locations of these bounding points is informative of dependence direction.

As such, we define the aggregation function $\pi$ to incorporate these construction points (where available) into the definition of the random rectangle as
\begin{equation}
\label{eq:pi_mult_int}
S = \pi (\bX): \real^{d \times N} \rightarrow \cS = \{ (a_1, a_2) \in \real^2: a_1 \leq a_2 \}^d 
\times \{ 2, \ldots, \max(2d,n) \} \times \cT \times \natural
\end{equation}
so that $\bx\mapsto \left( (x_{(1),i}, x_{(n),i})_{i=1,\ldots,d}, p, I(p), N \right),$
where $\bx = (x_1, \ldots, x_n)$,  $x_j = (x_{j,1}, \ldots, x_{j,d})^\top$ and $x_{(k),i}$ is the $k$-th order statistic of the $i$-th marginal component of $\bx$.  
The quantities $p$ and $I(p)$ represent the number of points involved in constructing the  rectangle, and   their locations (taking values in $\cT$), respectively.
In this context a symbol is written as $S = (S_{\min}, S_{\max}, S_p, S_{I_p}, N)$, where $S_{\min}$ and $S_{\max}$ are respectively the $d$-vectors corresponding to the marginal minima and maxima.

\begin{lem}
	\label{lem:mult_int}
	Consider a multivariate random rectangle $S \in \cS$, obtained through \eqref{eq:pi_mult_int} and assume that $g_\bX(\bx; \theta) = \prod_{i=1}^n g_X(x_i; \theta), \bx \in \real^{n \times d}$. Then the symbolic likelihood function is given by
	\begin{align}
	\label{eq:lik_mult_int}
	L(s; \theta) =
	\frac{n!}{(n-s_p)!}  \left[ \int_{s_{\min}}^{s_{\max}} g_X(z; \theta) \der z \right]^{n-s_p} \times \ell_{s_p},
	\end{align}
	where the multivariate integral is taken over the rectangular region defined by $s_{\min}$ and $s_{\max}$, and
	where $\ell_{s_p}$ is defined as follows.
	If $s_p=2$ then $s_{I_p}=(s_a,s_b)$ is the two co-ordinates of $d$-dimensional space which define the bounding rectangle, and
	$\ell_2 = g_X(s_a; \theta) g_X(s_b; \theta).$
	If $s_p=2d$ then $s_{I_p} = \emptyset$ and
	\begin{align}
	\ell_{2d} =& \prod_{i=1}^d \left[ 
	G_{X_{-i}| X_i = s_{\min,i}} (s_{\max, -i}; \theta) 
	- G_{X_{-i}| X_i = s_{\min,i}} (s_{\min, -i}; \theta)
	\right] g_{X_i} (s_{\min,i}) \nonumber\\
	& \times \prod_{i=1}^d \left[ 
	G_{X_{-i}| X_i = s_{\max,i}} (s_{\max, -i}; \theta) 
	- G_{X_{-i}| X_i = s_{\max,i}} (s_{\min, -i}; \theta)
	\right] g_{X_i} (s_{\max,i}),  \label{l2d}
	\end{align}
	where $X_i$ is the $i$-th component of $X$, $X_{-i} = X \backslash X_i$ and similarly for $s_{\min, -i}, s_{\max, -i}, s_{\min, i}$ and $s_{\max, i}$, and $G_{X_{-i}| X_i} $ is the conditional distribution function of $X_{-i}$ given $X_i$.
\end{lem}

\noindent In (\ref{l2d}) the product terms 
represent the joint distributions of $X_{-i}$ being between $s_{\min, -i}$ and $s_{\max, -i}$ given that $X_i$ is equal to $s_{\min, i}$ or $s_{\max, i}$. When $s_p=2$, (\ref{l2d}) 
reduces to $\ell_{2d}=\ell_2$.
General expressions for $\ell_{s_p}$ for $p\neq 2$ or $2d$ can be complex.
Simple expressions are available for $s_p=3$ when  $d=2$.

\begin{cor}
	\label{cor:biv_int}
	For a bivariate random rectangle, 
	if $s_p=3$ then $S_{I_p} = s_c \in \real^2$ is the co-ordinate of the point defining the bottom-left, top-left, top-right or bottom-right corner of the rectangle.
	In this case, if $\bar{s}_c$ is the element-wise complement of $s_c$, i.e. $\bar{s}_{c,i} = \{s_{\min,i}, s_{\max,i}\} \backslash \{s_{c,i}\}, i=1,2$, then
	
	\begin{equation}
	\ell_3 = 
	g_X(s_c; \theta) 
	\times \prod_{i=1}^2 
	\left[  
	G_{X_{-i} | X_i =  \bar{s}_{c,i}} ( s_{\max, -i}; \theta) 
	- G_{X_{-i} | X_i = \bar{s}_{c,i}} (s_{\min, -i}; \theta)
	\right] g_{X_i}(  \bar{s}_{c,i} ; \theta). \label{ell3}
	\end{equation}
	E.g.~if 
	$s_c = (s_{\min,1}, s_{\min,2})$  is in the bottom-left corner, then $\bar{s}_c = (s_{\max,1}, s_{\max,2})$.
\end{cor}

\noindent The first term in (\ref{ell3}) is the density of the point in the corner of the rectangle, and the other terms are the
probabilities of the two points on the edges being between two interval values given that the other component is fixed.
Qualitatively similar expressions can be derived for $d$-dimensional random rectangles in the cases where $s_p\neq2$ or $2d$, although there is no simple general expression.

%%%%%%%%%%%%%%%%%%%%%%%%%%%%%%%%%%%%%%%%%
\subsubsection{Using marginal order statistics}
%%%%%%%%%%%%%%%%%%%%%%%%%%%%%%%%%%%%%%%%%
\label{ssect:order}

\begin{figure}[h]
	$$
	\begin{array}{c}
	\includegraphics[width=0.25\textwidth]{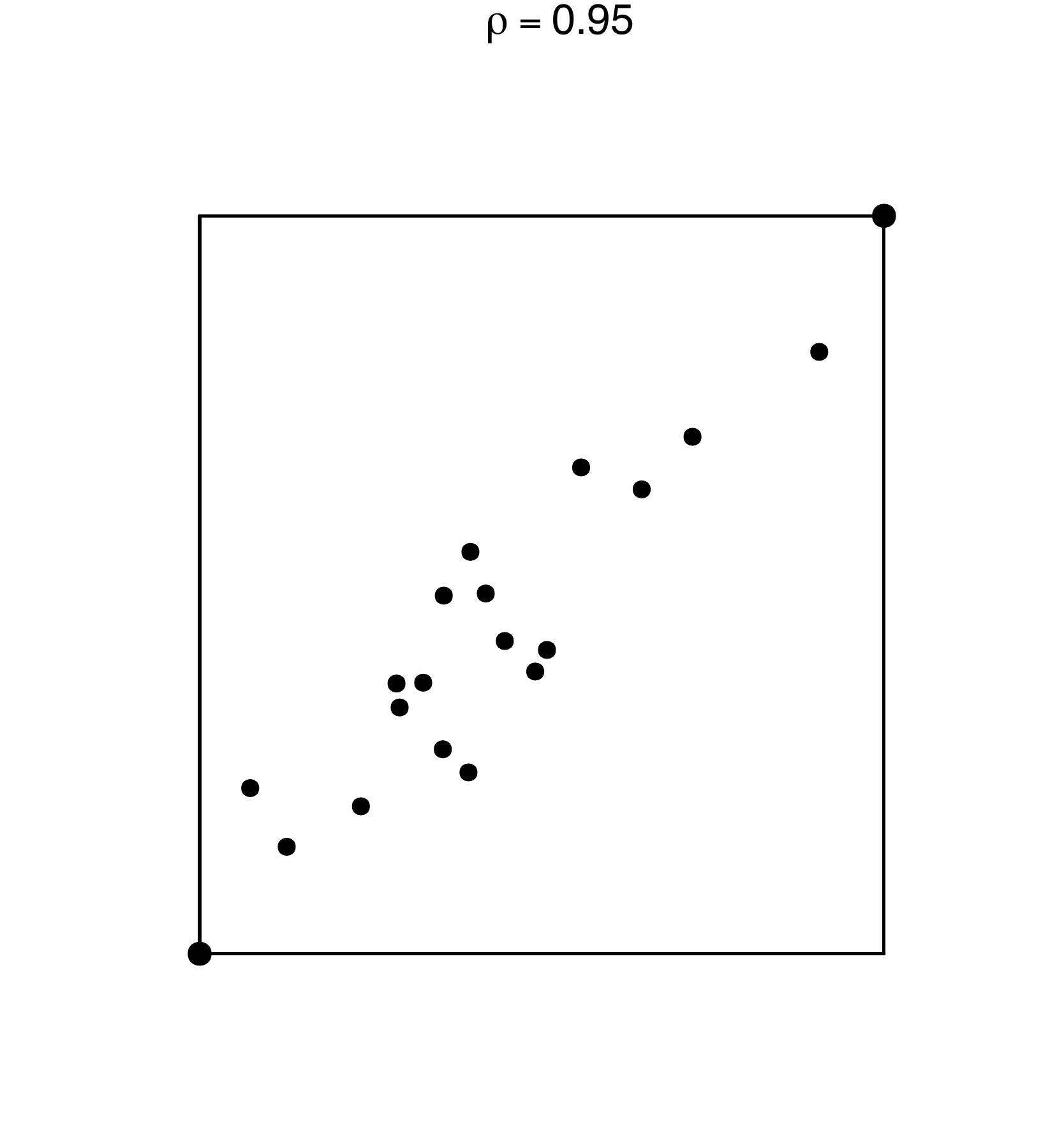}
	\includegraphics[width=0.25\textwidth]{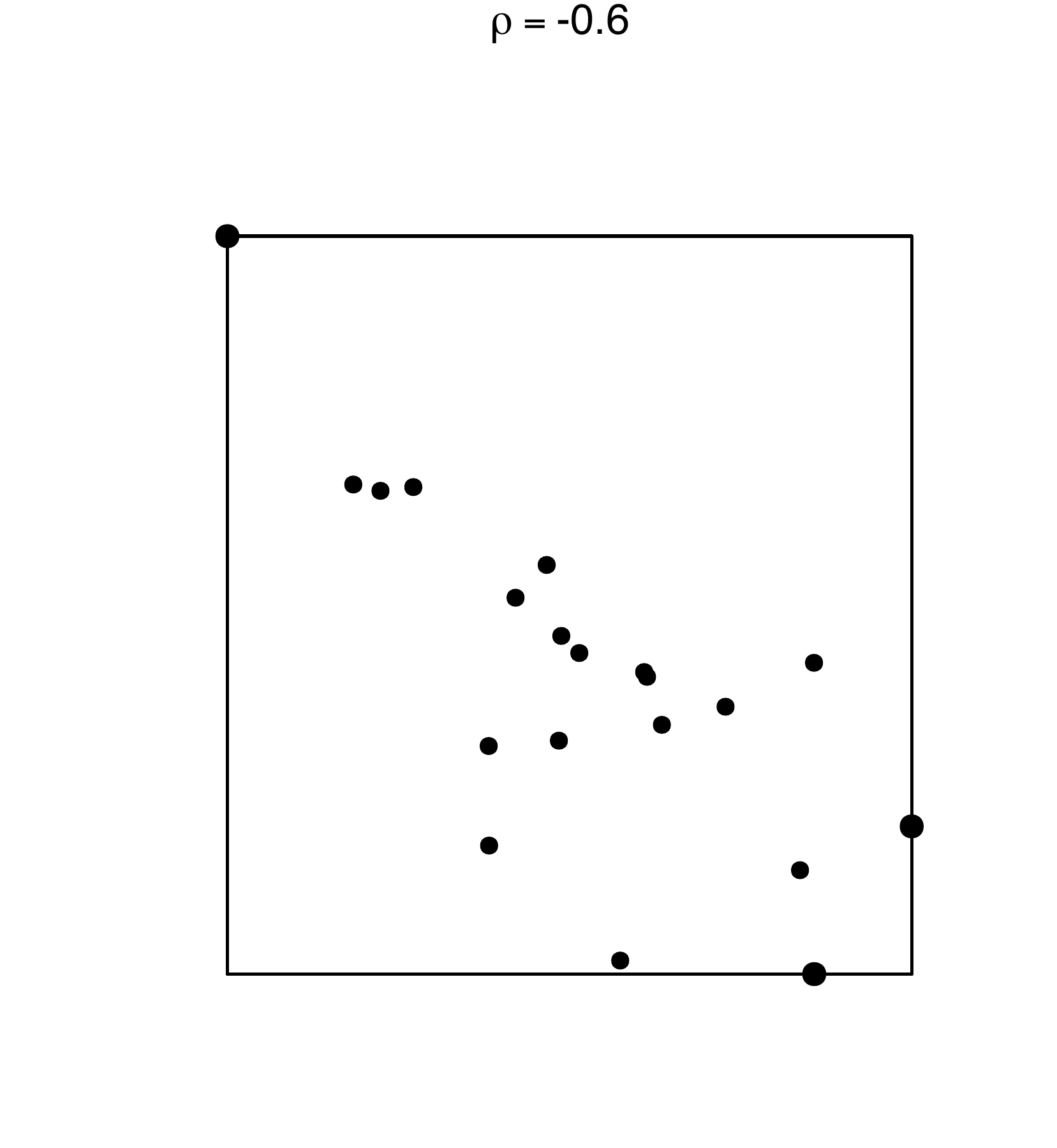} 
	\includegraphics[width=0.25\textwidth]{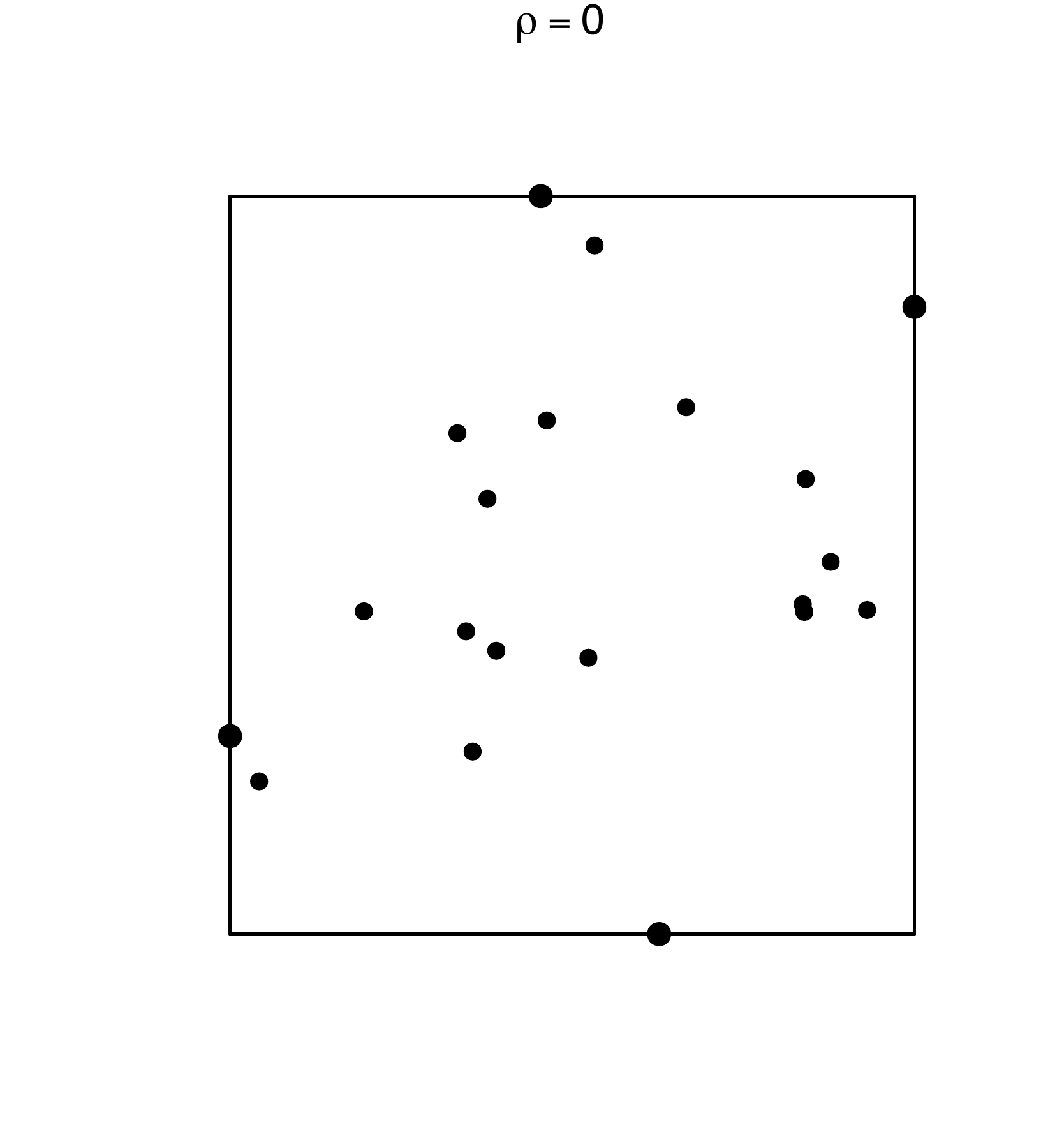} \\
	\includegraphics[width=0.25\textwidth]{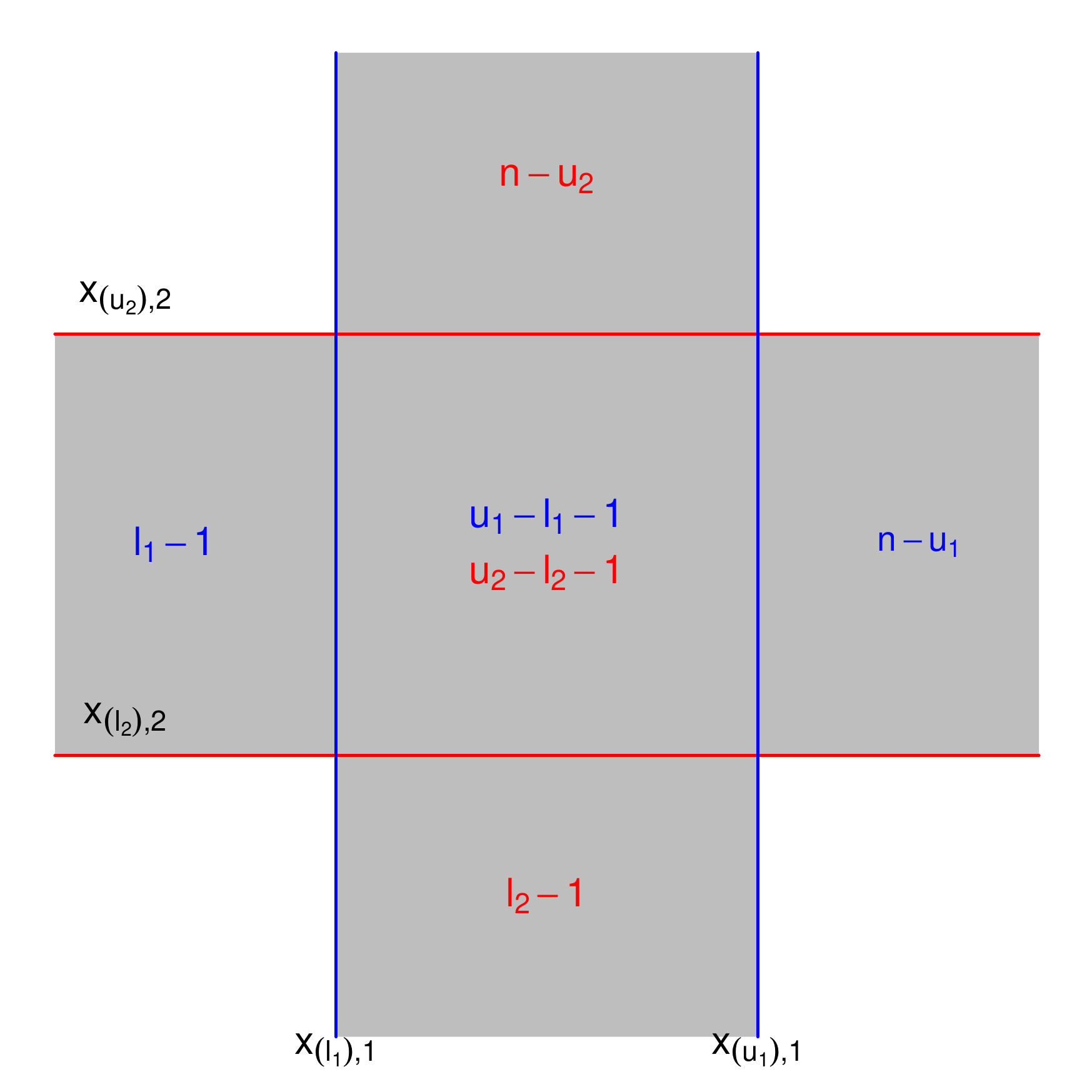} 
	\includegraphics[width=0.25\textwidth]{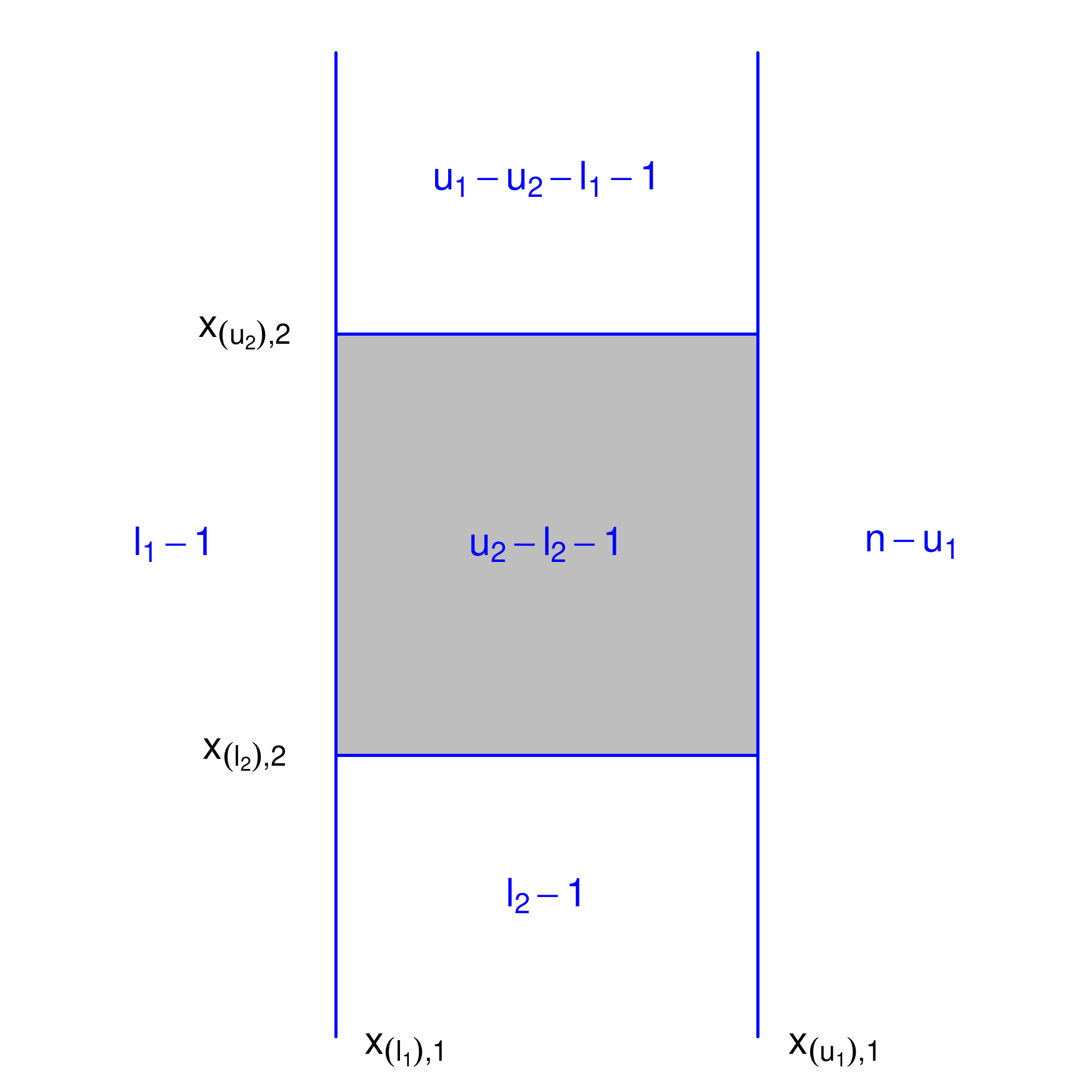}
	\includegraphics[width=0.25\textwidth]{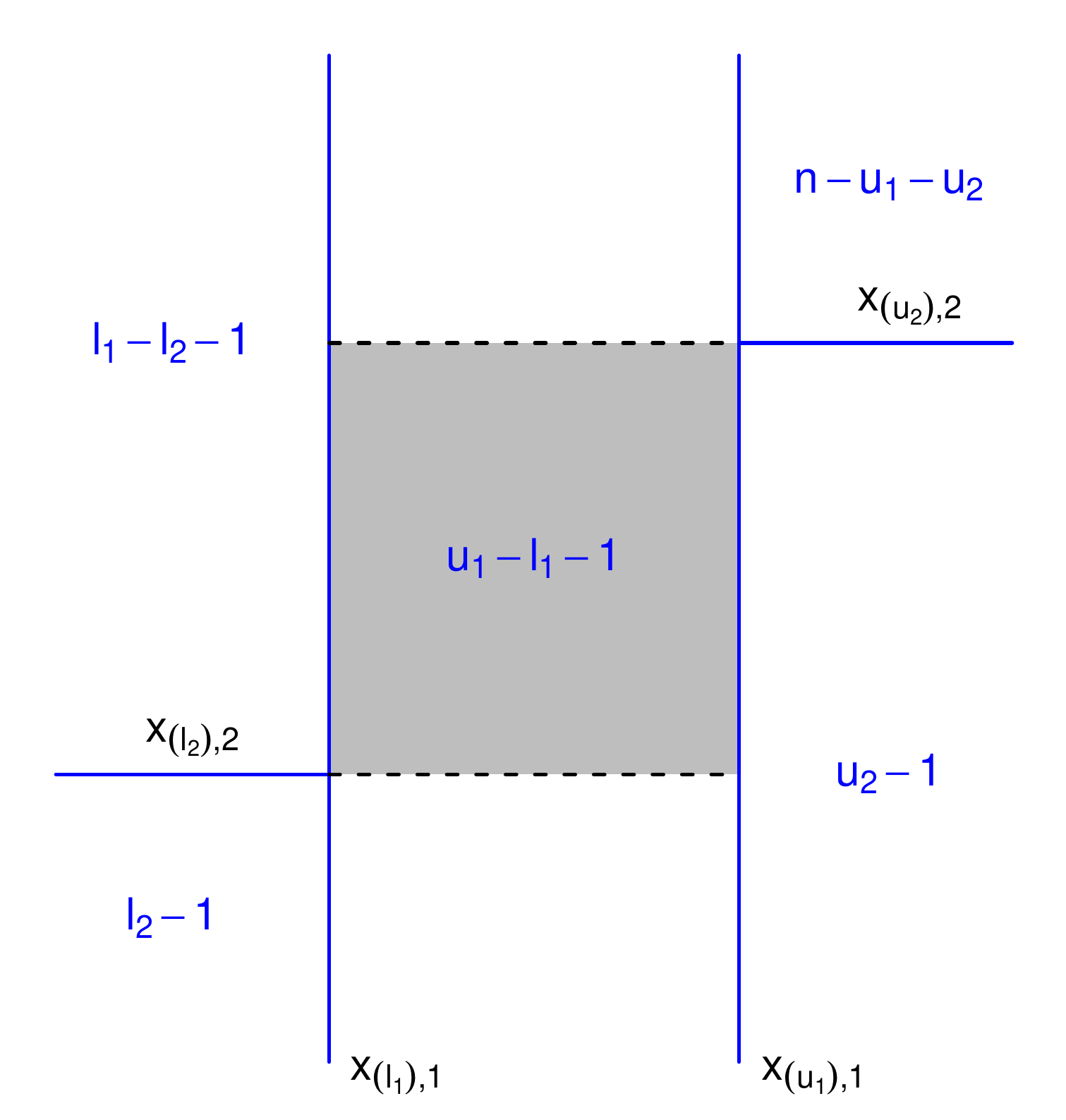}
	\end{array}
	$$
	\caption{\label{NeedNewFig}\footnotesize Construction methods for bivariate intervals using marginal minima/maxima (top panels) or marginal order statistics (bottom).
		Top panels: Illustrative random rectangles constructed from 2 points (high correlation), 3 points (moderate correlation) and 4 points (low/no correlation).
		Bottom panels: Three alternative construction methods: marginal only (left panel),  sequential nesting (centre; equation \eqref{eq:pi_mult_os_int2}) and  iterative segmentation (right; equation \eqref{eq:pi_mult_os_int}).
		Values in blue (red) denote the number of observations in the area bounded by blue (red) lines.
}
\end{figure}

As order statistics are defined in the univariate setting, there are a number of methods to use fixed vectors of lower $l=(l_1,\ldots,l_d)^\top$ and upper $u=(u_1,\ldots,u_d)^\top$ order statistic values, with $1\leq l_i<u_i\leq N$, to define a $d$-dimensional random rectangle. The simplest takes the cross product of the $d$-univariate marginal quantiles as suggested by \citet{neto+cc11}, where the authors indicate that the lower and upper values can take any pair of interval feature possible. Here the aggregation function $\pi$ is
\begin{eqnarray}
S = \pi (\bX): \real^{d \times N}  &\rightarrow &\cS = \{ (a_1, a_2) \in \real^2: a_1 \leq a_2 \}^d \times \natural \label{eq:marginalQuantile}\\
\bx & \mapsto & \left( \left( x_{(l_i),i} , x_{(u_i),i} \right)_{i=1,\ldots,d}, N \right). \label{eq:simpleQuant}
\end{eqnarray}
In this context the symbol is written as $S=(S_l, S_u,N)$, where $S_l$ and $S_u$ are respectively the $d$-vectors corresponding to the marginal lower and upper order statistics. 
This process is illustrated in Figure \ref{NeedNewFig} (bottom left panel) in the $d=2$ setting. For fixed $l$ and $u$, the observed counts in each region are then known as a function of the construction (\ref{eq:simpleQuant}).
The resulting symbolic likelihood function is then
\[
L(s; \theta) = \prod_{i=1}^d L(s_{l_i}, s_{u_i}, n; \theta_i)
\]
where $L(s_{l_i}, s_{u_i}, n; \theta_i)$ is as obtained in Lemma~\ref{lem:univ_int} using the $i$-th marginal distribution with parameter $\theta_i\in\Theta$. 
However, as the construction (\ref{eq:simpleQuant}) only contains marginal information, such a symbol will fail to adequately capture dependence between variables. As an alternative, we introduce two new order-statistic based representations of random rectangles that do account for such dependence.

The first, {\em sequential nesting} (Figure \ref{NeedNewFig}, bottom centre panel), constructs the order statistics within dimension $i$ conditionally on already being within the random rectangle in dimensions $j<i$.
The aggregation function $\pi$ is given by (\ref{eq:marginalQuantile}) as before, but where now
\begin{equation}
\label{eq:pi_mult_os_int2}
\bx  \mapsto \left( \left(  \left( x_{(l_i),i}, x_{(u_i),i} \right) |  \{ x_{(l_j), j} < x_j < x_{(u_j), j}; j <i \}  \right)_{i=1,\ldots,d}, N \right).
\end{equation}
As before, $S=(S_l,S_u,N)$, but where the known observed counts now lie in different regions (Figure \ref{NeedNewFig}), and with the additional constraints of 
$2 \leq u_{i+1} \leq u_i - l_i - 1$. 
\begin{lem}
	\label{lem:mult_os_int2}
	Consider a multivariate random rectangle $S \in \cS$, constructed via \eqref{eq:pi_mult_os_int2} and suppose that $g_\bX(\bx; \theta) = \prod_{i=1}^n g_X(x_i; \theta), \bx \in \real^{n \times d} $. The symbolic likelihood function is then given by
	\begin{equation}
	\label{eq:SeqNestLik}
	L(s; \theta) \propto
	\prob ( s_{\boldsymbol{l}} < X <  s_{\boldsymbol{u}} )^{u_d - l_d - 1} 
	\der \prob ( X_1 < s_{l,1} ) \der \prob ( X_1 <  s_{u,1} )
	\prod_{i=1}^{d} p_i( s_{\boldsymbol{l}} ) q_i( s_{\boldsymbol{u}} ),
	\end{equation}
	where $p_1(s_{\boldsymbol{l}}) = \prob (X_1 <  s_{l,1})^{l_1 -1} $, 
	$q_1(s_{\boldsymbol{u}}) = \prob (X_1 > s_{u,1})^{n-u_1}$ and for $i=2, \ldots, d$,
	\begin{align*}
	p_i(s_{\boldsymbol{l}}) =& \prob( s_{l,j} < X_j < s_{u,j}; j < i | X_i = s_{l,i}  ) 
	\der \prob ( X_i < s_{l,i} ) 
	\prob ( X_i < s_{l,i} | s_{l,j} < X_j < s_{u,j} ; j < i )^{l_i -1}  \\
	q_i( s_{\boldsymbol{u}} ) = & \prob( s_{l,j} < X_j < s_{u,j}; j < i | X_i = s_{u,i}  ) 
	\der \prob ( X_i <  s_{u,i} ) \\
	& \qquad \qquad \qquad \qquad \qquad \qquad \qquad \times 
	\prob ( X_i > s_{u,i} | s_{l,j} < X_j < s_{u,j} ; j < i )^{u_{i-1} - u_i - l_{i-1} -1} .
	\end{align*}
\end{lem}
\begin{cor}
	\label{cor:biv_os_int2}
	With $d=2$, the symbolic likelihood function in Lemma~\ref{lem:mult_os_int2} is given by
	\begin{align*}
	L(s; \theta) \propto&
	\left( G_{X} ( s_{\boldsymbol{u}} ) - G_{X} ( s_{\boldsymbol{l}} ) \right) ^{u_2 - l_2 - 1} 
	g_{X_1}(s_{l,1}) g_{X_1}(s_{u,1})  g_{X_2}(s_{l,2}) g_{X_2}(s_{u,2}) \\
	& \times G_{X_1}(s_{l,1})^{l_1 -1} \left[ 1 - G_{X_1}(s_{u,1}) \right]^{n-u_1}
	\left[ G_{X_1 | X_2 = s_{l,2}}(s_{u,1}) - G_{X_1 | X_2 = s_{l,2}}(s_{l,1}) \right] \\
	& \times \left[ G_{X_1 | X_2 = s_{u,2}}(s_{u,1}) - G_{X_1 | X_2 = s_{u,2}}(s_{l,1}) \right]
	\left[ G_X ( (s_{u,1}, s_{l,2}) ) - G_X( s_{\boldsymbol{l}} ) \right]^{l_2 - 1} \\
	& \times   
	\left[ G_{X_1}(s_{u,1}) - G_X( s_{\boldsymbol{u}} ) - G_{X_1}(s_{l,1}) +  G_X ( (s_{l,1}, s_{u,2}) ) \right]^{u_1 - u_2 - l_1 - 1},
	\end{align*}
	where $G_{X_i}(\cdot) \equiv G_{X_i}(\,\cdot\,; \theta)$ and 
	$G_{X_i | X_j}(\cdot) \equiv G_{X_i | X_j}(\,\cdot\,; \theta); i \neq j$ respectively denote the marginal and conditional distribution functions of $g_\bX(\bx; \theta)$.
\end{cor}

\noindent An alternative to sequential nesting is an {\em iterative segmentation} construction (Figure \ref{NeedNewFig}, bottom right). 
As before, for fixed vectors $l$ and $u$, the aggregation function $\pi$ is given by (\ref{eq:marginalQuantile}) but where
\begin{equation}
\bx  \mapsto \left( \left( x_{(l_i),i} |  \{ x_j < x_{(l_j), j}; j <i \}, x_{(u_i),i} |  \{ x_j > x_{(u_j), j}; j < i \} \right)_{i=1,\ldots,d}, N \right).\label{eq:pi_mult_os_int}
\end{equation}
Again $S=(S_l,S_u,N)$,
but now
where $S_{l,i}$, the $l_i$-th order statistic of the $i$-th margin, is restricted to the area where the  previous margins $j<i$ are all below their respective lower ($l_j$-th) order statistic. 
Similarly, 
$S_{u,i}$ is restricted to the area where the previous margins $j<i$ are all above their respective upper order statistic.
For fixed $l$ and $u$ the observed counts are then known (Figure \ref{NeedNewFig}, bottom right) but are attributed to different regions than for sequential nesting. Iterative segmentation
implies the additional constraints $l_{i+1} < l_i -1$ and $u_{i+1} < N - \sum_{j =1}^{i} u_j$ for $i=1,\ldots, d-1$.
\begin{lem}
	\label{lem:mult_os_int}
	Consider a multivariate random rectange $S \in \cS$, constructed via \eqref{eq:pi_mult_os_int} and suppose that $g_\bX(\bx; \theta) = \prod_{i=1}^n g_X(x_i; \theta), \bx \in \real^{n \times d}$. The symbolic likelihood function is then given by
	\begin{equation}
	\label{eq:IterSegLik}
	L(s; \theta) \propto
	\prob ( s_{l,1} < X_1 < s_{u,1} )^{u_1 - l_1 - 1} 
	\der \prob ( X_1 < s_{l,1} ) \der \prob ( X_1 < s_{u,1} )
	\prod_{i=2}^{d+1} p_i( s_{\boldsymbol{l}} ) q_i( s_{\boldsymbol{u}}),
	\end{equation}
	where $p_{d+1}( s_{\boldsymbol{l}} ) = \prob(X_1 < s_{l,1}, \ldots, X_d < s_{l,d})^{l_d - 1}$,
	$q_{d+1}( s_{\boldsymbol{u}}) = \prob(X_1 > s_{u,1}, \ldots, X_d > s_{u,d})^{n - \sum_{i=1}^d u_i}$ 
	and for $i=2, \ldots, d$
	\begin{align*}
	p_i( s_{\boldsymbol{l}} ) =& \prob( X_j <  s_{l,j}; j <i | X_i = s_{l,i}  ) 
	\der \prob ( X_i <  s_{l,i} ) 
	\left[  \prob ( X_j < s_{l,j}; j <i ) - \prob ( X_j < s_{l,j}; j \leq i ) \right]^{l_i - l_{i-1} -1} \\
	q_i( s_{\boldsymbol{u}} ) = & \prob( X_j > s_{u,j}; j <i | X_i = s_{u,i} ) 
	\der \prob ( X_i < s_{u,i} ) 
	\left[  \prob ( X_j > s_{u,j}; j <i ) - \prob ( X_j > s_{u,j}; j \leq i ) \right]^{u_i -1}.
	\end{align*}
	
\end{lem}

\begin{cor}
	\label{cor:biv_os_int}
	With $d=2$, the symbolic likelihood function in Lemma~\ref{lem:mult_os_int} is given by
	\begin{align*}
	L(s; \theta) \propto&
	\left( G_{X_1} ( s_{u,1} ) - G_{X_1} ( s_{l,1} ) \right) ^{u_1 - l_1 - 1} 
	g_{X_1}(s_{l,1}) g_{X_1}(s_{u,1})  g_{X_2}(s_{l,2}) g_{X_2}(s_{u,2})
	\\
	& \times G_{X_1 | X_2 = s_{l,2}}(s_{l,1})
	(1- G_{X_1 | X_2 = s_{u,2}}(s_{u,1})  )
	\left[ G_{X_1}(s_{l,1}) - G_X( s_{\boldsymbol{l}} ) \right]^{l_2 - l_1 - 1} \\
	& \times   
	\left[ G_{X_2}(s_{u,2}) - G_X( s_{\boldsymbol{u}} ) \right]^{u_2 - 1} 
	G_X(s_{\boldsymbol{l}})^{l_2 -1}
	\left( 1 - G_{X_1} (s_{u,1}) - G_{X_2} (s_{u,2}) - G_X ( s_{\boldsymbol{u}} ) \right)^{n - u_1 - u_2}, 
	\end{align*}
	where $G_{X_i}(\cdot) \equiv G_{X_i}(\,\cdot\,; \theta)$ and 
	$G_{X_i | X_j}(\cdot) \equiv G_{X_i | X_j}(\,\cdot\,; \theta); i \neq j$ respectively denote the marginal and conditional distribution functions of $g_\bX(\bx; \theta)$.
\end{cor}

\noindent When $l_1 = \cdots = l_d = 1$ and $u_i = n - 2(i-1)$ (so that the marginal minima and maxima are selected), the sequential nesting random interval construction \eqref{eq:pi_mult_os_int2} approximately reduces to the rectangle construction \eqref{eq:pi_mult_int} based on univariate marginal maxima and minima, indicating some degree of construction consistency. That is,  $S = (S_l,S_u,N)$ contains nearly the same information as the symbol $S = (S_{\min}, S_{\max}, S_p, S_{I_p},N)$ when $S_p = 2d$, and so the symbolic likelihood function (\ref{eq:SeqNestLik}) approximately reduces to (\ref{eq:lik_mult_int}). 
For highly correlated data $S = (S_l,S_u,N)$ is slightly more informative as the lower and upper bounds of each dimension $i$ are calculated on a set from which the $(i-1)$ lowest and largest observations are removed. The approximation improves as the correlation decreases until both symbols become identical when the random variables are completely independent.
A similar reduction cannot be obtained for the iterative segmentation construction.

%%%%%%%%%%%%%%%%%%%%%%%%%%%%%%%%%%%%%%%%%%%%%
\subsection{Modelling histograms with random counts}
%%%%%%%%%%%%%%%%%%%%%%%%%%%%%%%%%%%%%%%%%%%%%
\label{ssec:hist}

Histograms are a popular and typically univariate SDA tool to represent the distribution of continuous data.
%, 
They are commonly constructed as a set of fixed consecutive intervals for which random relative frequencies (or counts) are reported  
\citep[e.g.][]{bock2000, billard2006}.
Following \citet{lerademacher2011}, a histogram-valued random variable may be defined as a set of counts associated with a deterministic partition of the domain $\cX= \real^d$.
Suppose that the $i$-th margin of $\cX$ is partitioned into $B^i$ bins, so that $B^1 \times \cdots \times B^d$ bins are created in $\cX$ through the $d$-dimensional intersections of each marginal bin.
Index each bin by $\boldsymbol{b} = (b_1, \cdots, b_d)$, $ b_j=1, \ldots, B^j$ as the vector of co-ordinates of each bin in the histogram.
Each bin $\bb$ may then be constructed as
$$
\cB_{\bb} =  \cB_{b_1}^1 \times \cdots \times \cB_{b_d}^d
\quad  \mbox{where}\quad\mathcal{B}_{b_j}^j = ( y_{b_j-1}^j, y_{b_j}^j ], \, j=1, \ldots,d,
$$
where for each $j$, the marginal sequences $-\infty < y_0^j < y_1^j < \ldots < y_{B^j}^j <\infty$ are fixed.
We assume that all data counts outside of the constructed histogram are zero.
A $d$-dimensional histogram-valued random variable is constructed through the aggregation function $\pi$ where
\begin{align}
\label{eq:pi_hist}
\begin{array}{rcl}
S = \pi (\bX): & \real^{d \times N} & \rightarrow \cS = \{ 0, \ldots, N \}^{B^1 \times \cdots \times B^d} \\
& \bx & \mapsto \left( \sum_{i=1}^n \indic \{ x_i \in \cB_{\boldsymbol{1}} \}, \ldots, 
\sum_{i=1}^n \indic \{ x_i \in \cB_{\bB} \} \right),
\end{array}
\end{align}
where $\boldsymbol{1} = (1, \ldots, 1)$ and $\bB = (B^1, \ldots, B^d)$, and $\indic$ is the indicator function. 
The symbol $S = (S_{\boldsymbol{1}}, \ldots, S_\bB)$ is a vector of counts, $\sum_{\bb} S_{\bb} = N$, where $S_{\bb}$ denotes the frequency of data in bin $\mathcal{B}_{\bb}$.
\begin{lem}
	\label{lem:mult_hist}
	Consider a multivariate histogram-valued random variable $S \in \cS$, constructed via \eqref{eq:pi_hist} and suppose that $g_\bX(\bx; \theta) = \prod_{i=1}^n g_X(x_i;\theta), \bx \in \real^{n \times d}$. The symbolic likelihood function is given by
	\begin{equation}
	\label{eq:histRandCounts}
	L(s; \theta) =
	\frac{n!}{s_{\boldsymbol{1}}! \cdots s_\bB!}   
	\prod_\bb
	\left( \int_{\cB_\bb} g_X (z; \theta) \der z \right)^{s_\bb},
	\end{equation}
	where the integral denotes the probability that data $x\in\cX$ falls in bin $\cB_{\bb}$ under the model.
\end{lem}
\noindent In the univariate setting, this multinomial likelihood
coincides with the likelihood function for binned and truncated data 
introduced by \citet{mclachlan1988}.
It also extends the 
method of \cite{heitjan1991} who build corrected likelihood functions for 
coarsened data,
where the authors highlight the need to account for both the grouping and the 
stochastic nature of the coarsening.

In the limit as the histogram is reduced to its underlying classical data, the likelihood (\ref{eq:histRandCounts}) reduces to the classical data likelihood.
As the number of bins becomes large 
each bin of the histogram reduces in size and approaches a single point $\mathcal{B}_\bb \rightarrow x_b \in \real^d$. In the limit as the number of bins $\rightarrow\infty$,
only those $n$ coinciding with the underlying data points will have a count of $1$, while the others will have a count of $0$.
The likelihood contribution of the non-empty bins
$\mathcal{B}_\bb$ is then 
$g_X(x_b;\theta)$.
This is equivalent to specifying $f_{S|\bX = \bz} (s; \vartheta) = \prod_{i=1}^n \delta_{z_i}(x_i)$ in (\ref{eq:symb_lik}). 
Consequently 
$
L(\bs; \theta) \propto \prod_{i=1}^n g_{X}(x_i; \theta).
$

Finally, note that 
when the classical data are 
only observed on a subset of the domain $\cX$,
$g_\bX(\bx;\theta)$ should be truncated and rescaled over the same subdomain.

%%%%%%%%%%%%%%%%%%%%%%%%%%%%%%%%%%%%%%%%%%%%%
\subsection{Modelling histograms with random bins}
%%%%%%%%%%%%%%%%%%%%%%%%%%%%%%%%%%%%%%%%%%%%%

A common alternative to histograms with random counts over fixed bins is constructing histograms with fixed counts within random bins
\citep[e.g.][]{mousavi2011, ioannidis2003}.
Such random histograms can be seen as a generalisation of interval-valued random variables (Sections~\ref{ssec:interv}--\ref{ssect:rectangles}). 
In particular, random intervals can be viewed as histograms with the number of bins ranging from $1$ (all margins are intervals calculated from sample minima and maxima; Figure \ref{NeedNewFig}, top) to $3d$ (all margins are intervals calculated from order statistics $l>1$ and $u<n$; Figure \ref{NeedNewFig} bottom left).
In the following we focus on the univariate setting ($\cX = \real$) since extension to $d$-dimensions is challenging.
E.g.~given a matrix of counts, then a simply constructed grid matching these counts does not necessarily exist.

For a vector of orders $k = (k_1, \ldots, k_B)$, such that $1 \leq k_1 \leq \cdots \leq k_B \leq N$,
a univariate random histogram is constructed through the aggregation function $\pi$ where
\begin{align}
\label{eq:pi_os_hist}
\begin{array}{rcl}
S = \pi (\bX): & \real^{N} & \rightarrow \cS = \{ (a_1, \ldots, a_B) \in \real^B: a_1 \leq \cdots \leq a_B \} \times \natural \\
& \bx & \mapsto (x_{(k_1)}, \ldots, x_{(k_B)}, N).
\end{array}
\end{align}
This defines a histogram with bin $b$ located at $(s_{b-1}, s_b]$ with fixed count $k_b - k_{b-1}$, for $b = 1, \ldots, B+1$, where $s_{0} = - \infty, s_{B+1} = + \infty, k_0 = 0$ and $k_{B+1} = N +1$, and knowledge that there is a point located at each $s_b$, $b = 1, \ldots, B$. The symbol $S=((S_1,\ldots,S_B),N)$ is a $B$-vector of order statistics, plus $N$.
\begin{lem}
	\label{lem:os_hist}
	Consider a univariate random histogram $S \in \cS$, obtained through \eqref{eq:pi_os_hist} and assume that $g_\bX(\bx; \theta) = \prod_{i=1}^n g_X(x_i), \bx \in \real^{n \times d}$. Then the symbolic likelihood function is given by
	\begin{equation}
	\label{eq:randhistorder}
	L(s; \theta) = n!
	\prod_{b=1}^{B} g_X (s_b; \theta)
	\prod_{b=1}^{B+1}
	\frac{ \left( G_X(s_{b} ; \theta) - G_X(s_{b-1}; \theta) \right)^{k_b - k_{b-1} - 1}}
	{(k_b - k_{b-1} - 1)!}. 
	\end{equation}
\end{lem}
\noindent When $B=2$, $k_1 = l$ and $k_2 = u$ with $l,u=1, \ldots,n; l  < u$, then (\ref{eq:randhistorder}) reduces to the likelihood function in Lemma~\ref{lem:univ_int} (see Appendix~\ref{app:os_hist}).
Further, under this construction it is straightforward to show that if $B=N$ then the symbolic likelihood (\ref{eq:randhistorder}) recovers the classical data likelihood. 
Specifically this implies $k_b =b$ for all $b=1, \ldots, B$ so that the aggregation function \eqref{eq:pi_os_hist} is $S = \pi(\bX) = ((X_{(1)}, \ldots, X_{(n)}),N)$, 
$k_b - k_{b-1} = 1$ for all $b$ and so 
$L(s; \theta) \propto \prod_{b=1}^N g_X (x_b; \theta)$.

%%%%%%%%%%%%%%%%%%%%%%%%%%%%%%%%%%%%%%%%%%%%%%%%
%
% SECTION: ILLUSTRATIVE EXAMPLES 
%
%%%%%%%%%%%%%%%%%%%%%%%%%%%%%%%%%%%%%%%%%%%%%%%%
%
\section{Illustrative analyses}
\label{sec:examples}

Our symbolic likelihood function resolves many of the conceptual and practical issues with current SDA methods, opens the door for new classes of symbol design and construction, and positions SDA as a viable tool to enable and improve upon classical data analyses. We explore these benefits below.

%%%%%%%%%%%%%%%%%%%%%%%%%%%%%%%%%%%%%%%
\subsection{Effect reconstruction for meta-analyses}
%%%%%%%%%%%%%%%%%%%%%%%%%%%%%%%%%%%%%%%

In medical research, meta-analyses are often implemented to systematically examine the clinical effects of certain treatments, and typically use the effect sample mean and standard deviation from each individual study. However it is common practice that such studies only report certain quantile statistics, namely the sample minimum ($q_0$), maximum ($q_4$) and the sample quartiles ($q_1,q_2, q_3$). This establishes the problem of estimating a sample mean and standard deviation from quantiles.

The most sophisticated practiced method to estimate the sample mean was developed by \cite{luo2018} based on previous work by \citet{hozo2005} and \citet{wan2014}, whereby
\begin{equation}
\label{eq:luo18}
\hat{\bar{x}}_L = w_1 \left( \frac{q_0 + q_4}{2}\right) + w_2 \left( \frac{q_1+q_3}{2}\right)
+ (1-w_1-w_2) q_2,
\end{equation}
with $w_1 = 2.2 /(2.2 + n^{0.75})$ and $w_2 = 0.7 - 0.72 / n^{0.55}$.
Based on previous work by 
\citet{hozo2005} and \cite{bland2015} 
the best performing estimators of the sample standard deviation are due to
\citet{wan2014} and \citet{shi2018}, which are respectively given by
\begin{align}
\label{eq:sd_shi18} 
\hat{s}_W =  \frac{1}{2} \left( \frac{q_4-q_0}{\zeta(n) } + \frac{q_3-q_1}{\eta(n) } \right)
\quad \textrm{ and } \quad
\hat{s}_S =  \frac{q_4-q_0}{\theta_1(n)} + \frac{q_3-q_1}{\theta_2(n)},
\end{align}
where $\zeta(n) = 2\Phi^{-1} \left(\frac{n-0.375}{n+0.25}\right)$,
$\eta(n) =2\Phi^{-1} \left(\frac{0.75n - 0.125}{n+0.25}\right)$,
$\theta_1(n) = (2+0.14n^{0.6}) \Phi^{-1} (\frac{n-0.375}{n+0.25})$, 
$\theta_2(n) = (2+\frac{2}{0.07n^{0.6}}) \Phi^{-1} (\frac{0.75n-0.125}{n+0.25})$, and 
$\Phi^{-1} \left(\cdot\right)$ is the inverse of the standard normal c.d.f.
Each estimator in \eqref{eq:luo18} and \eqref{eq:sd_shi18} assumes the underlying data are normally distributed.

In the context of the symbolic random variables developed in Section~\ref{sec:gen_sym_lik}, this setting corresponds to constructing the symbolic variable $S$ defined through \eqref{eq:pi_os_hist} with $n=4Q+1, Q\in\mathbb{N}$ where $k = (1, Q+1, 2Q+1, 3Q+1, n)$ i.e.~a histogram with $B=4$ random bins and equal counts. 
If we make the same assumption of i.i.d.~normality of the underlying data, then maximising the symbolic likelihood (\ref{eq:randhistorder}) with $g_X(x;\theta)=\phi(x;\mu,\sigma)$ will yield maximum likelihood estimators $\hat{\theta}=(\hat{\mu},\hat{\sigma})\approx(\bar{x},\sqrt{(n-1)/n}s)$
which provide direct estimates $(\hat{\bar{x}}_*, \hat{s}_*)=(\hat{\mu},\sqrt{n/(n-1)}\hat{\sigma})$ of the sample mean $\bar{x}$ and standard deviation $s$ of the underlying data. 
Of course, other distributional assumptions can easily be made.

\begin{figure}[tbh]
\centering
	$
	\begin{array}{cc}
	\includegraphics[width=0.35\textwidth]{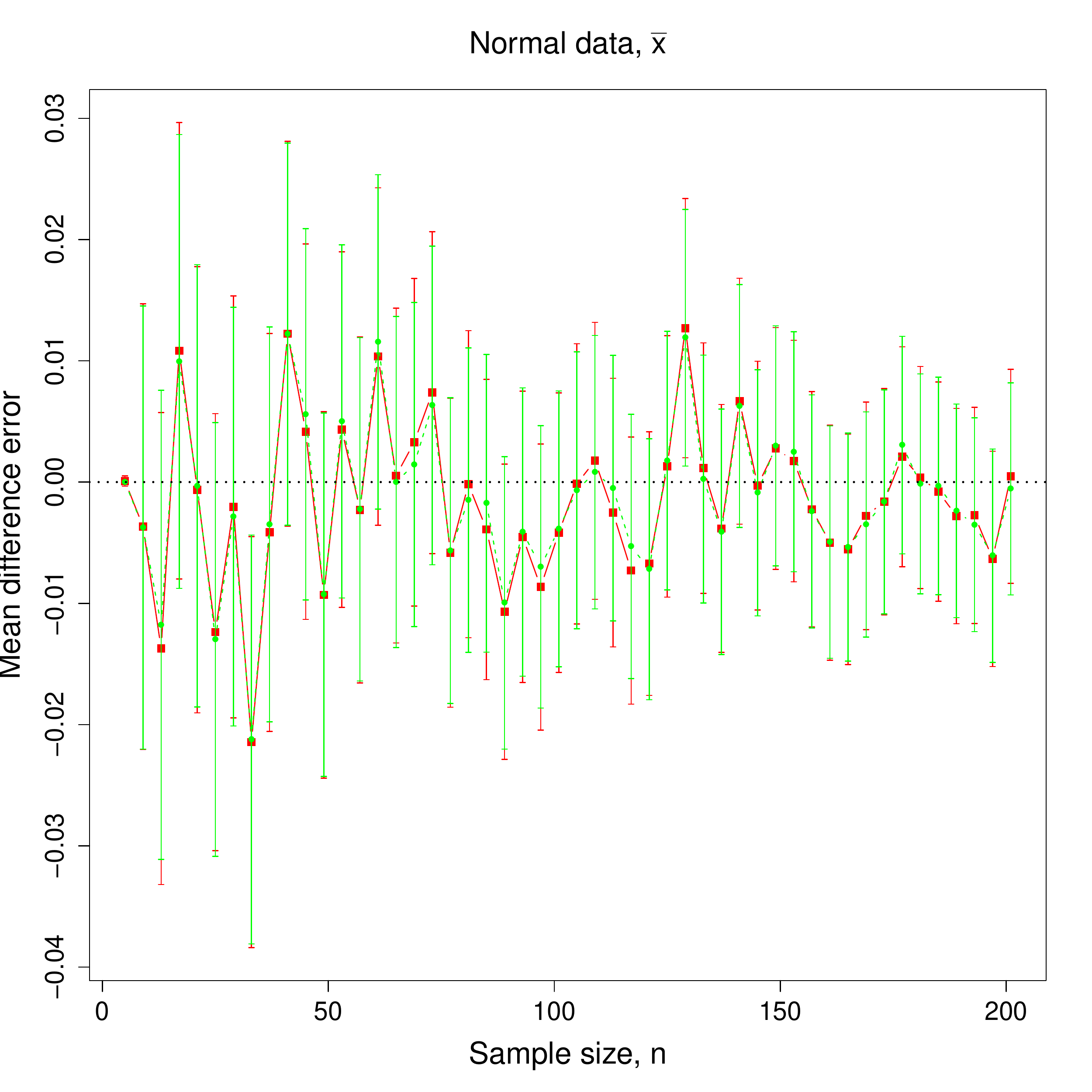} &
	\includegraphics[width=0.35\textwidth]{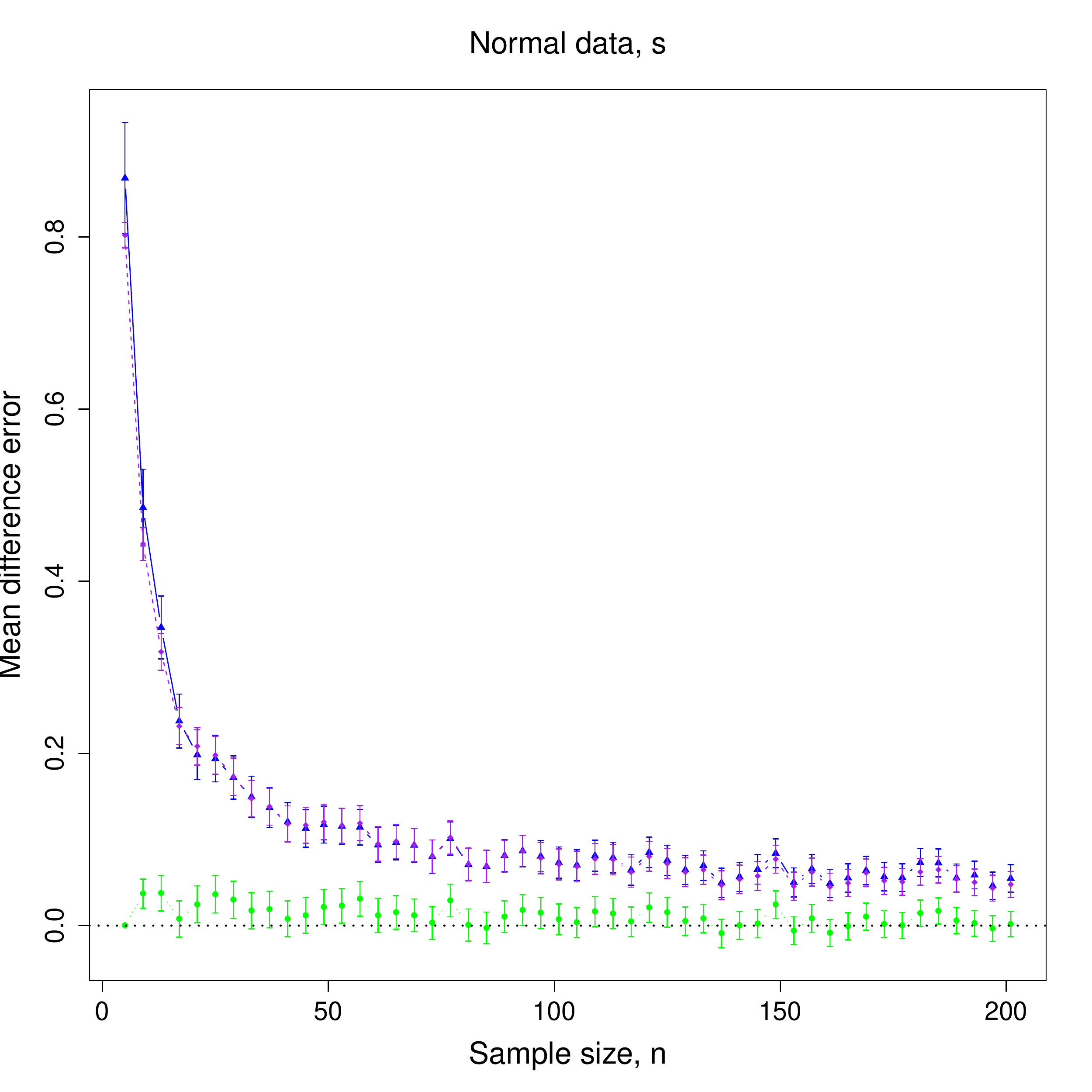} \\
	\includegraphics[width=0.35\textwidth]{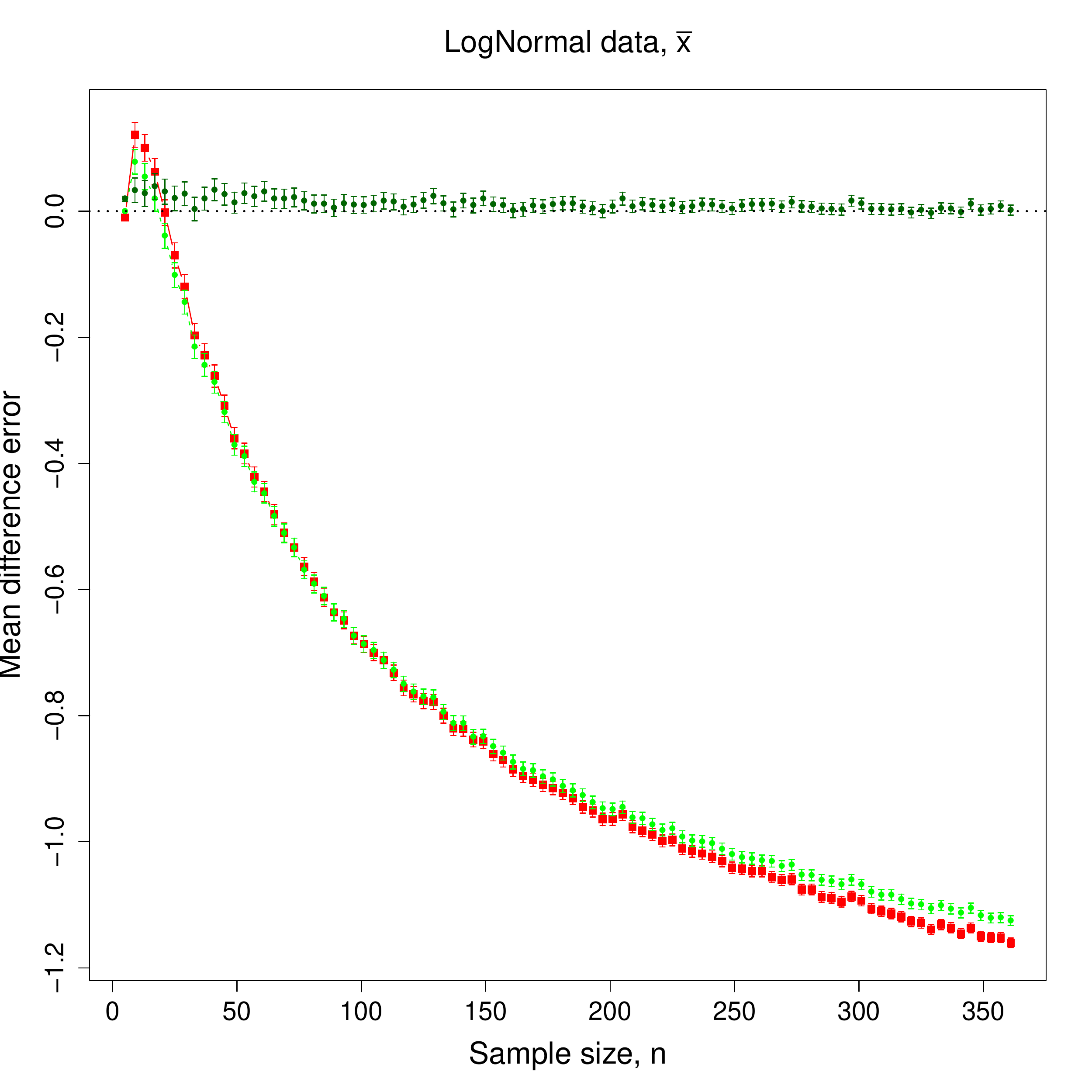} &
	\includegraphics[width=0.35\textwidth]{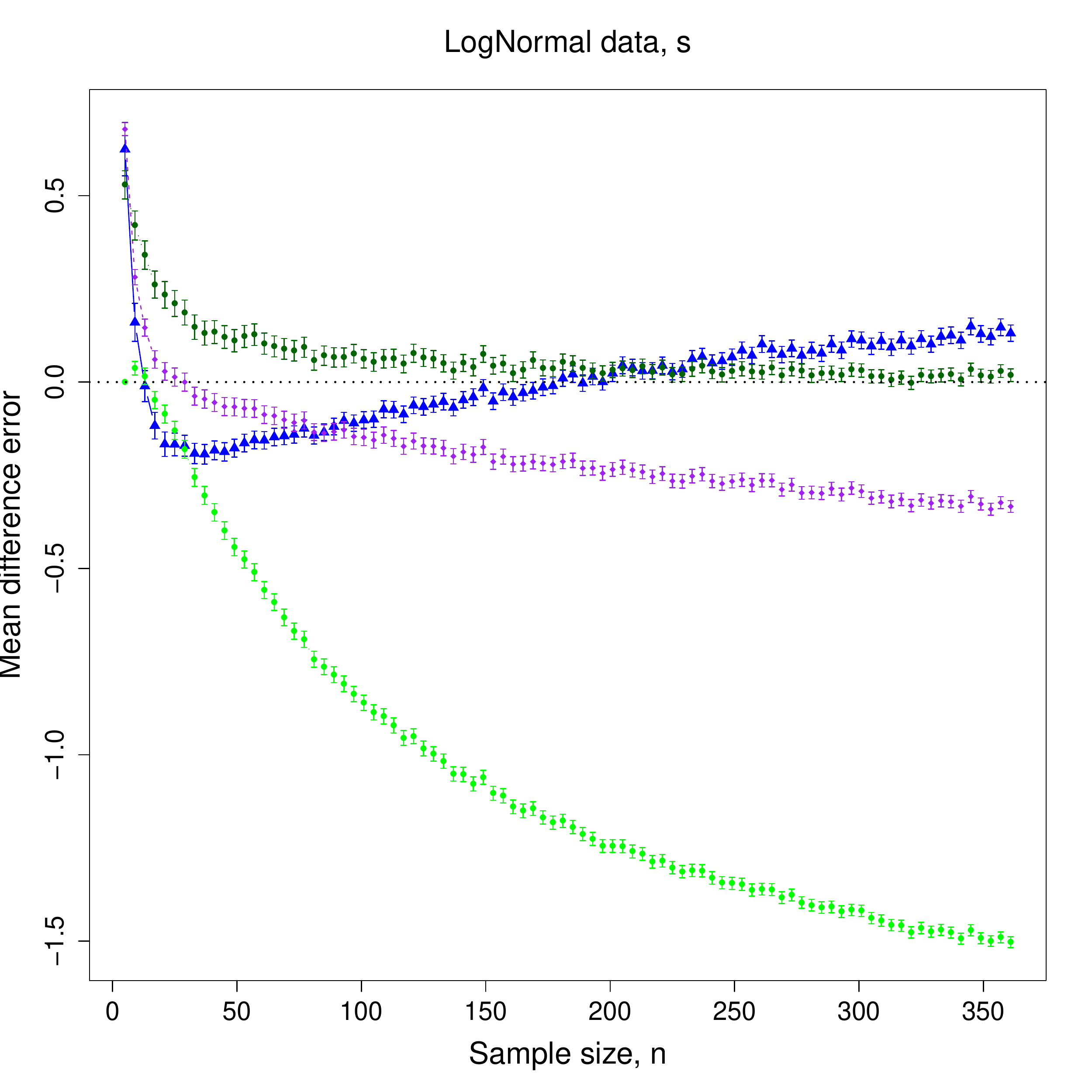}
	\end{array}
	$
	\caption{\footnotesize 
		Mean difference errors, $(\hat{\bar{x}}-\bar{x}_0)$ and $(\hat{s}-s_0)$,  of various estimates of the sample mean (left panels) and standard deviation (right) as a function of sample size $n=4Q+1, Q=1,\ldots,50$ or 90, for both normally (top panels) and log-normally (bottom) distributed data. $\bar{x}_0$ and $s_0$ denote the true sample mean and standard deviation for each dataset. Errors are averaged over $T=10,000$ dataset replicates generated from $\theta_0=(\mu_0,\sigma_0)=(50,17)$ (normal data) and 
		$\theta_0=(\mu_0,\sigma_0)=(4,0.3)$ following  \citet{hozo2005} and \citet{luo2018}.
		Colouring indicates the SDA estimates (light and dark green), $\hat{\bar{x}}_L$ (red), $\hat{s}_W$ (blue) and $\hat{s}_S$ (purple).
		Confidence intervals indicate $\pm 1.96$ standard errors. 
	}
	\label{fig:sample_mean_sd}
\end{figure}

Figure~\ref{fig:sample_mean_sd} illustrates the performance of each estimator compared to the true sample values (i.e. $(\hat{\bar{x}}-\bar{x}_0)$ and $(\hat{s}-s_0)$) based on data generated from normal (top panels) and lognormal (bottom) distributions, averaged over 10,000 replicates, and for a range of sample sizes $n$.
For normal data, the sample mean estimator  $\hat{\bar{x}}_L$ by \cite{luo2018} (red) and the symbolic likelihood-based estimator (green) perform comparably (top left).
Identifying performance differences of  the sample standard deviation estimators is much clearer (top right), with the symbolic estimator strongly outperforming the discipline-standard estimators of \citet{wan2014} and \citet{shi2018} (blue and purple, respectively). The differences are particularly stark for low $n$. As $\hat{s}_W$ and $\hat{s}_S$  substantially overestimate the true standard deviation, their usage 
will systematically undervalue the contribution of each study in any larger analysis, potentially weakening the power of the meta-analysis to detect significant clinical effects.
Note that for $n=5$, the symbolic estimator of the sample standard deviation is exact (i.e.~zero error) as the symbolic likelihood (\ref{eq:randhistorder})  reduces to the classical likelihood in this case.

When the sample data are lognormal (bottom panels), both symbolic (light green) and the industry-standard estimators perform poorly. This is unsurprising 
given the common normality assumption. While estimators equivalent to those in \eqref{eq:luo18} and \eqref{eq:sd_shi18} but for lognormally distributed data could in principle be derived, it is trivial to achieve this for the symbolic estimator by substituting the lognormal density (or any other distribution) for $g_X(\,\cdot\,;\theta)$ in (\ref{eq:randhistorder}). The resulting sample mean and standard deviation estimators assuming the lognormal distribution are illustrated in dark green. The lognormal-based symbolic likelihood estimator performance is clearly excellent in comparison.

One factor influencing the efficiency of the symbolic mle is the form and specification of the symbol as a summary representation of the underlying data. While a histogram with more bins should be more informative than one with less, for a fixed number of bins, sensible choice of  location can result in increased mle performance. This idea of {\em symbol design} has been largely ignored in the SDA literature e.g.~with random intervals routinely constructed from sample minima and maxima.

Consider the simplified setting of the univariate random interval $S=(s_l,s_u,n)$ defined in Lemma~\ref{lem:univ_int} constructed using symmetric upper and lower order statistics, and the associated 2-bin random histogram \eqref{eq:pi_os_hist} that results by additionally including the sample median, $q_2$.
I.e.~for sample sizes $n = 4Q+1, Q \in \mathbb{N}$ we have $l=i,u=n+1-i$ for the interval and $k=(i, 2Q +1, n+1-i)$ for the histogram. We examine the efficiency of the symbolic mle for the symbols defined by $i=1,\ldots,2Q$. For each of $t=1,\ldots,T=10,000$ replicate datasets of size $n=21, 81$ and $201$ (i.e.~$Q=5, 20$, $50$) drawn from a $N(\mu_0,\sigma_0)$ distribution with $(\mu_0,\sigma_0)=(50,17)$, we compute the rescaled symbolic mle $(\hat{\mu}_t, \tilde{\sigma}_t)$ where $\tilde{\sigma}_t=\sqrt{n/(n-1)}\hat{\sigma}_t$, and calculate the relative mean square errors (RMSE) defined by
\begin{equation*}
\textrm{RMSE}_{\hat{\mu}} = \frac{\sum_{t=1}^T ({\hat{\mu}}_t - \mu_0)^2} {\sum_{t=1}^T (\bar{x}_t-\mu_0)^2}
\quad \textrm{ and } \quad\textrm{RMSE}_{\tilde{\sigma}} = \frac{\sum_{t=1}^T (\tilde{\sigma}_t - \sigma_0)^2} {\sum_{t=1}^T (s_t-\sigma_0)^2},
\end{equation*}
where $\bar{x}_t$ and $s_t$ denote the sample mean and standard deviation of the $t$-th replicate.

Figure~\ref{fig:RMSE_quantiles} shows the RMSEs as function of the quantile $q=(n+1-i)/n$ used to construct the symbol. 
As expected, using a histogram (dark lines) provides more information about $\mu$ than the associated random interval (grey), as the extra information contained in the median is informative for this parameter. In contrast, the median  provides no information about $\sigma$ in addition to the two bounding quantiles, for the normal distribution.
Including alternative quantiles would be informative.

The convex shape of each RMSE curve indicates that the prevailing SDA practice of constructing intervals from sample minima and maxima ($i=1,q=n$) is highly inefficient for parameter estimation. Greater precision for both location and scale parameters is achieved by using less extreme quantiles, in this setting around the $q=0.85$-$0.90$ range (balancing optimal minimum RMSE values between the two parameters). There is also a severe penalty for using too low quantiles when estimating $\sigma$, as the data scale is not easily estimated using overly central quantities. Estimating $\mu$ is less sensitive in this regard. These conclusions are robust to sample size, $n$.
Overall this analysis indicates that  substantial efficiency gains should be possible in standard SDA with more informed symbol design.

%%%%%%%%%%%%%%%%%%%%%%%%%%%%%%%%%%%%%%
\subsection{Information content in multivariate random rectangles}
%%%%%%%%%%%%%%%%%%%%%%%%%%%%%%%%%%%%%%
\label{ssec:biv_int}

In Sections \ref{sssect:minmax} and \ref{ssect:order} we introduced two new symbolic constructions to increase the information content within multivariate random rectangles. We now examine the performance of each of these representations and contrast them with standard SDA  constructions. While we focus on bivariate intervals for clarity, extension of the results to higher dimensions is immediate.

When constructing random rectangles from marginal minima and maxima, Lemma~\ref{lem:mult_int} and Corollary~\ref{cor:biv_int} provide an expression for the symbolic likelihood that incorporates full knowledge of the number and location of unique points from which the interval is constructed (e.g.~Figure \ref{NeedNewFig}). We denote the resulting likelihood function \eqref{eq:lik_mult_int} by $L_\textrm{full}(s;\theta)$. 
Existing SDA definitions of random rectangles do not use this information. In its absence, the best likelihood model that can be constructed is by averaging the likelihood $L_\textrm{full}$ over all possible combinations of the unique point constructions, weighted according to the probability of that configuration arising under the classical data model. That is,
$$
L_{\emptyset}(s; \theta) = \sum_{t_p} \sum_{t_{I_p}} L_\textrm{full} ( (s_{\min},s_{\max}, t_p, t_{I_p},n); \theta ) \prob(S_p = t_p, S_{I_p} = t_{I_p}; \theta),
$$
where
\begin{equation}
\label{eq:mixProb}
\prob(S_p = t_p, S_{I_p} = t_{I_p}; \theta) 
=  \int \int L_\textrm{full}( (a, b, t_p, t_{I_p}, n); \theta ) \prod_{i=1}^dI(a_i\leq b_i)\,\der a \der b,
\end{equation}
where $a=(a_1,\ldots,a_d)$ and $b=(b_1,\ldots,b_d)$.
While not generally viable, below we estimate the probabilities \eqref{eq:mixProb} to high accuracy using Monte Carlo with a large number of samples, each time $L_\emptyset$ is evaluated. One alternative is to assume each random rectangle is constructed by the maximum number of unique points ($2d$), which is perhaps realistic when the number of points $n_c$ underlying a symbol is large and the dependence between the variables not too strong.
We denote the particular case of $L_\textrm{full}$ with $S_p=2d$ as $L_{2d}(s;\theta)$.
Here, $L_{2d}$ effectively represents the current state-of-the-art in SDA methods, $L_\emptyset$ represents the best that can likely be done with the existing constructions of random rectangles in the SDA literature (although it is likely impractical), and $L_\textrm{full}$ is our construction.

We assume $m=20,50$ classes, for each of which a random sample of size $n_c=5,10,50,100$ is drawn from a $N_2(\mu_0,\Sigma_0)$ distribution  ($d=2$) with $\mu_0 =  (2,5)^\top$, 
$\diag(\Sigma_0)=(\sigma^2_{0,1},\sigma^2_{0,2})=(0.5,0.5)$
and correlation $\rho_0=0,0.3,0.5,0.7,0.9$. 
The $m$ random rectangles are then constructed, retaining the information $(s_p,s_{I_p})$ required to maximise $L_\mathrm{full}$ but which is ignored when maximising $L_\emptyset$ and $L_4$.
For each of $T=100$ replicate datasets, the symbolic mle $\hat{\theta}=(\hat{\mu},\hat{\Sigma})$ is computed.

Table~\ref{tab:cor_biv_int} reports the mean and standard deviation of $\hat{\rho}$ over the replicate datasets under each likelihood. The marginal parameters ($\mu$, $\sigma_{1}$ and $\sigma_2$) are well estimated in each case (see Supplementary Material \ref{suppmat:biv_int_est}).
The main conclusion  from Table~\ref{tab:cor_biv_int} is that
only $L_\textrm{full}$, which incorporates full information of the number and location of the unique points that define the random rectangle, is able to accurately estimate dependence between the variables. For $L_4$ and $L_\emptyset$ the mle's are either zero (no dependence can be estimated) or they are biased upwards.
Note that for $L_\textrm{full}$, variability of the mle mostly increases  as $n_c$ increases, and is  more variable for lower correlation values. This can be explained as dependence information is contained in the proportion of rectangles constructed from 2 and 3 unique points (and their locations). For a fixed correlation, as $n_c$ gets large it is increasingly likely that the rectangles will be generated by 4 unique points, thereby weakening the dependence information that the sample of random rectangles can contain. 
This weakening naturally occurs more slowly for higher correlations, and so the correlation mle has greater accuracy and precision for stronger dependence.
As $n_c\rightarrow\infty$ all rectangles will be generated from 4 unique points, and it will not be possible to accurately estimate within-rectangle dependence. This effect can be seen for $n_c=1,000$ and $100,000$ for $\rho=0.3,0.5,0.7$ but not yet for $\rho=0.9$.
 This insight identifies clear limits on the dependence information content that this (discipline standard) interval construction can possess.

\begin{figure}[tbh]
\centering
	$
	\begin{array}{cc}
	\includegraphics[width=0.35\textwidth]{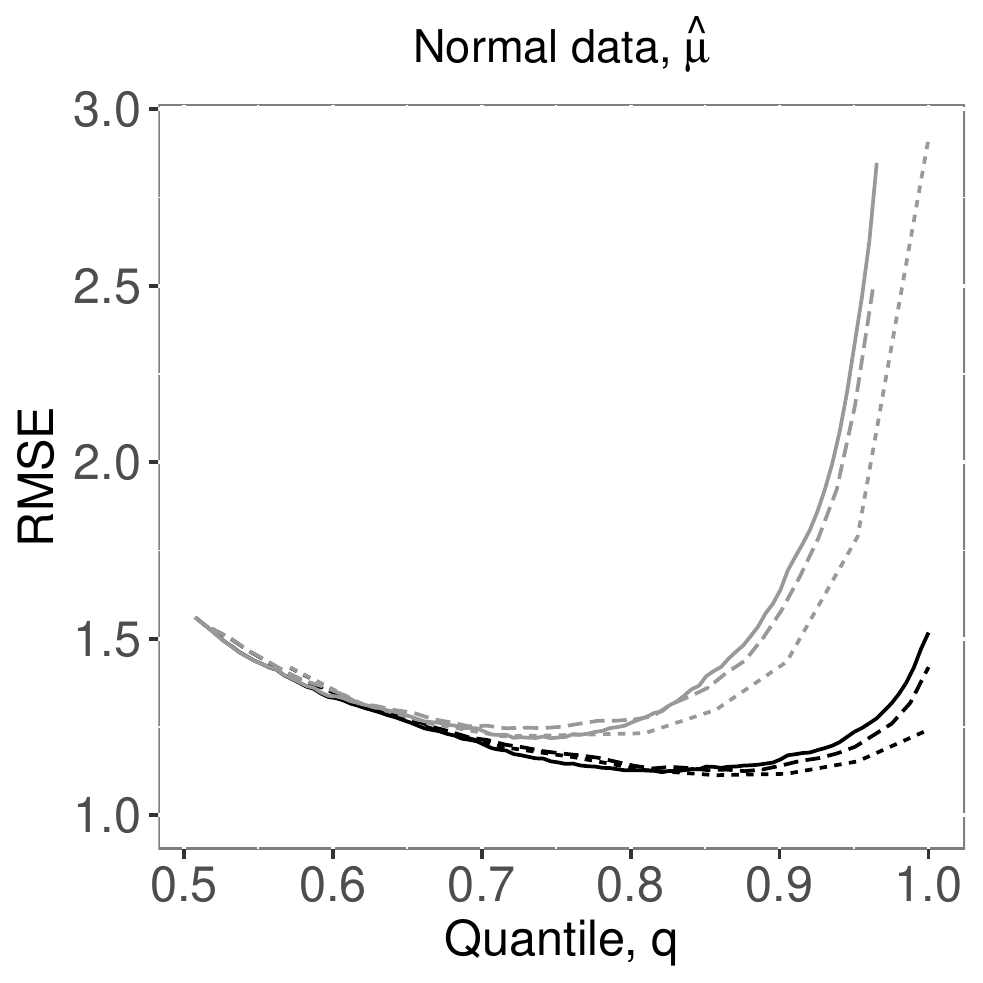} & 
	\includegraphics[width=0.35\textwidth]{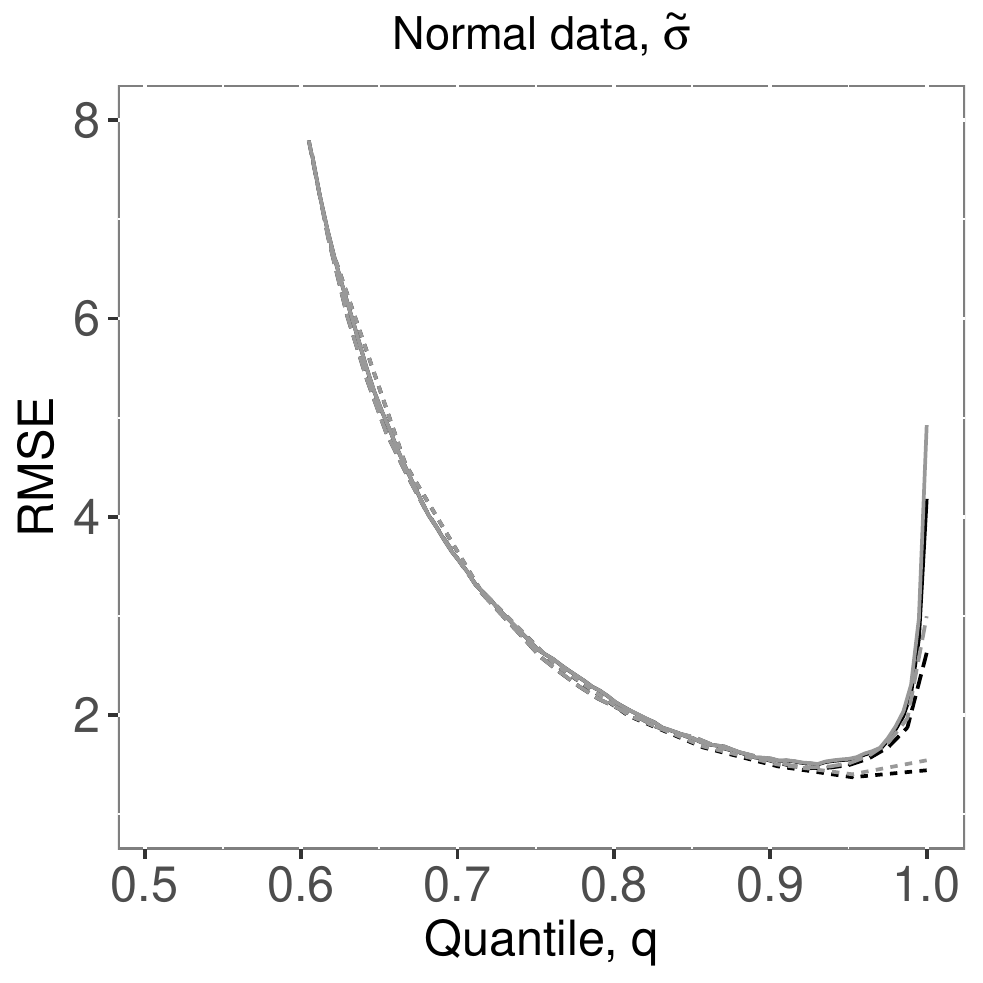}
	\end{array}
	$
	\caption{\footnotesize $\textrm{RMSE}_{\hat{\mu}}$ (left) and $\textrm{RMSE}_{\tilde{\sigma}}$ (right) as a function of quantile $q=(n+1-i)/n$ for $i=1,\ldots,(n+1)/2$. 
		Grey and black lines respectively denote random intervals and histograms.
		Solid, long-dashed and short-dashed lines indicate samples of size $n=21, 81$ and  $201$ respectively.
	}\label{fig:RMSE_quantiles}
\end{figure}

Given the statistical inefficiency of intervals constructed from minima and maxima (Figure \ref{fig:RMSE_quantiles}) and their informational limits,
a sensible alternative is to construct random rectangles using marginal order statistics (Section \ref{ssect:order}), which should be robust to these limitations. Given that such intervals constructed from independent marginal quantiles (\eqref{eq:marginalQuantile} and \eqref{eq:simpleQuant}) will not contain dependence information, we examine the performance of the sequential nesting \eqref{eq:pi_mult_os_int2} and iterative segmentation \eqref{eq:pi_mult_os_int} constructions, for which we  denote the respective likelihood functions  as $L_\textrm{sn}(s;\theta)$ and $L_\textrm{is}(s;\theta)$.

For each of $T=100$ replicate datasets, we generate $m=20$ classes, each constructed from $n=60$ and $300$ draws from a bivariate ($d=2$) $N_2(\mu_0,\Sigma_0)$ distribution with $\mu_0=(2,5)^\top$, $\sigma_{0,1}=\sigma_{0,2}=0.5$ and correlation $\rho_0=-0.7,0,0.7$. The symbols are constructed in four ways: $L_{\textrm{sn},x}$ using sequential nesting \eqref{eq:pi_mult_os_int2}; 
$L_{\textrm{sn},y}$ using sequential nesting but by exchanging the conditioning order of the $x$ and $y$ margins for symbol constuction;
$L_{\textrm{is},x}$ using iterative segmentation \eqref{eq:pi_mult_os_int}; 
$L_{\textrm{is},y}$ using iterative segmentation  but again by exchanging the conditioning order of the $x$ and $y$ margins.

Table~\ref{tab:biv_os_int_rho07} reports the mean (and standard deviation) of $\sigma_1,\sigma_2,\rho$
under each experimental setup when $\rho_0=0.7$ (results for $\rho_0=-0.7$ and $0$ are in Supplementary Material \ref{suppmat:biv_int_est2}). 
Estimates of $\sigma_1$ and $\sigma_2$ are unbiased for any rectangle configuration. However the standard deviations of the estimates are smaller for components which are conditioned on first in the symbol construction e.g.~$\sigma_1$ is more precisely estimated by $L_{sn,x}$ and $L_{is,x}$, and $\sigma_2$ by $L_{sn,y}$ and $L_{is,y}$. 
Constructing intervals using iterative segmentation produces more precise estimates of the correlation $\rho$ than using sequential nesting. This is because iterative segmentation provides more information about joint upper and lower values of the margins than nested segmentation, which provides stronger information about the centre of the marginal distributions (Figure \ref{NeedNewFig}).
Different axis constructions ($L_{\cdot,x}$ or $L_{\cdot,y}$) have little effect on the estimates in this case, due to the symmetry of the underlying Gaussian distribution.
As expected, increasing the amount of data per symbol, $n_c$, leads to more precise estimates of all parameters.

All estimates of $\rho$ are more precise than that obtained using  marginal minima and maxima, which gave a mle standard deviation of 0.0720 (for $n_c=50, m=20, \rho_0=0.7$ and using $L_\textrm{full}$ in Table~\ref{tab:cor_biv_int}).
Similar to  Figure \ref{fig:RMSE_quantiles}, within any method of symbol construction, the choice of order statistics has an impact on the performance of the mle. Clearly there is an important optimal symbol design question to be addressed, that goes beyond the scope of this paper. However, the iterative segmentation approach appears to be more informative for all parameters, for reasons described above. It is likely that there are other random rectangle constructions that would be even more informative.

%%%%%%%%%%%%%%%%%%%%%%%%%%%%%%%%%%%%%%%%%%%%%%%%
%
% SECTION: REAL DATA ANALYSIS
%
%%%%%%%%%%%%%%%%%%%%%%%%%%%%%%%%%%%%%%%%%%%%%%%%

\subsection{Peer-to-peer loan data analysis}
\label{section:CreditAnalysis}

We analyse data from the U.S.~peer-to-peer lending company {\em LendingClub}
available from the Kaggle platform (\url{https://www.kaggle.com/wendykan/lending-club-loan-data}). 
After removing missing values, it comprises $887,373$ loans issued during 2007--2015,
each with an associated grade, from A1 (least risky) to G5 (most risky),  based on risk and market conditions, which defines the interest rate.
We examine the link between the borrower's $\log$ annual income (in \$US) and loan grade via a highly computational analysis of the full dataset (providing a gold standard), a symbolic analysis using \eqref{eq:pi_hist} based on aggregating the income data in each risk group into a 5-bin histogram, and a  reference SDA analysis following \citet{lerademacher2011} (denoted LRB).

Denoting $X_{ij}$ as the log-income for individual $j$ in grade $i=1,\ldots,35$,
we consider both normal  
$X_{ij} \sim N(\mu_i, \sigma_i^2)$ and skew-normal $X_{ij} \sim SN(\mu_i, \sigma_i^2, \gamma_i)$ models for each grade (with the skew-normal parameterised in terms of mean $\mu_i$ and variance $\sigma^2_i)$, given that standard likelihood ratio tests identify the presence of asymmetry in 34/35 groups ($\alpha=0.05$). Within-grade sample sizes range from 576 (G5) to 56,323 (B3).
Coding the ordered grades A1--G5 as the numbers 1--35, each model specifies
\begin{equation}
\label{eq:hierar_mod}
\mu_i \sim T_3( c_0 + c_1 i + c_2 i^2, \tau^2) 
\quad\mbox{and}\quad
\sigma_i^2 \sim IG(\alpha, \beta),
\end{equation}
where $T_\nu(m, v)$ denotes a $t$-distribution with mean $m$, variance $v$ and $\nu$ degrees of freedom, and $IG(\alpha,\beta)$ the inverse-Gamma distribution with shape $\alpha$ and scale $\beta$. For the skew-normal model we additionally specify
$
\gamma_i \sim N(\eta, \epsilon)
$
for the skewness parameter. For the reference LRB analysis, we implement the model \eqref{eq:hierar_mod} where $\mu_i$ and $\sigma^2_i$ correspond to the mean and variance of the histogram of the $i$-th group \citep[Section~2.3]{lerademacher2011}.

\begin{figure}[tbh]
	\centering
	$
	\begin{array}{cc}
	\includegraphics[width=0.35\textwidth]{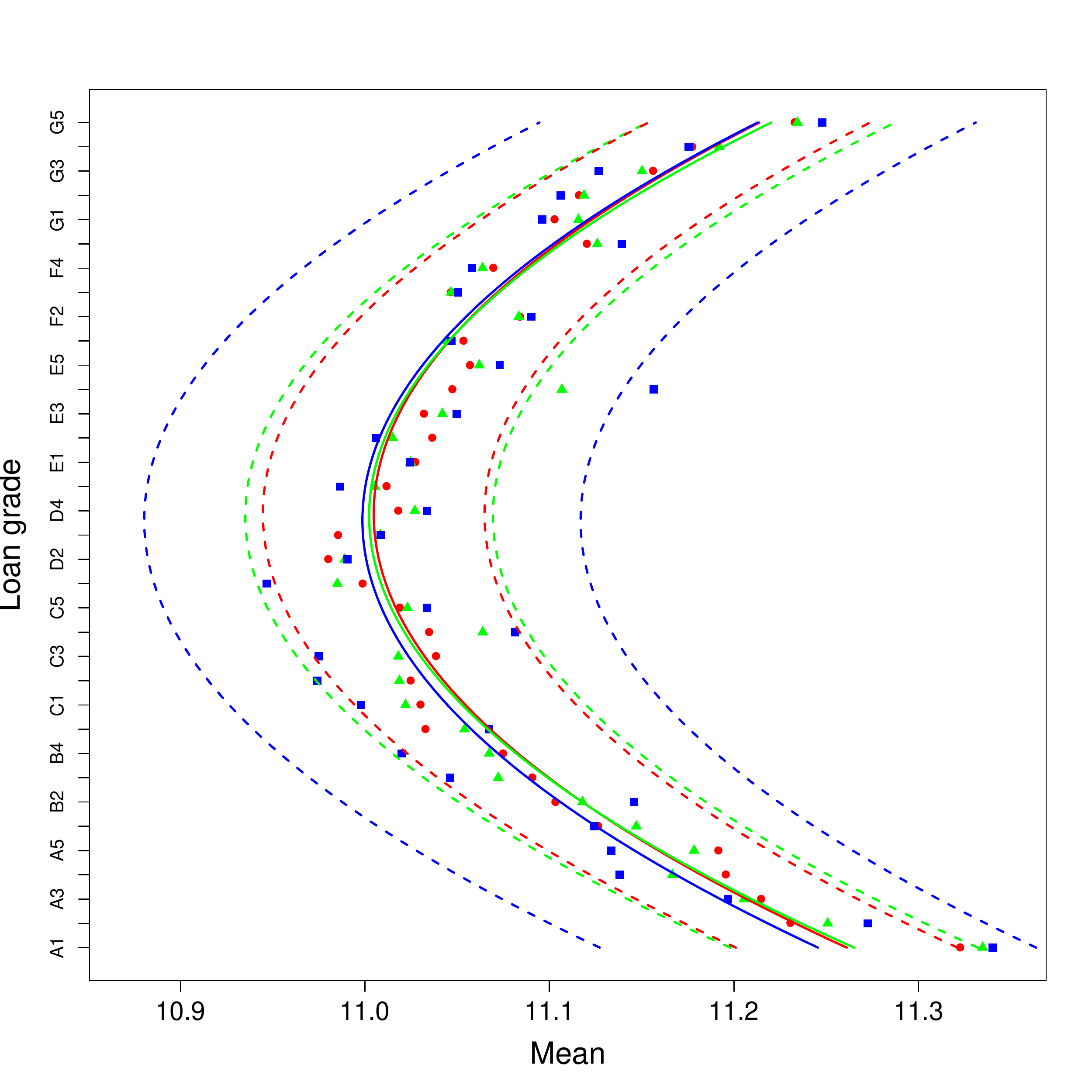} &
	\includegraphics[width=0.35\textwidth]{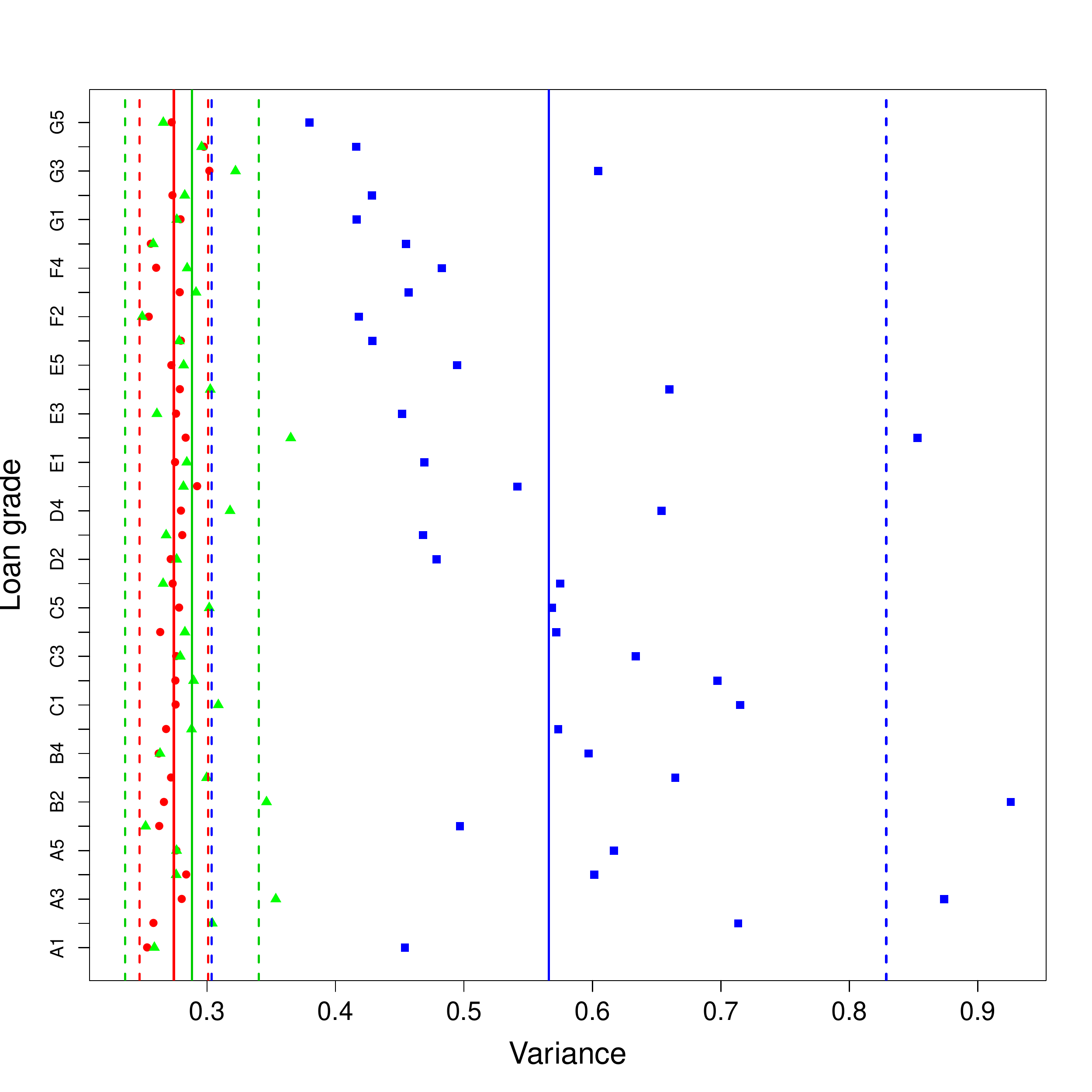} \\
	\includegraphics[width=0.35\textwidth]{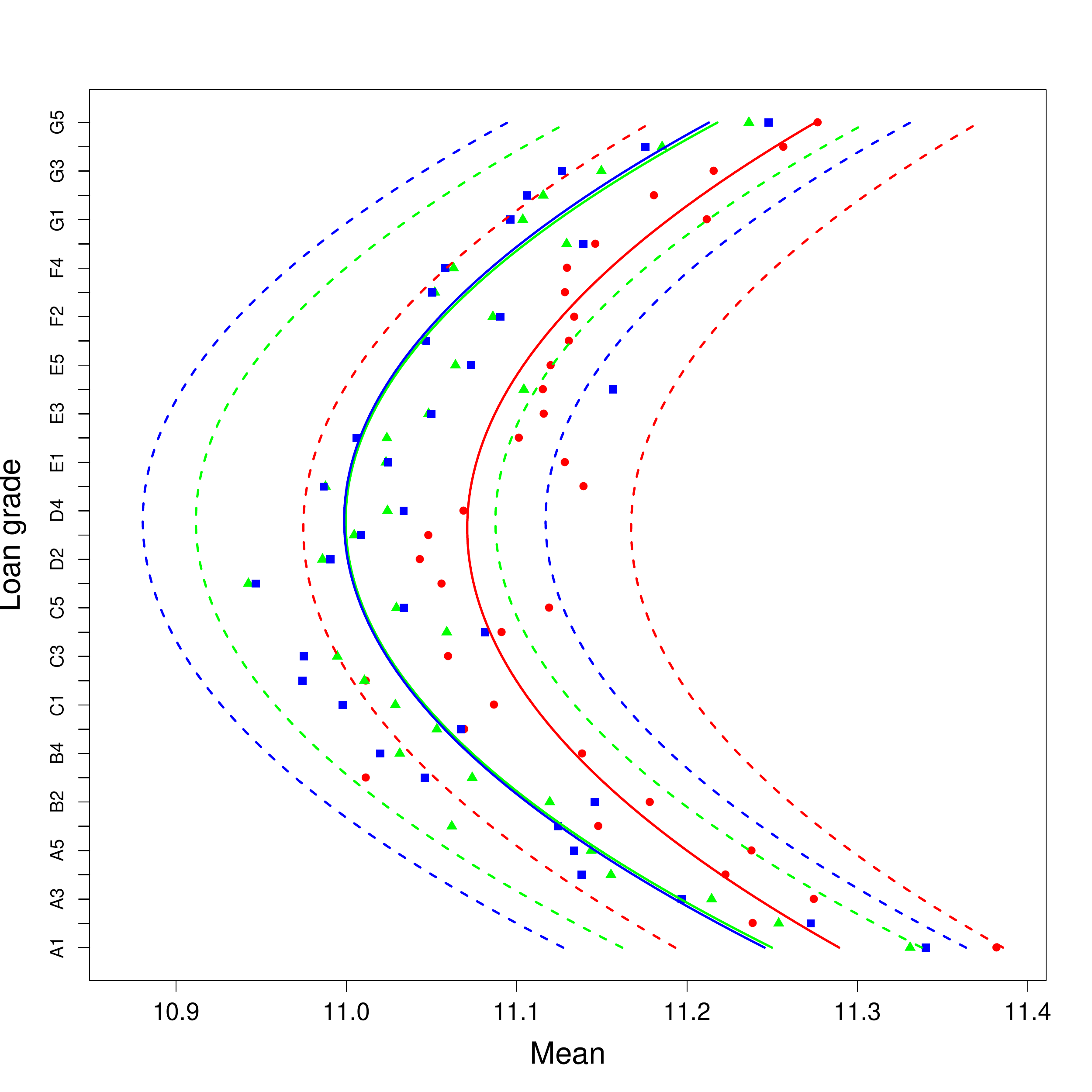} &
	\includegraphics[width=0.35\textwidth]{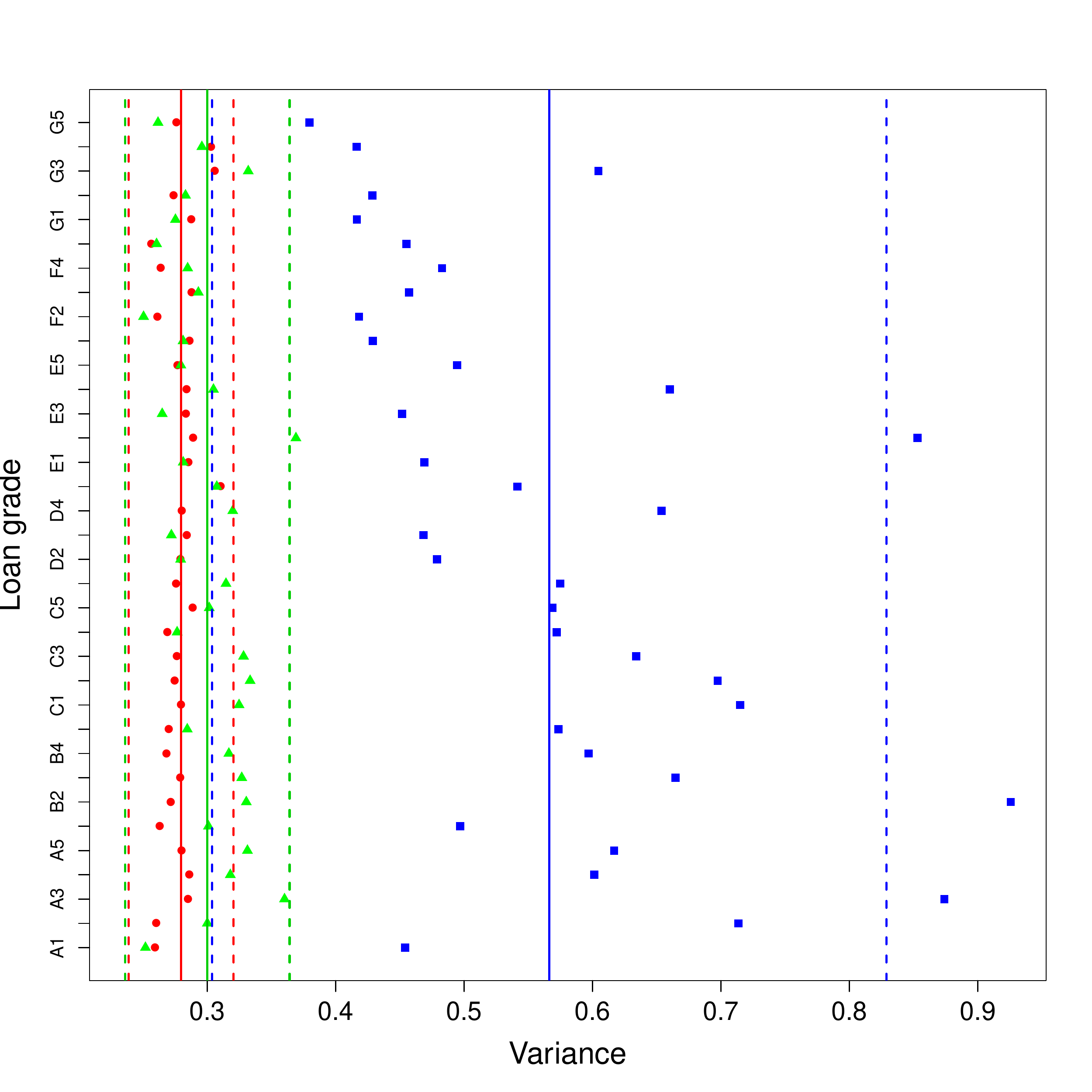} \\
	\end{array}
	$
	\caption{\label{fig:mle.plot} \small 
		Fitted group means and variances (solid lines) when the underlying distribution is Normal (top) and skew-Normal (bottom), using the classical (red) and symbolic (green) likelihoods and the LRB model (blue).
		Dashed lines indicate pointwise 95\% confidence intervals.
		Points denote $\hat{\mu}_i$ and $\hat{\sigma}^2_i$ under the classical and symbolic models, and the sample mean and variance of each grade histogram for the LRB model.
		} 
\end{figure}

Figure~\ref{fig:mle.plot} presents the fitted group means and variances obtained through each method.
The grade specific means under the Normal (top row) model are uniformly well estimated. Our symbolic model produces standard errors only slightly larger than the classical ones, while those from the LRB model are about twice as large. The means under the skew-Normal (bottom) model are less well estimated, but remain, for the majority, within the classical 95\% confidence band. 
The right panels highlight the inability of the LRB method to correctly replicate the classical analysis.
This is essentially because the LRB approach models the variances of a histogram generated from the underlying data (assuming uniformity within bins), rather than modelling the variance of the underlying data. However, given the same histograms, our symbolic approach approximates the classical analysis well.

\begin{figure}[t!]
	\centering
	$
	\begin{array}{ccc}
	\includegraphics[width=0.3\textwidth]{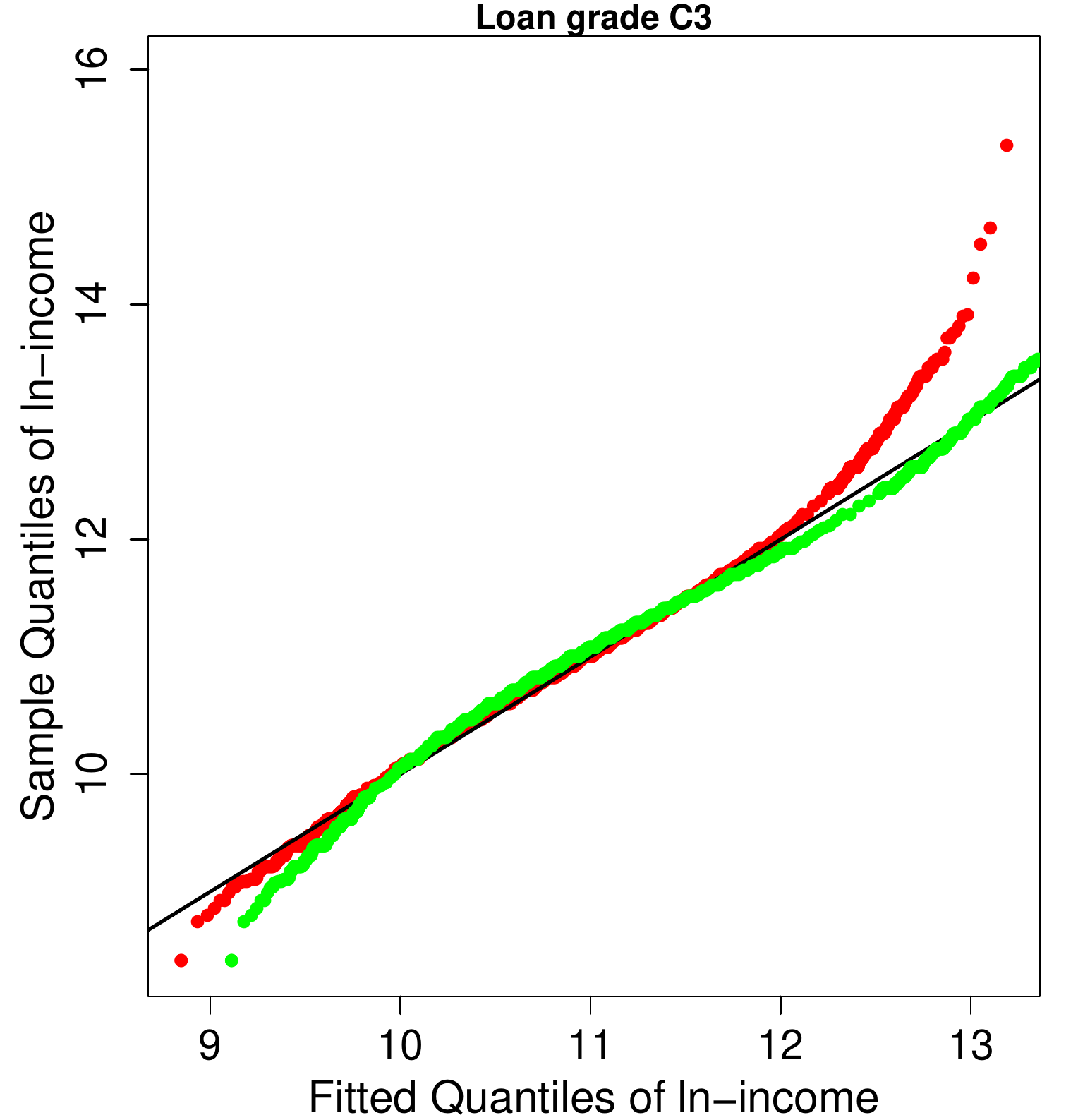} & 
	\includegraphics[width=0.3\textwidth]{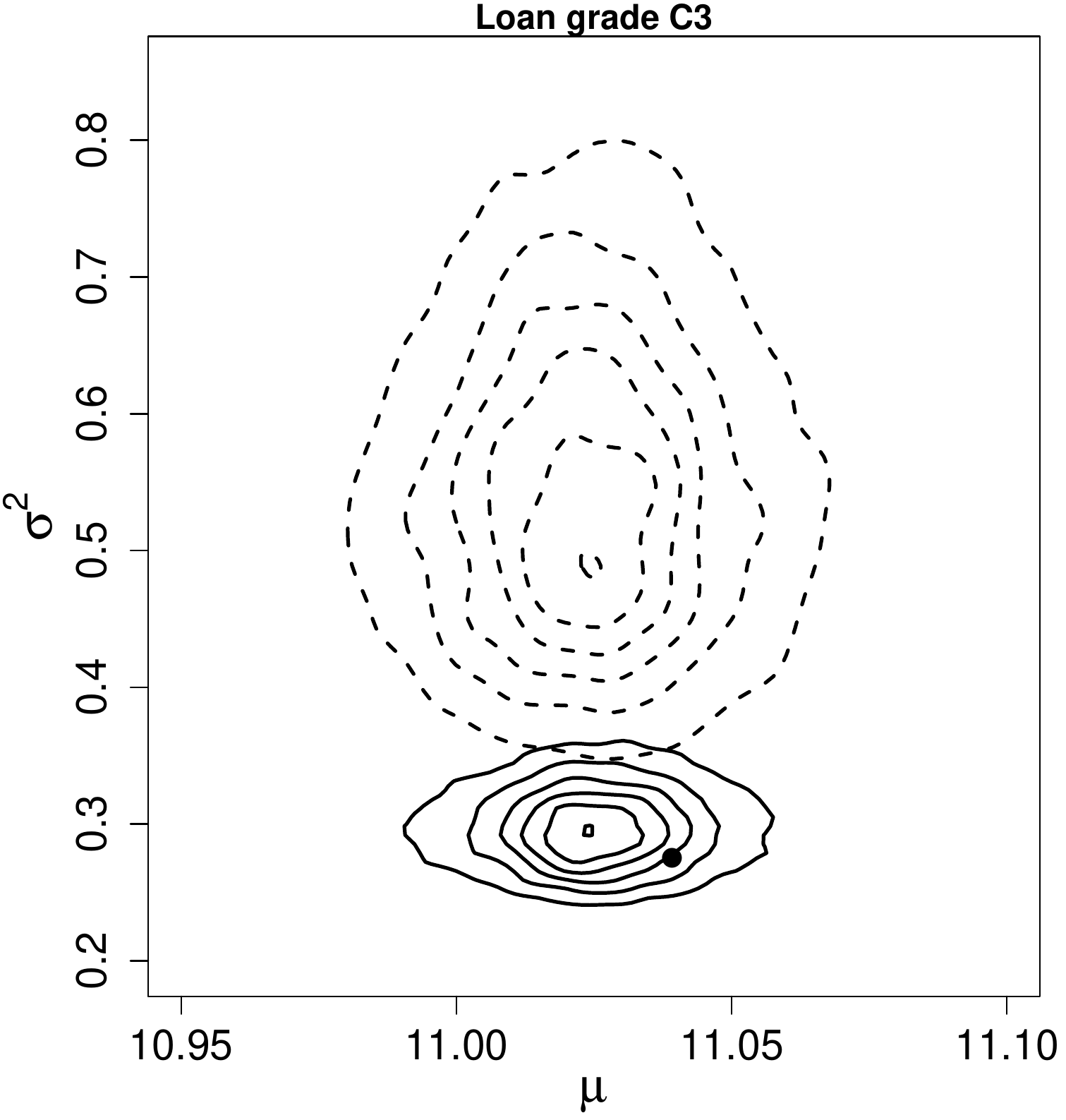}  &
	\includegraphics[width=0.3\textwidth]{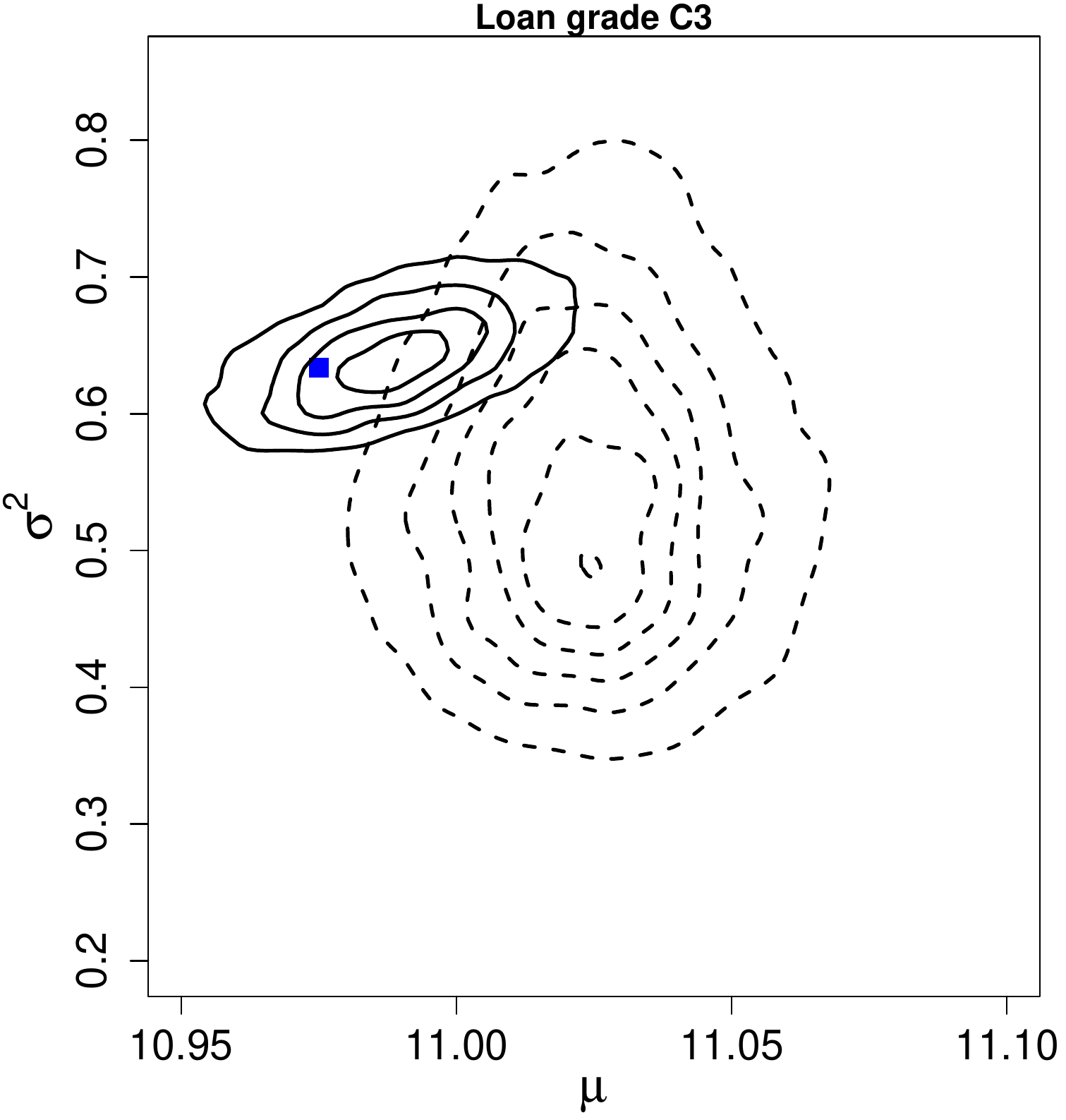} 
	\end{array}
	$
	\caption{\label{qq_plot} \small 
	Predictive inference for loan grade C3 ($n_{C3}=50,161$). Left panel: qq-plot of fitted vs.~empirical quantiles for our symbolic likelihood under Normal (red) and skew-Normal (green) distributions. 
	Predictive distributions of the group mean ($\mu_i$) and variance ($\sigma^2_i$) at the underlying data (centre panel; obtained by computing sample means and variances of data generated from the fitted model) and symbol (right panel; obtained by computing means and variances of histograms constructed from data generated from the fitted model) levels. Solid and dashed contours respectively denote the symbolic and LRB predictive distributions. 
	The LRB model predicts directly at the symbol level (not via underlying data).
	Black dot and blue square denote the underlying-data-based and histogram-based mean and variance for group $C3$. 
	%}
	}
	\label{fig:qqplots}
\end{figure}

By design, the LRB approach cannot discriminate between normal and skew-normal models
(the LRB fits in Figure~\ref{fig:mle.plot} for both models are the same), unlike our symbolic analysis which approximates the full classical analysis.
This means that we are able to make inference at both the level of the underlying data as well as the symbol level (LRB is restricted to the latter).
This is illustrated for the distribution of loan grade C3 in Figure~\ref{fig:qqplots}. Qq-plots for both models (left panel) suggest the skew-normal model appears to be a better fit in the upper tail of the $\log$ income distribution, and slightly worse in the lower tail. This could  be tested formally. 
As the LRB approach cannot make such judgements of model adequacy, it is confined to predictions about the mean and variance of (histograms constructed from data generated by) this underlying process. 
These predictive distributions are shown in the right panel. Even when considered on its own terms, the LRB method produces less accurate and precise predictions than our symbolic approach (the dot indicates the observed histogram mean/variance).
Beyond this, our symbolic approach can produce the equivalent predictive distributions for the sample mean and variance of the underlying predicted data, without first producing histograms (centre), which is perhaps more useful in an analysis as it captures knowledge of the underlying data generation process. The LRB method cannot produce these predictions.

Finally, Table~\ref{tab:times} provides the mean time to evaluate each likelihood function under each model, averaged over
1,000 randomly generated parameter vectors. 
The LRB analysis is most efficient as it is based on a likelihood with 35 bivariate points.
The classical analysis is efficient for the normal model given the available sufficient statistic for each loan grade, however the  the skew-normal likelihood requires iteration over all 887,373 records. In contrast, our symbolic likelihood \eqref{eq:histRandCounts} requires 6 cdf evaluations per loan grade. It is slower than the normal classical analysis, but 14 times faster than the skew-normal classical analysis, with  comparable model fit. The symbolic computational times will remain roughly constant as the dataset size increases (right column), generating increasingly large computational savings for the skew-normal model compared to the classical analysis.

%%%%%%%%%%%%%%%%%%%%%%%%%%%%%%%%%%%%%%%%
%%%%%%%%%%%%%%%%%%%%%%%%%%%%%%%%%%%%%%%%
\section{Discussion}
%%%%%%%%%%%%%%%%%%%%%%%%%%%%%%%%%%%%%%%%
%%%%%%%%%%%%%%%%%%%%%%%%%%%%%%%%%%%%%%%%
\label{sec:discussion}

In this article we have introduced a new framework for the analysis of data that have been summarised into distributional forms.
For the general statistical analyst, this method opens up the use of SDA as a broadly applicable statistical technique for analysing large and complex datasets with the potential for large data-storage and computational savings.
Within the SDA setting, the fundamentally different approach taken -- that of specifying probability models for the data underlying a symbol and deriving the resulting model at the symbolic level, rather than direct model specification at the symbolic level -- has introduced a new research direction in the field of SDA.
The proposed framework resolves open and new problems including the difficulty of specifying meaningful models at the symbolic level, avoidance of the routinely violated uniformity-within-symbols assumption, the ability to perform accurate inference at the level of the underlying data, including model choice, and providing a means to construct and analyse multivariate symbols.
We have exposed some weaknesses of current symbol design, and have introduced several new, more efficient symbol constructions.

While providing a step forwards, our approach is not without some caveats.
Most obviously, the symbolic likelihood function (\ref{eq:symb_lik}) requires enumeration of the integral over the underlying data space, which may be problematic in high dimensions. For many standard classes of models, 
distribution functions $G_\bX(\bx;\theta)$ are available in closed form. In other cases, numerical or approximate methods may be required, such as quadrature, Monte Carlo techniques \citep{andrieu+r09}, or factorisation of $g_\bX(\bx;\theta)$ to reduce the dimension of the integral. Alternatively, composite likelihood solutions \citep{whitaker2019} can be considered.

The symbolic likelihood is clearly an approximation of the classical likelihood as it is based on summary data, and so there will be some information loss. While the accuracy of the classical data model can be approached by letting the symbols approach the classical data (e.g.~by letting the number of  histogram bins $B\rightarrow\infty$), this may not be viable in practice, and in the extreme (e.g.~with huge numbers of bins) the computational costs could exceed those of the classical analysis.
It is therefore of interest, and the subject of future research, to understand the quality of the approximation. It is possible that some of the theory supporting approximate Bayesian computation
\citep[e.g.][]{ABChandbook}, which is also based on computation via summary statistics, could be useful here.

Within this context there is immense scope for optimum symbol design, whereby  symbols are constructed to provide maximal information for a specific or more general analyses 
that may be performed in the future.
New symbolic types could be developed such as 
continuous distribution-based symbols, which may additionally enable direct integration of the integral in (\ref{eq:symb_lik}) through conjugacy.

Since \cite{schweizer84}'s 35-year old prediction that ``distributions are the numbers of the future'', the explosive emergence of the data-rich biome -- the {\em infome} -- in which we now reside,
clearly substantiates the potential for
symbolic data analysis to become a powerful everyday tool for the statistical analyst.
\cite{schweizer84}'s future is very much here.

%%%%%%%%%%%%%%%%%%%%%%%%%%%%%%%%%%%%%%%%
%%%%%%%%%%%%%%%%%%%%%%%%%%%%%%%%%%%%%%%%
\section*{Acknowledgements}
%%%%%%%%%%%%%%%%%%%%%%%%%%%%%%%%%%%%%%%%
%%%%%%%%%%%%%%%%%%%%%%%%%%%%%%%%%%%%%%%%

%The authors are supported by the Australian Centre of Excellence for Mathematical and Statistical Frontiers (ACEMS, CE140100049). BB and SAS are also supported by the Australian Research Council Discovery Project Scheme (DP160102544 and FT170100079). 
The authors thank Felix Camirand Lemyre and Tom Whitaker for many useful conversations.

\bibliographystyle{chicago}
\bibliography{biblio}

%%%%%%%%%%%%%%%%%%%%%%%%%%%%%%%%%%%%%%%%%%%%%%%%
%
% SECTION: APPENDIX
%
%%%%%%%%%%%%%%%%%%%%%%%%%%%%%%%%%%%%%%%%%%%%%%%%

\appendix
\section{Proofs}
%

%
% UNIVARIATE INTERVALS
%

\subsection{Univariate intervals - Proof of Lemma~\ref{lem:univ_int}}
\label{app:univ_int}

Based on the the aggregation function \eqref{eq:pi_univ_int} we have $S = (S_l, S_u, N) = (X_{(l)}, X_{(u)}, N)$ and thus the role of $f_{S| \bX = \bz} (s; \vartheta)$ in Proposition~\ref{prop:symb_lik_fun}
is to ensure that $s_l = z_{(l)}$ and $s_u = z_{(u)}$. Consequently we can write
$$
f_{S| \bX = \bz} (s; \vartheta) 
= \delta_{z_{(l)}, z_{(u)}} \left( s_l, s_u \right)
= \delta_{z_{(l)}} \left( s_l \right) \delta_{z_{(u)}} \left( s_u \right), 
$$
so that $l-1$ points of $\bz$ belong to $(-\infty, s_l)$, one is at $s_l$, $u-l-1$ belong to $(s_l, s_u)$, one is at $s_u$ and $n-u$ belong to $(s_u, \infty)$.
As there are $n! / ((l-1)! (u-l-1)! (n-u)!)$ possible combinations to arrange $n$ points in such a way, the likelihood function can then be written as
\begin{align*}
L(s_l,s_u, n; \theta)=
& \frac{n!}{ (l-1)! (u-l-1)! (n-u)! } 
\left( \int_{-\infty}^{s_l} g_X(z; \theta) \der z \right)^{l-1} 
\int_{-\infty}^{+\infty} g_X(z; \theta) \delta_{z}(s_l) \der z
\\
& \times \left( \int_{s_l}^{s_u} g_X(z; \theta) \der z \right)^{u-l-1} 
\int_{-\infty}^{+\infty} g_X(z; \theta) \delta_{z}(s_u) \der z
\left( \int_{s_u}^\infty g_X(z; \theta) \der z \right)^{n-u}  \\
=& \frac{n!}{ (l-1)! (u-l-1)! (n-u)! } 
\left[ G_X(s_l; \theta) \right]^{l-1} 
\left[ G_X(s_u; \theta) - G_X(s_l; \theta) \right]^{u-l-1} \\
& \times \left[ 1- G_X(s_u; \theta) \right]^{n-u} g_X(s_l; \theta) g_X(s_u; \theta),
\end{align*}
using the independence between the $n$ replicates $X_1, \ldots, X_n$.

%
% MULTIVARIATE INTERVALS
%

\subsection{Multivariate intervals - Proof of Lemma~\ref{lem:mult_int} and Corollary~\ref{cor:biv_int}}
\label{app:mult_int}

Consider bivariate intervals for simplicity (with identical arguments  providing a full multivariate extension), so that
$X$ is  a bivariate random vector with pdf $g_X(\,\cdot\,; \theta)$ and  marginal and conditional pdfs respectively denoted by $g_{X_i}(\,\cdot\,; \theta), i=1,2$ and
$g_{X_i |X_j}(\,\cdot\,; \theta), i,j=1,2; i \neq j$.
The conditional distribution of $S$ given $\bX = \bz \in \real^2$ is obtained from the aggregation function \eqref{eq:pi_mult_int}.
When $S_p = 2$, $S_{I_p}=(s_a,s_b)$. Now
$s_a = (s_{a_1}, s_{a_2})$ and $s_b = (s_{b_1}, s_{b_2})$ which take values $s_a=(s_{\min,1}, s_{\min,2})$ and $s_b=(s_{\max,1}, s_{\max,2})$  if 
the rectangle constructed from top right and bottom left points,
or
$s_a=(s_{\min,1}, s_{\max,2})$ and $s_b=(s_{\max,1}, s_{\min,2})$ 
if  from the top left, bottom right points. 
Then
$$
f_{S| \bX = \bz} (s; \vartheta) = 
\left\{
\begin{array}{c}
\delta_{z_{(1),1}, z_{(1),2},  z_{(n),1}, z_{(n),2}} 
\left( s_{a_1}, s_{a_2}, s_{b_1}, s_{b_2}  \right) \\
\textrm{ or } \\
\delta_{z_{(1),1}, z_{(n),2},  z_{(n),1}, z_{(1),2} } 
\left( s_{a_1}, s_{a_2}, s_{b_1}, s_{b_2} \right).
\end{array}
\right.
$$
Straightforwardly, this ensures that two points give the marginal minima and maxima and the remaining points are within the interval. There are $n (n-1)$ possible combinations to arrange $n$ points in such a way and so the likelihood function is
\begin{align*}
L(s; \theta)=
& n (n-1) 
\left( \int_{s_{\min}}^{s_{\max}} g_X(z; \theta) \der z \right)^{n-2} 
\int_{\real^2} g_X(z; \theta) \delta_{s_a}(z) \der z
\int_{\real^2} g_X(z; \theta) \delta_{s_b}(z) \der z  \\
=& n (n-1) 
\left( \int_{s_{\min}}^{s_{\max}} g_X(z; \theta) \der z \right)^{n-2} 
g_X(s_a; \theta) g_X(s_b; \theta).
\end{align*}
When $S_p = 3$ 
so that
a single point $S_{I_p}=s_c = s_{\min}$ defines the bottom left rectangle corner,
then
$$
f_{S| \bX = \bz} (s; \vartheta) = 
\delta_{ z_{(1),1}, z_{(1),2}}(s_c)
\delta_{ (s_{\min,1}, s_{\max,1}),  (s_{\min,2}, s_{\max,2})  }
\left(  z_{j,1} | z_{j,2} = s_{\max,2},  z_{j,2} | z_{j,1} = s_{\max,1} \right).
$$
There are $n(n-1)(n-2)$ possible combinations to arrange $n$ points such that one is at a corner, two are on two different edges and the rest are inside the interval.
The likelihood is then
\begin{align*}
L(s; \theta)=
& n (n-1) (n-2) 
\int_{\real^2} g_X(z; \theta) \delta_{s_{\min}}(z) \der z
\left( \int_{s_{\min,1}}^{s_{\max,1}} g_{X_1 | X_2 = s_{\max,2} }( z_1; \theta) \der z_1 \right) 
g_{X_2}(s_{\max,2}; \theta) \\
& \times 
\left( \int_{s_{\min,2}}^{s_{\max,2}} g_{X_2 | X_1 = s_{\max,1} }( z_2; \theta) \der z_2 \right) 
g_{X_1}(s_{\max,1}; \theta)
\left( \int_{s_{\min}}^{s_{\max}} g_X(z; \theta) \der z \right)^{n-3} \\
=& n (n-1) (n-2)  
g_X(s_{\min}; \theta)
\left( \int_{s_{\min}}^{s_{\max}} g_X(z; \theta) \der z \right)^{n-3} \\
& \times \left[ G_{X_1 | X_2 = s_{\max,2} }( s_{\max,1}; \theta) 
- G_{X_1 | X_2 = s_{\max,2} }( s_{\min,1}; \theta)
\right] g_{X_2}(s_{\max,2}; \theta) \\
& \times \left[ G_{X_2 | X_1 = s_{\max,1} }( s_{\max,2}; \theta) 
- G_{X_2 | X_1 = s_{\max,1} }( s_{\min,2}; \theta)
\right] g_{X_1}(s_{\max,1}; \theta).
\end{align*}
Finally when $S_p = 4$ then 
\begin{align*}
f_{S| \bX = \bz} (s; \vartheta) = &
\delta_{ (s_{\min,1}, s_{\max,1}), (s_{\min,1}, s_{\max,1})} 
\left(  z_{j,1} | z_{j,2} = s_{\min,2}, z_{j,1} | z_{j,2} = z_{\max,2}  \right) \\
& \times
\delta_{ (s_{\min,2}, s_{\max,2}), (s_{\min,2}, s_{\max,2})} 
\left(  z_{j,2} | z_{j,1} = s_{\min,1}, z_{j,2} | z_{j,1} = s_{\max,1}  \right), 
\end{align*}
and there are $n (n-1)(n-2) (n-3)$ possible combinations to arrange four points on different 
edges and the rest inside the interval. The likelihood is then
\begin{align*}
L(s; \theta)=
& n (n-1) (n-2) (n-3) \left( \int_{s_{\min}}^{s_{\max}} g_X(z; \theta) \der z \right)^{n-4} \\
& \times
\left( \int_{s_{\min,1}}^{s_{\max,1}} g_{X_1 | X_2 = s_{\min,2} }( z_1; \theta) \der z_1 \right) 
g_{X_2}(s_{\min,2}; \theta) \\
& \times 
\left( \int_{s_{\min,1}}^{s_{\max,1}} g_{X_1 | X_2 = s_{\max,2} }( z_1; \theta) \der z_1 \right) 
g_{X_2}(s_{\max,2}; \theta) \\
& \times 
\left( \int_{s_{\min,2}}^{s_{\max,2}} g_{X_2 | X_1 = s_{\min,1} }( z_2; \theta) \der z_2 \right) 
g_{X_1}(s_{\min,1}; \theta) \\
& \times 
\left( \int_{s_{\min,2}}^{s_{\max,2}} g_{X_2 | X_1 = s_{\max,1} }( z_2; \theta) \der z_2 \right) 
g_{X_1}(s_{\max,1}; \theta) \\
=& n (n-1) (n-2) (n-3) 
\left( \int_{s_{\min}}^{s_{\max}} g_X(z; \theta) \der z \right)^{n-4} \\
& \times \left[ G_{X_1 | X_2 = s_{\min,2} }( s_{\max,1}; \theta) 
- G_{X_1 | X_2 = s_{\min,2} }( s_{\min,1}; \theta)
\right] g_{X_2}(s_{\min,2}; \theta) \\
& \times \left[ G_{X_1 | X_2 = s_{\max,2} }( s_{\max,1}; \theta) 
- G_{X_1 | X_2 = s_{\max,2} }( s_{\min,1}; \theta)
\right] g_{X_2}(s_{\max,2}; \theta) \\
& \times \left[ G_{X_2 | X_1 = s_{\min,1} }( s_{\max,2}; \theta) 
- G_{X_2 | X_1 = s_{\min,1} }( s_{\min,2}; \theta)
\right] g_{X_1}(s_{\min,1}; \theta) \\
& \times \left[ G_{X_2 | X_1 = s_{\max,1} }( s_{\max,2}; \theta) 
- G_{X_2 | X_1 = s_{\max,1} }( s_{\min,2}; \theta)
\right] g_{X_1}(s_{\max,1}; \theta).
\end{align*}

%
% MULTIVARIATE HISTOGRAMS
%

\subsection{Multivariate histograms with fixed bins -- Proof of Lemma~\ref{lem:mult_hist}}
\label{app:mult_hist}

As $S = \pi(\bX)$ is given by (\ref{eq:pi_hist}), then
$s_\bb = \sum_{i=1}^n \indic \{ z_i \in \cB_\bb \}$ for $\bb = \boldsymbol{1}, \ldots, \bB$, which is equivalent to 
\begin{align*}
f_{S| \bX = \bz}(s; \vartheta) =  \prod_{\bb = \boldsymbol{1}}^{ \bB} 
\delta_{\sum_{i=1}^n \indic \{ z_i \in \cB_\bb \}}  \left( s_\bb \right).
\end{align*}
The number of combinations to arrange $z_1, \ldots, z_n$ into the $B_1 \times \cdots \times B_\bB$ bins is the multinomial coefficient  $n! / \prod_\bb s_\bb !$, and sp the likelihood function \eqref{eq:symb_lik} becomes
\begin{align*}
L(s; \theta)=
& \frac{n!}{ s_{\boldsymbol{1}}! \cdots s_\bB } 
\int_{\real^{n \times d}} 
\delta_{z_1} (\cB_{\boldsymbol{1}}) \cdots 
\delta_{z_{s_{\boldsymbol{1}}}} (\cB_{\boldsymbol{1}})  \cdots 
\delta_{z_{n - s_\bB +1}} (\cB_{\bB}) \cdots 
\delta_{z_{n}} (\cB_{\bB})
\prod_{i=1}^n g_X(z_i; \theta) \der \bz
\\
=& \frac{n!}{ s_{\boldsymbol{1}}! \cdots s_\bB } 
\left( \int_{\real^d} g_X(z; \theta) \delta_{z}(\cB_{\boldsymbol{1}}) \der z \right)^{s_{\boldsymbol{1}}} 
\cdots
\left( \int_{\real^d} g_X(z; \theta) \delta_{z}(\cB_\bB) \der z \right)^{s_\bB} 
\\
=& \frac{n!}{ s_{\boldsymbol{1}}! \cdots s_\bB } 
\prod_{\bb = \boldsymbol{1}}^\bB \left( \int_{\cB_\bb} g_X (z; \theta) \der z \right)^{s_{\bb}}.
\end{align*}
%

%
% ORDER STATISTICS HISTOGRAMS
%

\subsection{Histograms with fixed counts -- Lemma~\ref{lem:os_hist}}
\label{app:os_hist}

The aggregation function \eqref{eq:pi_os_hist}
ensures that the $B$ bins are defined as order statistics, so that the symbol S provides the location of $B$ out of $n$ points, with the number of points between these fixed and known through $k=(k_1, \ldots, k_B)$. As a consequence the conditional density $f_{S| \bX = \bz} (s; \vartheta)$ is
\begin{align*}
f_{S| \bX = \bz}(s; \vartheta) =  \prod_{b = 1}^B 
\delta_{z_{(k_b)}}  \left( s_b \right) 
\prod_{b = 1}^{B+1}  \prod_{j = k_{b-1}}^{k_b - 1} \delta_{z_{(j)}} \left( (s_{b-1}, s_b) \right),
\end{align*}
for which there are $n! / \prod_{b=1}^{B+1} (k_b - k_{b-1} - 1)!$ possible combinations to arrange $n$ points.  Hence
\begin{align*}
\cL(s; \theta)=
& \frac{n!}{ \prod_{b=1}^{B+1} (k_b - k_{b-1} - 1)! } 
\int_{\real^n } 
\left( \prod_{b = 1}^B \delta_{z_{(k_b)}}  \left( s_b \right) \right)
\prod_{b = 1}^{B+1} \left( \prod_{j = k_{b-1}}^{k_b - 1} \delta_{z_{(j)}} \left( (s_{b-1}, s_b) \right) \right)
\prod_{i=1}^n g_X(z_i; \theta) \der \bz
\\
=& \frac{n!}{ \prod_{b=1}^{B+1} (k_b - k_{b-1} - 1)! } 
\prod_{b=1}^B  \left( \int_\real \delta_z  \left( s_b \right) g_X(z; \theta) \der z \right)
\prod_{b = 1}^{B+1} \left( \int_{s_{b-1}}^{s_b} g_X(z; \theta) \der z \right)^{k_b - k_{b-1} - 1} \\
=& \frac{n!}{ \prod_{b=1}^{B+1} (k_b - k_{b-1} - 1)! } 
\prod_{b=1}^B  g_X(s_b; \theta) 
\prod_{b = 1}^{B+1} \left( G_X(s_b; \theta) - G_X(s_{b-1}; \theta) \right)^{k_b - k_{b-1} - 1}.
\end{align*}
%

%%%%%%%%%% FIGURES AND TABLES

\begin{table}[tbh]
	\footnotesize
	\[\arraycolsep=2.4pt\def\arraystretch{1.1}
	\begin{array}{r l ccccc c ccccc}
	\multicolumn{1}{c}{} & & \multicolumn{5}{c}{m=20} & & \multicolumn{5}{c}{m=50} \\ 
	\cline{3-7} \cline{9-13}  
	\multicolumn{2}{c}{n_c} & \multicolumn{1}{c}{5} & \multicolumn{1}{c}{10} & \multicolumn{1}{c}{100} & \multicolumn{1}{c}{1,000} & \multicolumn{1}{c}{100,000} & & \multicolumn{1}{c}{5} & \multicolumn{1}{c}{10} & \multicolumn{1}{c}{100} & \multicolumn{1}{c}{1,000} & \multicolumn{1}{c}{100,000} \\
	\cline{1-13}
	\rho_0=0.0 & L_4 & \phantom{-}0.004 & -0.001 & \phantom{-}0.002 & \phantom{-}0.000 & \phantom{-}0.000 &  & -0.004 & -0.003 & \phantom{-}0.001 & \phantom{-}0.000 & \phantom{-}0.000 \\
	& & (0.071) & (0.054) & (0.020) & (0.012) & (0.004) &  & (0.056) & (0.032) & (0.015) & (0.008) & (0.004) \\
	& L_{\emptyset} & -0.018 & -0.017 & -0.006 & -0.001 & -^{\textrm{a}} &  & -0.055 & -0.018 & -0.009 & -0.005 & -^{\textrm{a}}\\
	& & (0.476) & (0.059) & (0.017) & (0.009) & -^{\textrm{a}} &  & (0.399) & (0.029) & (0.016) & (0.008) & -^{\textrm{a}} \\
	& L_{\textrm{full}} & -0.001 & \phantom{-}0.015 & \phantom{-}0.006 & -0.003 & \phantom{-}0.000 &  & -0.009 & \phantom{-}0.001 & -0.001 & \phantom{-}0.011 & \phantom{-}0.000 \\
	& & (0.126) & (0.123) & (0.146) & (0.068) & (0.004) &  & (0.087) & (0.082) & (0.100) & (0.108) & (0.004) \\
	\cline{2-13}
	0.3 & L_4 & \phantom{-}0.082 & \phantom{-}0.034 & \phantom{-}0.006 & -0.002 & -0.001 &  & \phantom{-}0.089 & \phantom{-}0.041 & \phantom{-}0.007 & \phantom{-}0.000 & -0.002 \\ 
	& & (0.080) & (0.055) & (0.025) & (0.012) & (0.015) &  & (0.046) & (0.035) & (0.015) & (0.008) & (0.016)) \\
	& L_{\emptyset} & \phantom{-}0.499 & \phantom{-}0.014 & -0.005 & \phantom{-}0.002 & -^{\textrm{a}} &  & \phantom{-}0.523 & \phantom{-}0.034 & -0.004 & \phantom{-}0.000 & -^{\textrm{a}} \\ 
	& & (0.281) & (0.059) & (0.019) & (0.013) & -^{\textrm{a}} &  & (0.115) & (0.044) & (0.018) & (0.011) & -^{\textrm{a}} \\
	& L_{\textrm{full}} & \phantom{-}0.304 & \phantom{-}0.297 & \phantom{-}0.273 & \phantom{-}0.168 & \phantom{-}0.011 &  & \phantom{-}0.306 & \phantom{-}0.303 & \phantom{-}\phantom{-}0.289 & \phantom{-}0.249 & \phantom{-}0.041 \\ 
	& & (0.112) & (0.129) & (0.160) & (0.217) & (0.088) &  & (0.067) & (0.066) & (0.100) & (0.152) & (0.143) \\ 
	\cline{2-13}
	0.5 & L_4 & \phantom{-}0.147 & \phantom{-}0.073 & \phantom{-}0.014 & \phantom{-}0.001 & \phantom{-}0.023 &  & \phantom{-}0.157 & \phantom{-}0.082 & \phantom{-}0.015 & \phantom{-}0.002 & \phantom{-}0.029 \\
	& & (0.081) & (0.060) & (0.026) & (0.012) & (0.021) &  & (0.048) & (0.038) & (0.016) & (0.009) & (0.021) \\
	& L_{\emptyset} & \phantom{-}0.687 & \phantom{-}0.053 & -0.001 & \phantom{-}0.001 & -^{\textrm{a}} &  & \phantom{-}0.677 & \phantom{-}0.077 & \phantom{-}0.003 & -0.001 & -^{\textrm{a}} \\ 
	& & (0.117) & (0.069) & (0.018) & (0.013) & -^{\textrm{a}} &  & (0.067) & (0.049) & (0.017) & (0.012) & -^{\textrm{a}} \\
	& L_{\textrm{full}} & \phantom{-}0.505 & \phantom{-}0.499 & \phantom{-}0.490 & \phantom{-}0.426 & \phantom{-}0.224 &  & \phantom{-}0.508 & \phantom{-}0.503 & \phantom{-}\phantom{-}0.494 & \phantom{-}0.488 & \phantom{-}0.327 \\
	& & (0.094) & (0.105) & (0.134) & (0.204) & (0.290) &  & (0.058) & (0.055) & (0.083) & (0.076) & (0.259) \\
	\cline{2-13}
	0.7 & L_4 & \phantom{-}0.239 & \phantom{-}0.134 & \phantom{-}0.034 & \phantom{-}0.009 & \phantom{-}0.135 &  & \phantom{-}0.252 & \phantom{-}0.148 & \phantom{-}0.034 & \phantom{-}0.010 & \phantom{-}0.129 \\
	& & (0.083) & (0.071) & (0.031) & (0.014) & (0.047) &  & (0.051) & (0.044) & (0.019) & (0.010) & (0.050) \\
	& L_{\emptyset} & \phantom{-}0.821 & \phantom{-}0.202 & \phantom{-}0.007 & \phantom{-}0.009 & -^{\textrm{a}} &  & \phantom{-}0.819 & \phantom{-}0.155 & \phantom{-}0.011 & \phantom{-}0.005 & -^{\textrm{a}} \\ 
	& & (0.056) & (0.262) & (0.019) & (0.017) & -^{\textrm{a}} &  & (0.035) & (0.113) & (0.013) & (0.015) & -^{\textrm{a}} \\
	& L_{\textrm{full}} & \phantom{-}0.701 & \phantom{-}0.700 & \phantom{-}0.696 & \phantom{-}0.692 & \phantom{-}0.655 &  & \phantom{-}0.706 & \phantom{-}0.702 & \phantom{-}0.701 & \phantom{-}0.702 & \phantom{-}0.695 \\
	& & (0.077) & (0.074) & (0.079) & (0.081) & (0.194) &  & (0.044) & (0.039) & (0.047) & (0.045) & (0.055) \\
	\cline{2-13}
	0.9 & L_4 & \phantom{-}0.408 & \phantom{-}0.267 & \phantom{-}0.098 & \phantom{-}0.038 & \phantom{-}0.029 &  & \phantom{-}0.425 & \phantom{-}0.290 & \phantom{-}0.095 & \phantom{-}0.036 & \phantom{-}0.016 \\
	& & (0.093) & (0.096) & (0.060) & (0.033) & (0.074) &  & (0.055) & (0.060) & (0.034) & (0.016) & (0.036) \\
	& L_{\emptyset} & \phantom{-}0.936 & \phantom{-}0.933 & \phantom{-}0.282 & \phantom{-}0.023 & -^{\textrm{a}} &  & \phantom{-}0.935 & \phantom{-}0.937 & \phantom{-}0.188 & \phantom{-}0.025 & -^{\textrm{a}} \\ 
	& & (0.017) & (0.036) & (0.420) & (0.024) & -^{\textrm{a}} &  & (0.010) & (0.010) & (0.355) & (0.020) & -^{\textrm{a}} \\
	& L_{\textrm{full}} & \phantom{-}0.901 & \phantom{-}0.899 & \phantom{-}0.901 & \phantom{-}0.901 & \phantom{-}0.903 &  & \phantom{-}0.902 & \phantom{-}0.901 & \phantom{-}0.900 & \phantom{-}0.900 & \phantom{-}0.902 \\
	& & (0.029) & (0.026) & (0.025) & (0.028) & (0.023) &  & (0.017) & (0.014) & (0.016) & (0.016) & (0.015) \\
	\hline
	\end{array}
	\]
	\caption{\footnotesize Mean (and standard deviation) of the symbolic maximum likelihood estimate of the correlation, $\rho$, over $T=100$ replicate bivariate random rectangle datasets. The symbolic datasets vary in the number of symbols ($m$), the number of classical datapoints per symbol ($n_c$), and the strength of the correlation between the two variables ($\rho_0$). Estimates maximise the three symbolic likelihoods $L_\textrm{full}$, $L_\emptyset$ and $L_4$. $^\textrm{a}$ indicates that computation times were too high.}
	\label{tab:cor_biv_int}
\end{table}

\begin{table}[tbh]
	\footnotesize
	\[\arraycolsep=2.4pt\def\arraystretch{1.1}
	\begin{array}{r ll ccc c ccc}
	\multicolumn{1}{c}{} & & & \multicolumn{3}{c}{n_c=60} & & \multicolumn{3}{c}{n_c=300}  \\ 
	\cline{4-6} \cline{8-10} 
	\multicolumn{2}{c}{} & \textrm{Orders } (l,u) & \multicolumn{1}{c}{\sigma_1} & \multicolumn{1}{c}{\rho} & \multicolumn{1}{c}{\sigma_2} & & \multicolumn{1}{c}{\sigma_2} & \multicolumn{1}{c}{\rho_0} & \multicolumn{1}{c}{\sigma_2}  \\
	\cline{1-10}

	& L_{\textrm{sn},x} 
	& ((6, 5), (55, 35))  
	& 0.4992 & 0.6933 & 0.5050 &  & 0.4984 & 0.6772 & 0.5075 \\
	& & & (0.0019)&(0.0255)& (0.0054)& &(0.0004)&(0.0146)&(0.0024) \\
	& & ((16,6), (45,24)) 
	& 0.4981 & 0.6402 & 0.5043 &  & 0.4985 & 0.6739 & 0.5177 \\
	& & & (0.0021) &(0.0273) &(0.0107)& &(0.0005) &(0.0115) &(0.0048) \\
	& & ((20, 5),(41, 16))  
	& 0.4991 & 0.6396 & 0.5054 &  & 0.4981 & 0.6451 & 0.5141 \\
	& & &(0.0027) &(0.0256) &(0.0129)& &(0.0006) &(0.0127) &(0.0059) \\
	\cline{3-10}
	& L_{\textrm{sn},y} 
	& ((5,6),( 35, 55))
	& 0.5106 & 0.6912 & 0.4974 &  & 0.5082 & 0.6774 & 0.4998 \\
	& & & (0.0061) &(0.0339)& (0.0016)& &(0.0024) &(0.0156) &(0.0004) \\
	& & ((6, 16),(24, 45)) 
	& 0.5289 & 0.6933 & 0.4986 &  & 0.5088 & 0.6453 & 0.4994 \\
	& & & (0.0123) &(0.0239) &(0.0021)& &(0.0049)&(0.0129) &(0.0004) \\
	& & ((5,20),(16, 41))
	& 0.5231 & 0.6699 & 0.5004 &  & 0.5154 & 0.6702 & 0.4992 \\
	& & & (0.0127) &(0.0253) &(0.0024)& &(0.0053) &(0.0106) &(0.0005) \\
	\cline{3-10}
	& L_{\textrm{is},x} 
	& ((6, 3), (55, 3))   
	& 0.4993 & 0.7130 & 0.4900 &  & 0.4984 & 0.7124 & 0.4932 \\
	& & & (0.0019)& (0.0067) &(0.0037) &&(0.0004) &(0.0032)& (0.0019) \\
	& & ((16,10), (45, 2))  
	& 0.4981 & 0.7037 & 0.4806 &  & 0.4985 & 0.7051 & 0.4866 \\
	& & & (0.0021) &(0.0039) &(0.0064)& &(0.0005) &(0.0011) &(0.0025) \\
	& & ((20,7), (41,14)) 
	& 0.4993 & 0.7465 & 0.4871 &  & 0.4981 & 0.7169 & 0.4979 \\
	& & & (0.0027) &(0.0128) &(0.0037)& &(0.0006)&(0.0051) &(0.0013) \\
	\cline{3-10}
	& L_{\textrm{is},y} 
	& ((3,6), (3, 55))  
	& 0.4929 & 0.7133 & 0.4975 &  & 0.4896 & 0.7151 & 0.4998 \\
	& & & (0.0051) &(0.0064) &(0.0016)& &(0.0018)&(0.0032) &(0.0004)\\
	& & ((10,16), (2, 45)) 
	& 0.4868 & 0.7053 & 0.4986 &  & 0.4848 & 0.7066 & 0.4993 \\
	& & & (0.0068) &(0.0035) &(0.0021)& &(0.0026) &(0.0011)&(0.0004) \\
	& & ((7,20), (14, 41)) 
	& 0.4933 & 0.7311 & 0.5004 &  & 0.4947 & 0.7268 & 0.4993 \\
	& & & (0.0040) &(0.0115) &(0.0023)& &(0.0016)&(0.0057)& (0.0005) \\
	\hline
	\end{array}
	\]
	\caption{\footnotesize Mean (and standard deviation) of the symbolic maximum likelihood estimate of $\sigma_1,\rho$ and $\sigma_2$, over $T=100$ replicate bivariate random rectangle datasets containing $m=20$ symbols. The symbolic datasets vary in 
		the number of classical datapoints per symbol ($n_c$), the type of symbol construction (sn $=$ sequential nesting; is $=$ iterative segmentation), which axis is used first in the symbol construction ($x$ or $y$), and the vectors of lower ($l$) and upper ($u$) order statistics used.
		True parameter values are $\sigma_{0,1}=\sigma_{0,2}=0.5$ and $\rho_0=0.7$.
		For $L_{sn,x}$, orders $(l,u)=((6,5),(55,35))$ mean firstly take the (6,55) lower/upper order statistics on the $x$-axis, and then the $(5,35)$ $y$-order statistics of the remaining $n_c-12$ observations in the central $x$ range (Figure \ref{NeedNewFig}, bottom centre).
		For $L_{is,x}$, orders $(l,u)=((6,3),(55,3))$ mean firstly take the $(6,55)$ lower/upper order statistics on the $x$-axis, the $3$-rd $y$-order statistic of the remaining 5 observations below the lower $x$ quantile, and the $3$-rd $y$-order statistic of the remaining 5 observations above the upper $x$ order statistic (Figure \ref{NeedNewFig}, bottom right).
		For $L_{\cdot,y}$ the procedure is the same as for $L_{\cdot,x}$ but starting with the $y$-quantiles (the resulting 3 bivariate intervals for e.g.~$L_{sn,x}$ are identical to those for $L_{sn,y}$).
		The orders shown are for $n_c=60$. For $n_c=300$ the utilised orders are multiplied by 5 so that the intervals are directly comparable.
	}
	\label{tab:biv_os_int_rho07}
\end{table}

\begin{table}
\centering
\small{
\begin{tabular}{c c c c c |c}
& Normal & & Skew-Normal & & Skew-Normal ($n=1,000,000$) \\
\hline
Classical & 3.886 (0.478) &  & 90.754 (0.097) &  & 3533.900 (2.472) \\
New Symbolic & 1.551 (0.045) &  & 12.721 (0.034) &  & \phantom{00}11.487 (0.030) \\
LRB & 0.498 (0.001) &  & \phantom{0}0.476 (0.001) &  & \phantom{000}0.457 (0.001) \\
\hline
\end{tabular}
}
\caption{\small Mean (s.e.)~likelihood  evaluation times (seconds $\times 10^{-3}$) over $T=1,000$ parameter vector replicates, using the loan dataset (first two columns) and simulated data with 35 groups of $n=1,000,000$ observations.}
\label{tab:times}
\end{table}

%%%%%%%%%%%%%%%%%%%%%%%%%%%%%%%%%%%%%%%%
%%%%%%%%%%%%%%%%%%%%%%%%%%%%%%%%%%%%%%%%
%
% SUPPORTING INFORMATION
%
%%%%%%%%%%%%%%%%%%%%%%%%%%%%%%%%%%%%%%%%
%%%%%%%%%%%%%%%%%%%%%%%%%%%%%%%%%%%%%%%%

\newpage

\section{Supplementary Material}

%
% SECTION 1
%

\subsection{Estimates of the $\mu_1, \mu_2, \sigma_1$ and $\sigma_2$, from Section~\ref{ssec:biv_int}}
\label{suppmat:biv_int_est}

Tables \ref{tab:mean1_biv_int}--\ref{tab:sig2_biv_int}.

\begin{table}[h]
	\footnotesize
	\[\arraycolsep=2.4pt\def\arraystretch{1.1}
	\begin{array}{r l ccccc c ccccc}
	\multicolumn{1}{c}{} & & \multicolumn{5}{c}{m=20} & & \multicolumn{5}{c}{m=50} \\ 
	\cline{3-7} \cline{9-13}  
	\multicolumn{2}{c}{n_c} & \multicolumn{1}{c}{5} & \multicolumn{1}{c}{10} & \multicolumn{1}{c}{100} & \multicolumn{1}{c}{1,000} & \multicolumn{1}{c}{100,000} & & \multicolumn{1}{c}{5} & \multicolumn{1}{c}{10} & \multicolumn{1}{c}{100} & \multicolumn{1}{c}{1,000} & \multicolumn{1}{c}{100,000} \\
	\cline{1-13}
	\rho_0=0.0 & L_4 & 1.999 & 2.004 & 2.006 & 1.998 & 2.003 &  & 1.999 & 2.000 & 2.001 & 1.999 & 2.002 \\ 
	& & (0.051) & (0.045) & (0.035) & (0.025) & (0.018) &  & (0.031) & (0.027) & (0.021) & (0.015) & (0.012) \\ 
	& L_{\emptyset} & 1.999 & 2.004 & 2.006 & 1.998 & -^{\textrm{a}} &  & 1.999 & 2.000 & 2.001 & 1.999 & -^{\textrm{a}} \\
	& & (0.051) & (0.045) & (0.035) & (0.025) & -^{\textrm{a}} &  & (0.031) & (0.027) & (0.021) & (0.015) & -^{\textrm{a}} \\
	& L_{\textrm{full}} & 1.999 & 2.004 & 2.006 & 1.998 & 2.003 &  & 1.999 & 2.000 & 2.001 & 1.999 & 2.002 \\ 
	& & (0.051) & (0.045) & (0.035) & (0.025) & (0.018) &  & (0.031) & (0.027) & (0.021) & (0.015) & (0.012) \\
	\cline{1-13}
	0.3 & L_4 & 1.995 & 1.996 & 1.998 & 2.000 & 2.001 &  & 1.996 & 1.999 & 1.996 & 1.998 & 2.001 \\ 
	& & (0.052) & (0.044) & (0.034) & (0.024) & (0.016) &  & (0.034) & (0.028) & (0.020) & (0.016) & (0.011) \\
	& L_{\emptyset} & 1.995 & 1.996 & 1.998 & 2.000 & -^{\textrm{a}} &  & 1.996 & 1.999 & 1.996 & 1.998 & -^{\textrm{a}} \\ 
	& & (0.053) & (0.044) & (0.034) & (0.024) & -^{\textrm{a}} &  & (0.034) & (0.028) & (0.020) & (0.015) & -^{\textrm{a}} \\
	& L_{\textrm{full}} & 1.995 & 1.996 & 1.998 & 2.000 & 2.001 &  & 1.996 & 2.000 & 1.996 & 1.998 & 2.001 \\
	& & (0.053) & (0.044) & (0.034) & (0.024) & (0.016) &  & (0.034) & (0.028) & (0.020) & (0.016) & (0.011) \\  
	\cline{1-13}
	0.5 & L_4 & 1.995 & 1.995 & 1.998 & 2.000 & 2.000 &  & 1.996 & 1.999 & 1.996 & 1.998 & 2.001 \\
	& & (0.053) & (0.044) & (0.034) & (0.024) & (0.016) &  & (0.035) & (0.028) & (0.021) & (0.016) & (0.012) \\ 
	& L_{\emptyset} & 1.995 & 1.995 & 1.998 & 2.000 & -^{\textrm{a}} &  & 1.996 & 1.999 & 1.996 & 1.998 & -^{\textrm{a}} \\ 
	& & (0.054) & (0.044) & (0.034) & (0.024) & -^{\textrm{a}} &  & (0.034) & (0.028) & (0.021) & (0.016) & -^{\textrm{a}} \\
	& L_{\textrm{full}} & 1.996 & 1.995 & 1.998 & 2.000 & 2.000 &  & 1.996 & 2.000 & 1.996 & 1.998 & 2.001 \\ 
	& & (0.054) & (0.044) & (0.034) & (0.024) & (0.016) &  & (0.034) & (0.028) & (0.021) & (0.016) & (0.012) \\
	\cline{1-13}
	0.7 & L_4 & 1.994 & 1.996 & 1.997 & 2.001 & 2.000 &  & 1.997 & 1.999 & 1.995 & 1.999 & 2.001\\
	& & (0.054) & (0.043) & (0.034) & (0.023) & (0.017) &  & (0.035) & (0.027) & (0.020) & (0.017) & (0.012) \\
	& L_{\emptyset} & 1.994 & 1.996 & 1.997 & 2.001 & -^{\textrm{a}} &  & 1.997 & 1.999 & 1.995 & 1.998 & -^{\textrm{a}} \\ 
	& & (0.055) & (0.043) & (0.034) & (0.023) & -^{\textrm{a}} &  & (0.035) & (0.028) & (0.020) & (0.017) & -^{\textrm{a}} \\
	& L_{\textrm{full}} & 1.995 & 1.995 & 1.998 & 2.000 & 2.000 &  & 1.997 & 2.000 & 1.995 & 1.998 & 2.001 \\
	& & (0.054) & (0.043) & (0.033) & (0.023) & (0.017) &  & (0.034) & (0.027) & (0.020) & (0.017) & (0.012) \\
	\cline{1-13}
	0.9 & L_4 & 1.993 & 1.998 & 1.996 & 2.000 & 1.999 &  & 1.997 & 1.999 & 1.995 & 1.999 & 2.001 \\
	& & (0.054) & (0.043) & (0.033) & (0.024) & (0.019) &  & (0.035) & (0.027) & (0.020) & (0.018) & (0.012) \\
	& L_{\emptyset} & 1.993 & 1.997 & 1.996 & 2.000 & -^{\textrm{a}} &  & 1.997 & 1.999 & 1.996 & 1.999 & -^{\textrm{a}} \\ 
	& & (0.055) & (0.044) & (0.033) & (0.024) & -^{\textrm{a}} &  & (0.035) & (0.027) & (0.019) & (0.018) & -^{\textrm{a}} \\
	& L_{\textrm{full}} & 1.994 & 1.998 & 1.998 & 1.999 & 2.000 &  & 1.997 & 2.001 & 1.996 & 1.999 & 2.001 \\ 
	& & (0.054) & (0.041) & (0.032) & (0.022) & (0.018) &  & (0.035) & (0.026) & (0.019) & (0.017) & (0.011) \\
	\hline
	\end{array}
	\]
	\caption{
	\footnotesize 
	As for Table \ref{tab:cor_biv_int} but for estimates of the mean $\mu_1$.
	}
	\label{tab:mean1_biv_int}
\end{table}

\begin{table}[h]
	\footnotesize
	\[\arraycolsep=2.4pt\def\arraystretch{1.1}
	\begin{array}{r l ccccc c ccccc}
	\multicolumn{1}{c}{} & & \multicolumn{5}{c}{m=20} & & \multicolumn{5}{c}{m=50} \\ 
	\cline{3-7} \cline{9-13}  
	\multicolumn{2}{c}{n_c} & \multicolumn{1}{c}{5} & \multicolumn{1}{c}{10} & \multicolumn{1}{c}{100} & \multicolumn{1}{c}{1,000} & \multicolumn{1}{c}{100,000} & & \multicolumn{1}{c}{5} & \multicolumn{1}{c}{10} & \multicolumn{1}{c}{100} & \multicolumn{1}{c}{1,000} & \multicolumn{1}{c}{100,000} \\
	\cline{1-13}
	\rho_0=0.0 & L_4 & 4.993 & 5.000 & 4.997 & 5.002 & 4.998 &  & 4.998 & 5.000 & 4.996 & 5.001 & 5.000 \\ 
	& & (0.053) & (0.045) & (0.032) & (0.027) & (0.018) &  & (0.034) & (0.026) & (0.018) & (0.019) & (0.012) \\
	& L_{\emptyset} & 4.993 & 5.001 & 4.997 & 5.002 & -^{\textrm{a}} &  & 4.998 & 5.000 & 4.996 & 5.001 & -^{\textrm{a}} \\ 
	& & (0.053) & (0.045) & (0.032) & (0.027) & -^{\textrm{a}} &  & (0.035) & (0.026) & (0.018) & (0.019) & -^{\textrm{a}} \\
	& L_{\textrm{full}} & 4.993 & 5.001 & 4.997 & 5.002 & 4.998 &  & 4.998 & 5.000 & 4.997 & 5.001 & 5.000 \\
	& & (0.053) & (0.046) & (0.032) & (0.027) & (0.018) &  & (0.034) & (0.026) & (0.018) & (0.019) & (0.012) \\
	\cline{1-13}
	0.3 & L_4 & 4.999 & 5.001 & 5.004 & 4.997 & 4.999 &  & 5.000 & 5.002 & 5.001 & 5.000 & 5.001 \\ 
	& & (0.053) & (0.045) & (0.032) & (0.023) & (0.018) &  & (0.033) & (0.028) & (0.017) & (0.017) & (0.011) \\
	& L_{\emptyset} & 4.999 & 5.001 & 5.004 & 4.997 & -^{\textrm{a}} &  & 5.000 & 5.002 & 5.002 & 5.000 & -^{\textrm{a}} \\ 
	& & (0.053) & (0.045) & (0.032) & (0.023) & -^{\textrm{a}} &  & (0.033) & (0.027) & (0.017) & (0.017) & -^{\textrm{a}} \\
	& L_{\textrm{full}} & 4.999 & 5.001 & 5.004 & 4.997 & 4.999 &  & 5.000 & 5.003 & 5.001 & 5.000 & 5.001 \\
	& & (0.053) & (0.045) & (0.032) & (0.023) & (0.018) &  & (0.033) & (0.027) & (0.017) & (0.017) & (0.011) \\  
	\cline{1-13}
	0.5 & L_4 & 4.998 & 5.001 & 5.004 & 4.996 & 4.999 &  & 5.000 & 5.002 & 5.001 & 5.001 & 5.000 \\
	& & (0.053) & (0.047) & (0.032) & (0.024) & (0.018) &  & (0.033) & (0.028) & (0.018) & (0.017) & (0.012) \\ 
	& L_{\emptyset} & 4.998 & 5.001 & 5.004 & 4.996 & -^{\textrm{a}} &  & 5.000 & 5.002 & 5.001 & 5.001 & -^{\textrm{a}} \\ 
	& & (0.053) & (0.046) & (0.032) & (0.024) & -^{\textrm{a}} &  & (0.033) & (0.027) & (0.018) & (0.017) & -^{\textrm{a}} \\
	& L_{\textrm{full}} & 4.999 & 5.001 & 5.005 & 4.996 & 4.999 &  & 5.000 & 5.002 & 5.002 & 5.001 & 5.000 \\
	& & (0.052) & (0.046) & (0.032) & (0.024) & (0.018) &  & (0.032) & (0.027) & (0.018) & (0.017) & (0.012) \\
	\cline{1-13}
	0.7 & L_4 & 4.997 & 5.001 & 5.003 & 4.996 & 4.998 &  & 5.000 & 5.001 & 5.001 & 5.001 & 5.000 \\
	& & (0.053) & (0.047) & (0.032) & (0.024) & (0.018) &  & (0.033) & (0.028) & (0.018) & (0.018) & (0.012) \\
	& L_{\emptyset} & 4.997 & 5.001 & 5.003 & 4.996 & -^{\textrm{a}} &  & 5.000 & 5.001 & 5.001 & 5.001 & -^{\textrm{a}} \\
	& & (0.053) & (0.047) & (0.032) & (0.024) & -^{\textrm{a}} &  & (0.033) & (0.028) & (0.018) & (0.018) & -^{\textrm{a}} \\
	& L_{\textrm{full}} & 4.998 & 5.001 & 5.004 & 4.996 & 4.998 &  & 5.000 & 5.002 & 5.001 & 5.001 & 5.000 \\
	& & (0.052) & (0.046) & (0.031) & (0.024) & (0.018) &  & (0.032) & (0.027) & (0.018) & (0.017) & (0.012) \\
	\cline{1-13}
	0.9 & L_4 & 4.995 & 5.001 & 5.000 & 5.000 & 4.998 &  & 4.999 & 5.000 & 5.000 & 5.002 & 4.999 \\
	& & (0.053) & (0.047) & (0.031) & (0.024) & (0.018) &  & (0.034) & (0.028) & (0.017) & (0.017) & (0.011) \\
	& L_{\emptyset} & 4.994 & 5.001 & 5.000 & 5.000 & -^{\textrm{a}} &  & 4.999 & 5.000 & 5.000 & 5.002 & -^{\textrm{a}} \\ 
	& & (0.054) & (0.048) & (0.032) & (0.024) & -^{\textrm{a}} &  & (0.034) & (0.028) & (0.017) & (0.017) & -^{\textrm{a}} \\
	& L_{\textrm{full}} & 4.995 & 5.002 & 5.002 & 4.999 & 4.999 &  & 4.999 & 5.002 & 5.000 & 5.002 & 5.000 \\
	& & (0.052) & (0.045) & (0.029) & (0.021) & (0.018) &  & (0.033) & (0.027) & (0.018) & (0.016) & (0.011) \\
	\hline
	\end{array}
	\]
	\caption{
	\footnotesize 
	As for Table \ref{tab:cor_biv_int} but for estimates of the mean $\mu_2$.
	}
	\label{tab:mean2_biv_int}
\end{table}

\begin{table}[h]
	\footnotesize
	\[\arraycolsep=2.4pt\def\arraystretch{1.1}
	\begin{array}{r l ccccc c ccccc}
	\multicolumn{1}{c}{} & & \multicolumn{5}{c}{m=20} & & \multicolumn{5}{c}{m=50} \\ 
	\cline{3-7} \cline{9-13}  
	\multicolumn{2}{c}{n_c} & \multicolumn{1}{c}{5} & \multicolumn{1}{c}{10} & \multicolumn{1}{c}{100} & \multicolumn{1}{c}{1,000} & \multicolumn{1}{c}{100,000} & & \multicolumn{1}{c}{5} & \multicolumn{1}{c}{10} & \multicolumn{1}{c}{100} & \multicolumn{1}{c}{1,000} & \multicolumn{1}{c}{100,000} \\
	\cline{1-13}
	\rho_0=0.0 & L_4 & 0.247 & 0.247 & 0.250 & 0.250 & 0.251 &  & 0.247 & 0.247 & 0.249 & 0.250 & 0.250 \\ 
	& & (0.039) & (0.029) & (0.015) & (0.008) & (0.004) &  & (0.024) & (0.018) & (0.009) & (0.006) & (0.002) \\
	& L_{\emptyset} & 0.249 & 0.247 & 0.250 & 0.250 & -^{\textrm{a}} &  & 0.248 & 0.247 & 0.249 & 0.250 & -^{\textrm{a}} \\ 
	& & (0.039) & (0.029) & (0.015) & (0.008) & -^{\textrm{a}} &  & (0.024) & (0.018) & (0.009) & (0.006) & -^{\textrm{a}} \\
	& L_{\textrm{full}} & 0.247 & 0.247 & 0.250 & 0.250 & 0.251 &  & 0.247 & 0.247 & 0.249 & 0.250 & 0.250 \\
	& & (0.039) & (0.029) & (0.015) & (0.008) & (0.004) &  & (0.024) & (0.018) & (0.009) & (0.006) & (0.002) \\
	\cline{1-13}
	0.3 & L_4 & 0.250 & 0.255 & 0.248 & 0.250 & 0.250 &  & 0.246 & 0.251 & 0.249 & 0.250 & 0.250 \\ 
	& & (0.044) & (0.031) & (0.014) & (0.009) & (0.005) &  & (0.026) & (0.018) & (0.008) & (0.005) & (0.003) \\
	& L_{\emptyset} & 0.253 & 0.255 & 0.248 & 0.250 & -^{\textrm{a}} &  & 0.249 & 0.251 & 0.249 & 0.250 & -^{\textrm{a}} \\ 
	& & (0.044) & (0.031) & (0.014) & (0.009) & -^{\textrm{a}} &  & (0.026) & (0.018) & (0.008) & (0.005) & -^{\textrm{a}} \\
	& L_{\textrm{full}} & 0.250 & 0.255 & 0.248 & 0.250 & 0.250 &  & 0.246 & 0.251 & 0.249 & 0.250 & 0.250 \\
	& & (0.043) & (0.031) & (0.014) & (0.009) & (0.005) &  & (0.026) & (0.018) & (0.008) & (0.005) & (0.003) \\  
	\cline{1-13}
	0.5 & L_4 & 0.251 & 0.256 & 0.248 & 0.251 & 0.250 &  & 0.248 & 0.252 & 0.249 & 0.250 & 0.250 \\
	& & (0.043) & (0.031) & (0.013) & (0.008) & (0.004) &  & (0.025) & (0.018) & (0.008) & (0.005) & (0.003) \\ 
	& L_{\emptyset} & 0.257 & 0.255 & 0.248 & 0.251 & -^{\textrm{a}} &  & 0.252 & 0.252 & 0.249 & 0.250 & -^{\textrm{a}} \\ 
	& & (0.045) & (0.030) & (0.013) & (0.008) & -^{\textrm{a}} &  & (0.026) & (0.018) & (0.008) & (0.005) & -^{\textrm{a}} \\
	& L_{\textrm{full}} & 0.250 & 0.255 & 0.248 & 0.251 & 0.250 &  & 0.247 & 0.251 & 0.249 & 0.250 & 0.250 \\
	& & (0.042) & (0.030) & (0.013) & (0.008) & (0.004) &  & (0.025) & (0.018) & (0.008) & (0.005) & (0.003) \\
	\cline{1-13}
	0.7 & L_4 & 0.253 & 0.256 & 0.248 & 0.251 & 0.251 &  & 0.250 & 0.253 & 0.249 & 0.250 & 0.250 \\
	& & (0.042) & (0.029) & (0.013) & (0.008) & (0.004) &  & (0.025) & (0.019) & (0.008) & (0.004) & (0.003) \\
	& L_{\emptyset} & 0.259 & 0.255 & 0.248 & 0.251 & -^{\textrm{a}} &  & 0.258 & 0.252 & 0.249 & 0.250 & -^{\textrm{a}} \\ 
	& & (0.043) & (0.029) & (0.013) & (0.008) & -^{\textrm{a}} &  & (0.025) & (0.018) & (0.008) & (0.004) & -^{\textrm{a}} \\
	& L_{\textrm{full}} & 0.249 & 0.255 & 0.248 & 0.251 & 0.251 &  & 0.247 & 0.251 & 0.249 & 0.250 & 0.250 \\
	& & (0.041) & (0.029) & (0.012) & (0.008) & (0.004) &  & (0.024) & (0.018) & (0.008) & (0.004) & (0.003) \\
	\cline{1-13}
	0.9 & L_4 & 0.260 & 0.258 & 0.248 & 0.251 & 0.251 &  & 0.258 & 0.256 & 0.249 & 0.251 & 0.250 \\
	& & (0.041) & (0.029) & (0.013) & (0.008) & (0.004) &  & (0.024) & (0.020) & (0.008) & (0.005) & (0.003) \\
	& L_{\emptyset} & 0.257 & 0.253 & 0.247 & 0.251 & -^{\textrm{a}} &  & 0.253 & 0.251 & 0.249 & 0.251 & -^{\textrm{a}} \\ 
	& & (0.040) & (0.027) & (0.013) & (0.008) & -^{\textrm{a}} &  & (0.022) & (0.018) & (0.008) & (0.005) & -^{\textrm{a}} \\
	& L_{\textrm{full}} & 0.249 & 0.253 & 0.248 & 0.251 & 0.251 &  & 0.248 & 0.251 & 0.249 & 0.251 & 0.250 \\
	& & (0.039) & (0.027) & (0.012) & (0.008) & (0.004) &  & (0.023) & (0.018) & (0.008) & (0.004) & (0.003) \\
	\hline
	\end{array}
	\]
	\caption{
		\footnotesize 
		As for Table \ref{tab:cor_biv_int} but for estimates of the standard deviation $\sigma_1$.
			}
	\label{tab:sig1_biv_int}
\end{table}

\begin{table}[h]
	\footnotesize
	\[\arraycolsep=2.4pt\def\arraystretch{1.1}
	\begin{array}{r l ccccc c ccccc}
	\multicolumn{1}{c}{} & & \multicolumn{5}{c}{m=20} & & \multicolumn{5}{c}{m=50} \\ 
	\cline{3-7} \cline{9-13}  
	\multicolumn{2}{c}{n_c} & \multicolumn{1}{c}{5} & \multicolumn{1}{c}{10} & \multicolumn{1}{c}{100} & \multicolumn{1}{c}{1,000} & \multicolumn{1}{c}{100,000} & & \multicolumn{1}{c}{5} & \multicolumn{1}{c}{10} & \multicolumn{1}{c}{100} & \multicolumn{1}{c}{1,000} & \multicolumn{1}{c}{100,000} \\
	\cline{1-13}
	\rho_0=0.0 & L_4 & 0.251 & 0.251 & 0.250 & 0.251 & 0.251 &  & 0.251 & 0.251 & 0.250 & 0.250 & 0.251 \\ 
	& & (0.038) & (0.029) & (0.013) & (0.009) & (0.005) &  & (0.023) & (0.019) & (0.008) & (0.005) & (0.003) \\
	& L_{\emptyset} & 0.253 & 0.251 & 0.250 & 0.251 & -^{\textrm{a}} &  & 0.253 & 0.251 & 0.250 & 0.250 & -^{\textrm{a}} \\
	& & (0.039) & (0.029) & (0.013) & (0.009) & -^{\textrm{a}} &  & (0.023) & (0.019) & (0.008) & (0.005) & -^{\textrm{a}} \\
	& L_{\textrm{full}} & 0.251 & 0.251 & 0.250 & 0.251 & 0.251 &  & 0.251 & 0.251 & 0.250 & 0.250 & 0.251 \\
	& & (0.038) & (0.028) & (0.013) & (0.009) & (0.005) &  & (0.023) & (0.019) & (0.008) & (0.005) & (0.003) \\
	\cline{1-13}
	0.3 & L_4 & 0.250 & 0.247 & 0.250 & 0.251 & 0.251 &  & 0.254 & 0.250 & 0.251 & 0.251 & 0.250 \\ 
	& & (0.032) & (0.026) & (0.011) & (0.009) & (0.004) &  & (0.024) & (0.019) & (0.007) & (0.005) & (0.003) \\ 
	& L_{\emptyset} & 0.254 & 0.247 & 0.250 & 0.251 & -^{\textrm{a}} &  & 0.257 & 0.250 & 0.251 & 0.251 & -^{\textrm{a}} \\ 
	& & (0.033) & (0.026) & (0.011) & (0.009) & -^{\textrm{a}} &  & (0.025) & (0.020) & (0.007) & (0.005) & -^{\textrm{a}} \\
	& L_{\textrm{full}} & 0.250 & 0.246 & 0.250 & 0.251 & 0.251 &  & 0.253 & 0.250 & 0.251 & 0.251 & 0.250 \\
	& & (0.033) & (0.026) & (0.011) & (0.009) & (0.004) &  & (0.024) & (0.019) & (0.007) & (0.005) & (0.003) \\  
	\cline{1-13}
	0.5 & L_4 & 0.252 & 0.247 & 0.250 & 0.251 & 0.251 &  & 0.255 & 0.250 & 0.251 & 0.251 & 0.250 \\
	& & (0.033) & (0.027) & (0.011) & (0.009) & (0.004) &  & (0.024) & (0.020) & (0.007) & (0.005) & (0.002) \\ 
	& L_{\emptyset} & 0.257 & 0.247 & 0.250 & 0.251 & -^{\textrm{a}} &  & 0.259 & 0.250 & 0.251 & 0.251 & -^{\textrm{a}} \\
	& & (0.034) & (0.027) & (0.011) & (0.009) & -^{\textrm{a}} &  & (0.025) & (0.020) & (0.007) & (0.005) & -^{\textrm{a}} \\
	& L_{\textrm{full}} & 0.250 & 0.247 & 0.250 & 0.251 & 0.251 &  & 0.253 & 0.250 & 0.251 & 0.251 & 0.250 \\
	& & (0.033) & (0.027) & (0.011) & (0.009) & (0.004) &  & (0.024) & (0.020) & (0.007) & (0.005) & (0.002) \\
	\cline{1-13}
	0.7 & L_4 & 0.254 & 0.249 & 0.250 & 0.251 & 0.251 &  & 0.257 & 0.252 & 0.251 & 0.251 & 0.251 \\
	& & (0.033) & (0.028) & (0.011) & (0.008) & (0.004) &  & (0.024) & (0.020) & (0.007) & (0.005) & (0.003) \\
	& L_{\emptyset} & 0.260 & 0.248 & 0.250 & 0.251 & -^{\textrm{a}} &  & 0.264 & 0.251 & 0.251 & 0.251 & -^{\textrm{a}} \\ 
	& & (0.035) & (0.028) & (0.011) & (0.008) & -^{\textrm{a}} &  & (0.024) & (0.020) & (0.007) & (0.005) & -^{\textrm{a}} \\
	& L_{\textrm{full}} & 0.250 & 0.247 & 0.250 & 0.251 & 0.251 &  & 0.253 & 0.250 & 0.251 & 0.251 & 0.251 \\
	& & (0.033) & (0.027) & (0.011) & (0.008) & (0.004) &  & (0.024) & (0.020) & (0.007) & (0.005) & (0.003) \\
	\cline{1-13}
	0.9 & L_4 & 0.260 & 0.253 & 0.251 & 0.252 & 0.251 &  & 0.262 & 0.255 & 0.251 & 0.251 & 0.251 \\
	& & (0.036) & (0.030) & (0.011) & (0.008) & (0.004) &  & (0.024) & (0.021) & (0.007) & (0.005) & (0.003) \\
	& L_{\emptyset} & 0.257 & 0.248 & 0.249 & 0.252 & -^{\textrm{a}} &  & 0.257 & 0.250 & 0.250 & 0.251 & -^{\textrm{a}} \\ 
	& & (0.035) & (0.028) & (0.011) & (0.008) & -^{\textrm{a}} &  & (0.022) & (0.019) & (0.007) & (0.005) & -^{\textrm{a}} \\
	& L_{\textrm{full}} & 0.249 & 0.248 & 0.250 & 0.252 & 0.251 &  & 0.252 & 0.250 & 0.251 & 0.251 & 0.251 \\
	& & (0.034) & (0.028) & (0.011) & (0.008) & (0.004) &  & (0.023) & (0.019) & (0.007) & (0.005) & (0.003) \\
	\hline
	\end{array}
	\]
	\caption{
			\footnotesize 
			As for Table \ref{tab:cor_biv_int} but for estimates of the standard deviation $\sigma_2$.
			}
	\label{tab:sig2_biv_int}
\end{table}

\subsection{Estimates of $(\sigma_1, \rho, \sigma_2)$, from Section~\ref{ssec:biv_int}}
\label{suppmat:biv_int_est2}

Tables \ref{tab:biv_os_int_rho-07} and \ref{tab:biv_os_int_rho0}.

\begin{table}[h!]
	\footnotesize
	\[\arraycolsep=4pt\def\arraystretch{1.1}
	\begin{array}{r ll ccc c ccc}
	\multicolumn{1}{c}{} & & & \multicolumn{3}{c}{n_c=60} & & \multicolumn{3}{c}{n_c=300}  \\ 
	\cline{4-6} \cline{8-10} 
	\multicolumn{2}{c}{} & \textrm{Orders $(l,u)$} & \multicolumn{1}{c}{\sigma_1} & \multicolumn{1}{c}{\rho} & \multicolumn{1}{c}{\sigma_2} & & \multicolumn{1}{c}{\sigma_1} & \multicolumn{1}{c}{\rho} & \multicolumn{1}{c}{\sigma_2}  \\
	\cline{1-10}
	& L_{sn,x} 
	& ((6, 5),(55, 35)) &  0.4974 & -0.6912 &  0.5106 &  &  0.4998 & -0.6596 &  0.5040 \\
	& & & (0.0124) & (0.2625) & (0.0472) &  & (0.0060) & (0.2790) & (0.0410) \\
	& & ((16,6), (45,24)) &  0.4986 & -0.6933 &  0.5289 &  &  0.4994 & -0.6606 &  0.5144 \\ 
	& & & (0.0164) & (0.1854) & (0.0949) &  & (0.0075) & (0.2146) & (0.0856) \\
	& & ((20,5), (41,16)) &  0.5004 & -0.6699 &  0.5231 &  &  0.4993 & -0.6790 &  0.5201 \\
	& & & (0.0184) & (0.1963) & (0.0987) &  & (0.0080) & (0.1753) & (0.0919) \\
	\cline{3-10}
	& L_{sn,y} 
	& ((5,6), (35, 55)) &  0.4979 & -0.6423 &  0.4993 &  &  0.5006 & -0.6405 &  0.4984 \\
	& & & (0.0394) & (0.2486) & (0.0148) &  & (0.0364) & (0.2746) & (0.0061) \\
	& & ((6, 16),(24, 45)) &  0.5060 & -0.6447 &  0.4981 &  &  0.5223 & -0.6726 &  0.4985 \\
	& & & (0.0859) & (0.2168) & (0.0162) &  & (0.0910) & (0.2231) & (0.0078) \\
	& & (5, 20),(16, 41)) &  0.5054 & -0.6396 &  0.4991 &  &  0.5141 & -0.6451 &  0.4981 \\
	& & & (0.0999) & (0.1981) & (0.0206) &  & (0.1018) & (0.2205) & (0.0101) \\
	\cline{3-10}
	& L_{is,x} 
	& ((6,3), (55, 3)) &  0.4975 & -0.7133 &  0.4929 &  &  0.4999 & -0.7133 &  0.4896 \\ 
	& & & (0.0121) & (0.0497) & (0.0393) &  & (0.0060) & (0.0472) & (0.0320) \\ 
	& & ((16, 10),(45, 2)) &  0.4987 & -0.7325 &  0.4966 &  &  0.4994 & -0.7215 &  0.4983 \\ 
	& & & (0.0162) & (0.0932) & (0.0277) &  & (0.0075) & (0.1051) & (0.0248) \\
	& & ((20, 7),(41, 14)) &  0.5004 & -0.7108 &  0.4869 &  &  0.4993 & -0.7128 &  0.4771 \\
	& & & (0.0180) & (0.0363) & (0.0444) &  & (0.0080) & (0.0275) & (0.0453) \\
	\cline{3-10}
	& L_{is,y} 
	& ((3, 6),(3, 55)) &  0.4900 & -0.7130 &  0.4993 &  &  0.4915 & -0.7127 &  0.4984 \\ 
	& & & (0.0288) & (0.0517) & (0.0147) &  & (0.0326) & (0.0447) & (0.0061) \\ 
	& & ((10, 16),(2, 45)) &  0.4915 & -0.7327 &  0.4982 &  &  0.4955 & -0.7284 &  0.4985 \\
	& & & (0.0228) & (0.1020) & (0.0163) &  & (0.0238) & (0.0999) & (0.0077) \\
	& & ((7, 20),(14, 41)) &  0.4802 & -0.7155 &  0.4990 &  &  0.4850 & -0.7096 &  0.4981 \\
	& & & (0.0424) & (0.0335) & (0.0205) &  & (0.0401) & (0.0253) & (0.0101) \\
	\hline
	\end{array}
	\]
	\caption{\footnotesize 
	As for Table \ref{tab:biv_os_int_rho07} but with $\rho_0=-0.7$.}
	\label{tab:biv_os_int_rho-07}
\end{table}

\begin{table}[h!]
	\footnotesize
	\[\arraycolsep=2.4pt\def\arraystretch{1.1}
	\begin{array}{r ll ccc c ccc}
	\multicolumn{1}{c}{} & & & \multicolumn{3}{c}{n_c=60} & & \multicolumn{3}{c}{n_c=300}  \\ 
	\cline{4-6} \cline{8-10} 
	\multicolumn{2}{c}{} & \textrm{Orders $(l,u)$} & \multicolumn{1}{c}{\sigma_1} & \multicolumn{1}{c}{\rho} & \multicolumn{1}{c}{\sigma_2} & & \multicolumn{1}{c}{\sigma_1} & \multicolumn{1}{c}{\rho} & \multicolumn{1}{c}{\sigma_2}  \\
	\cline{1-10}
	& L_{sn,x} 
	& ((6,5),(55, 35)) & 0.4980 & \phantom{-}0.0183 & 0.5235 &  &  0.4998 & -0.0191 &  0.5216 \\
	& & & (0.0126) & (0.4156) & (0.0322) &  & (0.0059) & (0.3888) & (0.0301) \\
	& & ((16,6),(45, 24)) & 0.4968 & \phantom{-}0.0670 & 0.5329 &  &  0.5001 & -0.0172 &  0.5307 \\ 
	& & & (0.0157) & (0.3490) & (0.0612) &  & (0.0076) & (0.3375) & (0.0572) \\
	& & ((20,5), (41, 16)) & 0.4957 & \phantom{-}0.0847 & 0.5394 &  &  0.4990 & -0.0018 &  0.5355 \\
	& & & (0.0186) & (0.3747) & (0.0551) &  & (0.0099) & (0.3671) & (0.0508) \\
	\cline{3-10}
	& L_{sn,y} 
	& ((5,6), (35, 55)) & 0.5252 & \phantom{-}0.0024 & 0.4983 &  & 0.5235 & \phantom{-}0.0261 & 0.4995 \\ 
	& & & (0.0412) & (0.4303) & (0.0142) &  & (0.0306) & (0.4018) & (0.0058) \\
	& & ((6, 16),(24, 45)) &  0.5382 & -0.0048 &  0.4983 &  & 0.5359 & \phantom{-}0.0240 & 0.4986 \\
	& & & (0.0532) & (0.3863) & (0.0151) &  & (0.0558) & (0.3647) & (0.0063) \\ 
	& & ((5, 20),(16, 41)) &  0.5343 & -0.0024 &  0.5005 &  & 0.5434 & \phantom{-}0.0080 & 0.4984 \\ 
	& & & (0.0586) & (0.3645) & (0.0174) &  & (0.0569) & (0.3855) & (0.0089) \\
	\cline{3-10}
	& L_{is,x} 
	& ((6,3), (55, 3)) &  0.4980 & -0.0048 &  0.4856 &  & 0.4998 & \phantom{-}0.0008 & 0.4838 \\
	& & & (0.0126) & (0.0519) & (0.0546) &  & (0.0059) & (0.0205) & (0.0584) \\
	& & ((16, 10),(45, 2)) &  0.4968 & -0.0353 &  0.4777 &  &  0.5001 & -0.0260 &  0.4881 \\
	& & & (0.0157) & (0.0828) & (0.0520) &  & (0.0076) & (0.0653) & (0.0514) \\
	& & ((20,7),(41, 14)) & 0.4957 & \phantom{-}0.0214 & 0.4846 &  & 0.4990 & \phantom{-}0.0184 & 0.4829 \\
	& & & (0.0186) & (0.0618) & (0.0547) &  & (0.0099) & (0.0566) & (0.0527) \\ 
	\cline{3-10}
	& L_{i,sy} 
	& ((3, 6),(3,  55)) & 0.4830 & \phantom{-}0.0074 & 0.4984 &  & 0.4775 & \phantom{-}0.0004 & 0.4995 \\ 
	& & & (0.0538) & (0.0524) & (0.0141) &  & (0.0516) & (0.0272) & (0.0058) \\
	& & ((10,16), (2, 45)) &  0.5006 & -0.0055 &  0.4984 &  &  0.4762 & -0.0391 &  0.4986 \\ 
	& & & (0.0491) & (0.0752) & (0.0151) &  & (0.0563) & (0.0743) & (0.0063) \\
	& & ((7, 20),(14,  41)) & 0.4804 & \phantom{-}0.0270 & 0.5005 &  & 0.4852 & \phantom{-}0.0163 & 0.4984 \\
	& & & (0.0577) & (0.0697) & (0.0174) &  & (0.0494) & (0.0525) & (0.0089) \\ 
	\hline
	\end{array}
	\]
	\caption{\footnotesize 
	As for Table \ref{tab:biv_os_int_rho07} but with $\rho_0=0$.}
	\label{tab:biv_os_int_rho0}
\end{table}

\end{document}